\documentclass[11pt]{article}
\usepackage[utf8]{inputenc}
\usepackage[T1]{fontenc}
\usepackage[a4paper]{geometry}
\usepackage{microtype}
\usepackage{amsmath}
\usepackage{amssymb}
\usepackage{amsthm}
\usepackage{dsfont}
\usepackage{natbib}
\usepackage{hyperref}
\usepackage{bm}
\usepackage{tikz}
\usepackage{fourier}
\usepackage{booktabs}

\newtheorem{lemma}{Lemma}[section]
\newtheorem{proposition}[lemma]{Proposition}
\newtheorem{theorem}[lemma]{Theorem}
\newtheorem{assumption}[lemma]{Assumption}
\newtheorem{corollary}[lemma]{Corollary}

\theoremstyle{definition}
\newtheorem{definition}{Definition}[section]

\theoremstyle{remark}
\newtheorem{remark}[lemma]{Remark}

\newcommand{\prob}{\operatorname{P}}

\newcommand{\bW}{\bm{W}}
\newcommand{\bw}{\bm{w}}
\newcommand{\bx}{\bm{x}}
\newcommand{\bX}{\bm{X}}
\newcommand{\bY}{\bm{Y}}
\newcommand{\bZ}{\bm{Z}}

\newcommand{\bz}{\bm{z}}
\newcommand{\bzero}{\bm{0}}

\newcommand{\bbeta}{\bm{\beta}}

\newcommand{\reals}{\mathbb{R}}

\newcommand{\cbr}[1]{\left\{ {#1} \right\}}

\renewcommand{\ge}{\geqslant}
\renewcommand{\le}{\leqslant}
\renewcommand{\geq}{\geqslant}

\usepackage{xcolor}
\definecolor{royalblue(web)}{rgb}{0.25, 0.41, 0.88}
\definecolor{orange-red}{rgb}{1.0, 0.27, 0.0}

\newcommand \ag{\alpha}

\newcommand \Gg{\Gamma}

\newcommand \ep{\epsilon}

\newcommand \kg{\kappa}

\newcommand \sg{\sigma}
\newcommand \Sg{\Sigma}

\newcommand\bga{\mbox{\boldmath${\gamma}$}}

\newcommand\bSg{\mbox{\boldmath${\Sigma}$}}

 \newcommand\ba{{\bf a}}
 \newcommand\bb{{\bf b}}
 \newcommand\bc{{\bf c}}
 \newcommand\bd{{\bf d}}
\newcommand\mbE{{\mathbb E}}
\newcommand\mbR{{\mathbb R}}
 \newcommand\mbS{{\mathbb S}}

\newcommand\lp{\left (}
\newcommand\rp{\right )}
\newcommand\lb{\left [}
\newcommand\rb{\right ]}

\newcommand\lbr{\left \{}
\newcommand\rbr{\right \}}
\newcommand\lip{\left \langle}
\newcommand\rip{\right \rangle}

\def\convv{\stackrel{\mbox{$\scriptstyle v$}}{\rightarrow}}
\def\convP{\stackrel{\mbox{$\scriptstyle P$}}{\rightarrow}}

\newcommand{\cV}{\mathcal{V}}
\newcommand{\cS}{\mathcal{S}}
\newcommand\var{\hbox{\rm Var}}

\allowdisplaybreaks[4]

\begin{document}

\title{Hypothesis testing for partial tail correlation in multivariate extremes}

\author{Mihyun Kim\textsuperscript{1}\thanks{Department of Statistics, West Virginia University, Morgantown, WV 26506, USA.}
\and Jeongjin Lee\textsuperscript{1}\thanks{School of Mathematical Sciences, Lancaster University, Fylde College, Lancaster, LA1 4YF, United Kingdom. Corresponding author. E-mail: j.lee58@lancaster.ac.uk}
}

\maketitle

\begingroup
\renewcommand\thefootnote{1}
\footnotetext{Both authors, Mihyun Kim and Jeongjin Lee, contributed equally to this work.}
\endgroup

\begin{abstract}
Statistical modeling of high dimensional extremes remains challenging and has generally been limited to moderate dimensions.
Understanding structural relationships among variables at their extreme levels is crucial both for constructing simplified models and for identifying sparsity in extremal dependence. In this paper, we introduce the notion of partial tail correlation to characterize structural relationships between pairs of variables in their tails. To this end, we propose a tail regression approach for nonnegative regularly varying random vectors and define partial tail correlation based on the regression residuals. Using an extreme analogue of the covariance matrix, we show that the resulting regression coefficients and partial tail correlations take the same form as in classical non-extreme settings. For inference, we develop a hypothesis test to explore sparsity in extremal dependence structures, and demonstrate its effectiveness through simulations and an application to the Danube river network.
\smallskip

\end{abstract}

\section{Introduction} \label{s:int}

Characterizing and modeling high dimensional extremes is challenging and has largely been limited to moderate dimensions.
In the past few years,
there has been a concerted effort to develop tools to understand extremal structure among variables and to use this learned structure to develop simplified models for high dimensional extremes. An approach first studied by~\cite{gissibl2018max} constructs a max-linear model whose structure is given by a directed acyclic graph. Subsequent work has extended this approach in several directions:~\cite{gissibl2021identifiability} focus on estimation for these models;~\cite{amendola2022conditional} investigate conditional independence in max-linear networks;~\cite{kluppelberg2021estimating} explore its causal order; and~\cite{tran2021estimating} propose an efficient estimation algorithm. In another approach,~\cite{engelke2020graphical} develop the notion of conditional independence for a multivariate Pareto distribution and establish a connection to a graphical representation. They focus on the H{\" usler}-Reiss model in~\cite{husler1989} and show that its graphical structure coincides with sparsity in the inverse covariance matrices associated with this model. 
Focusing on extremal tree structures,~\cite{engelke2022structure} further propose a nonparametric estimation method to recover the underlying tree.

In this work, we develop the notion of {\it partial tail correlation}, a measure for characterizing and investigating structural relationships between the extremes of pairs of variables. We begin by introducing a new approach to regression for multivariate regularly varying random vectors $\bX$. It performs a tail regression of one component of $\bX$ on the others, generalizing the method with univariate covariates by Van Oordt and Zhou(~\citeyear{van2016systematic},~\citeyear{van2019estimating}). Our approach is developed within the framework of regular variation and does not impose structural assumptions, such as a space of transformed-linear combinations of regularly varying random variables as in~\cite{lee2021transformed}. Based on our regression formulation, we define {\it partial tail correlation} as the correlation between the tail residuals from regressing two components of $\bX$ on the remaining ones. 
This notion plays a crucial role in identifying sparsity of $X$ in the tail, as a zero partial tail correlation between two variables indicates no additional extremal relationship about one another, given the information provided by the other variables.
Our approach differs from the max-linear approach of~\cite{gissibl2018max} in that it is more closely linked to linear regression in the non-extreme context, and it is less model-based than~\cite{engelke2020graphical}, since we rely on the assumption of regular variation without specifying a model.

Our methodology has similar implications to the development of linear regression and partial correlation in non-extreme statistics, providing an intuitive analog for extremes. We briefly review the classical concept to provide context. For a comprehensive treatment on partial correlation, we refer to~\cite{anderson:2003},~\cite{whittaker:1990} and~\cite{lauritzen:1996}. When a distributional assumption is not made or one cannot fully characterize conditional distributions, partial correlation serves as a measure of the strength of the linear relationship between two variables after accounting for the remaining variables. Consider a centered random vector $\bX \in \mbR^p$ with covariance matrix $\Gg_{\bX}$. Partition to obtain $\bX=[\bX_K^{\top}, \bX_L^{\top}]^{\top}$, where $\bX_K=[X_i, X_j]^{\top}$ and $\bX_L = \bX_{\setminus (i,j)}$, and partition the covariance matrix accordingly
\[
\Gg_{\bX} = \begin{bmatrix}
\Gg_{KK} & \Gg_{KL} \\
\Gg_{LK} & \Gg_{LL}
\end{bmatrix} \in \mbR^{p \times p}, \quad\quad   
\Gg_{LK} = \begin{bmatrix}
\Gg_{L,i} & \Gg_{L,j}
\end{bmatrix} \in \mbR^{(p-2) \times 2}.
\]
The coefficient $\bbeta^{\ast}$ for the linear regression of $X_i$ on $\bX_L$, which minimizes the expected squared error $E[(X_i - \bbeta^{\top} \bX_L)^2]$, is given by
\begin{equation}\label{e:best-nonex}
\bbeta^{\ast} = \Gg_{LL}^{-1} \Gg_{L,i}.    
\end{equation}
Similarly, the best regression coefficient for $X_j$ on $\bX_L$ is given by $\bga^{\ast} = \Gg_{LL}^{-1} \Gg_{L,j}$. The partial correlation between $X_i$ and $X_j$ given $\bX_L$ is defined as the correlation of the residuals from their respective linear regressions on $\bX_L$. Denote the conditional covariance matrix of $\bX_K$ given $\bX_L$ by
\begin{align} \nonumber
\Gg_{K|L} &=  E \lb  (\bX_K - \Gg_{KL}\Gg_{LL}^{-1} \bX_L) (\bX_K - \Gg_{KL}\Gg_{LL}^{-1} \bX_L)^{\top} \rb \\ \label{e:cov_K|L_1}  
&= \Gg_{KK} - \Gg_{KL}\Gg_{LL}^{-1}\Gg_{LK}.
\end{align}
Then the partial correlation is defined as
\[
\eta_{ij \mid L} = \frac{[\Gg_{K|L}]_{12}}{\sqrt{[\Gg_{K|L}]_{11}[\Gg_{K|L}]_{22}}}.
\]
The range of $\eta_{ij \mid L}$ is $[-1,1]$. When $\eta_{ij \mid L}$ = 0, there is no additional linear relationship between $X_i$ and $X_j$ given $\bX_{L}$, while positive (negative) values indicate their positive (negative) association. The zero coefficient can be identified from the precision matrix via the relationship that $\eta_{ij \mid L}=0$ if and only if $[\Gg_{\bX}^{-1}]_{ij}=0$, provided that $\Gg_{\bX}^{-1}$ exists. Note that for  Gaussian random vectors $\bX$, a zero partial correlation between $X_i$ and $X_j$ implies their conditional independence, however this generally does not hold in non-Gaussian settings.

While the classical approach is developed for non-extreme data, our method focuses on high dimensional extremes and aims to identify their extremal sparsity structure. To begin, we introduce {\it tail regression} for multivariate extremes, a method designed to describe the extremal relationship between one component of $\bX$ and the others. In this regression, we assume that a nonnegative random vector $\bX \in \mathbb{R}_+^p := [0,\infty)^p$ is regularly varying, and we summarize the tail dependence of $\bX$ using tail pairwise dependence matrices (TPDMs), an extreme analogue of the covariance matrix introduced by~\cite{cooley:thibaud:2019}. Modeling within $\mbR_+^p$ provides a suitable framework for assessing risk in the direction of interest, but the challenge is that the least squares method is not directly applicable in $\mbR_+^p$. To address this, we perform regression using the transformed-linear operations introduced by~\cite{cooley:thibaud:2019}. This approach is equivalent to first mapping $\bX$ to $\mbR^p:=(-\infty, \infty)^p$ in a way that preserves its regular variation, performing linear regression in $\mbR^p$, and then mapping the results back to $\mbR_+^p$. We show that the regression coefficient obtained from this procedure attains~\eqref{e:best-nonex} using the TPDM of $\bX$.

Building on the residuals from the {\it tail regression}, we define the {\it partial tail correlation}, which quantifies the extremal association between two components of $\bX$ given the others. This measure is also described using the transformed-linear operations in $\mbR_+^p$. Similar to the non-extreme case, we prove that the conditional covariance matrix for extremes, which we use to define the {\it partial tail correlation}, achieves~\eqref{e:cov_K|L_1} with TPDMs. We then establish the relationship between the {\it partial tail correlation} and the corresponding coefficient from the {\it tail regression}. Furthermore, we connect the {\it partial tail correlation} to the inverse of the TPDM, showing that a zero element in the inverse matrix corresponds to a zero partial tail correlation. It is important to note that, as a trade-off for adopting a less model-based approach, we cannot go so far as to say that two variables with zero partial tail correlation are conditionally independent. 
Our method, however, allows partial tail relationships among variables in full generality, not limited to specific graphical representations, such as trees in~\cite{engelke2022structure}. This point is further illustrated in Section~\ref{sec:Danube}.

For statistical inference, we propose an estimator for the {\it partial tail correlation} and establish its asymptotic properties. Since the {\it partial tail correlation} is entirely defined through the TPDM, the estimator is constructed from an estimator of the TPDM. Its consistency under multivariate regular variation then follows from the consistency of the TPDM estimator, which is established via a series of weak convergence results for tail empirical measures. In addition, its asymptotic normality is proven under a second-order regular variation condition, which is a sufficient condition for that used in~\cite{larsson:resnick:2012}. Building on these theoretical results, we further develop a hypothesis test to assess whether partial tail correlations between variables are significantly different from zero, thus facilitating the statistical identification of tail associations in high-dimensional extremes. Additional and more closely related papers are introduced as our methods are developed.

We begin by reviewing the necessary theoretical background for our development, as outlined in Section~\ref{s:BG}. Section~\ref{s:TR} introduces the {\it tail regression} for multivariate extremes, and in Section~\ref{s:PTC}, we define the {\it partial tail correlation}. Section~\ref{s:E} introduces an estimator for the {\it partial tail correlation}, investigates its theoretical properties, and develops a hypothesis test for identifying sparsity patterns of $\bX$. Section~\ref{sec:simulation} provides simulation studies, and in Section~\ref{sec:Danube} we apply our analysis to extreme river discharges recorded at 31 gauging stations in the upper Danube basin. Theoretical justification of our approach requires additional background and technical derivations, which are provided in Sections~\ref{s:app} and~\ref{s:app2} of Appendix.

\section{Background} \label{s:BG}

Our methodology builds on the assumption that a random vector is regularly varying in the nonnegative orthant $\mbR_+^p$. The extremal dependence of such a vector can be summarized by its TPDM. For necessary background, we briefly review multivariate regular variation in Section~\ref{ss:MRV} and TPDM in Section~\ref{ss:TPDM}. Additionally, in Section~\ref{ss:TL} we review the transformed-linear operations, used to perform the {\it tail regression} and define the {\it partial tail correlation}.

\subsection{Multivariate Regular Variation} \label{ss:MRV}

Multivariate regular variation is fundamental to extreme value modeling, as it provides a mathematical framework for characterizing the asymptotic dependence structure of multivariate extremes, see e.g., Chapter 6 of~\cite{dehaan:ferreira:2006}. We refer to~\cite{resnick:2007} for a comprehensive treatment of regular variation.
Let $\mbE^p = [0,\infty]^p \setminus \{\bzero\}$. The space of Radon measures on $\mbE^p$ is denoted by $M_+(\mbE^p)$, with $\convv$ denoting vague convergence in $M_+(\mbE^p)$. A $p$--dimensional random vector $\bX = [X_1,\ldots,X_p]^{\top}$ in $\mbR_+^p$ is said to be regularly varying with index $-\ag$, $\ag >
0$, if there exists a sequence $a(s)\to\infty$ and an exponent measure $\nu_{\bX}$ in $M_+(\mbE^p)$ such that 
\begin{equation} \label{e:def1}
s P \lb \frac{\bX}{a(s)} \in \cdot \rb \convv \nu_{\bX},\quad\quad\quad s\to\infty.
\end{equation}
The tail index $\ag$ determines the power law of the tail, and the normalizing function is shown to be $a(s) = U(s)s^{1/\ag}$ where $U(s)$ is slowly varying at infinity.

The homogeneity property of $\nu_{\bX}$, i.e., $\nu_{\bX}(t \cdot) = t^{-\ag}\nu_{\bX}(\cdot)$, provides an equivalent formulation using a polar coordinate representation, see e.g., Theorem 6.1 of
\cite{resnick:2007}. Given any norm $\|\cdot\|$ on $\mbR^p$, define the unit sphere in the nonnegative orthant as $\Theta_+^{p-1} = \{ \bx \in \mbR^p : \| \bx \| = 1\} \cap \mbE^p$. Then, $\bX$ in $\mbR_+^p$ is regularly varying if
and only if there exists a sequence $b(s)\to\infty$ and a finite
measure $H_{\bX}$ on $\Theta_+^{p-1}$ such that in $M_+((0,\infty] \times \Theta_+^{p-1})$,
\begin{equation} \label{e:def2}
s P \lb \lp \frac{\|\bX\|}{b(s)}, \frac{\bX}{\|\bX\|} \rp  \in \cdot \rb \convv \nu_{\ag} \times H_{\bX},\quad\quad\quad s\to\infty,
\end{equation}
where $\nu_{\ag}(x,\infty] = x^{-\ag}$. The angular measure $H_{\bX}$ provides a full characterization of tail dependence in the limit, and the homogeneity property implies that $\nu_{\bX}(\{\bx \in \mbE^p: \|\bx\|>t, \bx/\|\bx\|\in S\}) = t^{-\ag} H_{\bX}(S)$, for measurable sets $S$ on $\Theta_+^{p-1}$. We denote the total mass of $H_{\bX}$ on $\Theta_+^{p-1}$ by
\begin{equation} \label{e:m}
m:=H_{\bX}(\Theta_+^{p-1}).    
\end{equation}

We write $\bX \in RV_+^p(\ag)$ if either~\eqref{e:def1} or~\eqref{e:def2} holds.



\subsection{Tail Pairwise Dependence Matrix} \label{ss:TPDM}

The angular measure $H_{\bX}$ fully captures the dependence structure in the joint upper tail of $\bX$; however, modeling $H_{\bX}$ is challenging in high dimensions. Rather than modeling $H_{\bX}$,~\cite{cooley:thibaud:2019} propose a matrix of bivariate measures that summarizes the strength of pairwise extremal dependence, providing attainable and useful information about tail dependence in high dimensions. Given $\bX = [X_1, \ldots, X_p]^{\top} \in RV_+^p(\ag)$ with angular measure $H_{\bX}$ on $\Theta_+^{p-1}$, the TPDM of $\bX$ is defined as
\begin{equation} \label{e:TPDM}
\Sg_{\bX} = \{\sg_{X, ij}\}_{i,j\in [p]}, \quad \quad \sg_{X, ij} = \int_{\Theta_+^{p-1}} w_iw_j H_{\bX}(d\bw),    
\end{equation}
where $[p]:=\{1, \ldots, p\}$. The element $\sg_{X, ij}$ quantifies the tendency of large values of $X_i$ and $X_j$ to occur simultaneously. It is based on the {\it extremal dependence measure}, introduced by~\cite{larsson:resnick:2012}, and has been generalized by~\cite{kiriliouk:chen:2022} through a modification of $\sg_{X, ij}$.

TPDMs can be viewed as an extreme analogue of covariance matrices, possessing several similar properties, as discussed in~\cite{cooley:thibaud:2019}.
First, $\Sg_{\bX}$ is positive semi-definite. In addition, the off-diagonal elements quantify the tail dependence between $X_i$ and $X_j$. Specifically, $\sg_{X, ij}=0$ if and only if (iff) $X_i$ and $X_j$ are asymptotically independent, and $\sg_{X, ij}>0$ iff they are asymptotically dependent. The diagonal elements $\sg_{X, ii}$ represent the `scale' of $X_i$, analogous to the diagonal entries of a covariance matrix representing individual variances. 

A useful property of TPDMs is that when $\alpha = 2$ and the $L_2$-norm is used to define $\Theta_+^{p-1}$, the sum of the diagonal entries of $\Sg_{\bX}$ equals the total mass of the angular measure $H_{\bX}$, i.e., 
\begin{equation}\label{e:mp}
\sum_{i=1}^p\sg_{X, ii}=m,    
\end{equation}
where $m$ is defined in~\eqref{e:def2}. This eliminates the need to separately estimate $m$ when estimating TPDMs. For example, when the marginals are scaled to have unit marginal scale, one can simply set $m = p$, as done in our data analysis.

\subsection{Transformed-Linear Operations} \label{ss:TL}

\cite{cooley:thibaud:2019} introduce transformed-linear operations that enable linear operations on the positive orthant $\mbR_+^p.$ Let $t$ be a monotone bijection from $\mbR$ to $\mbR_+$, where $t$ is understood to act elementwise when applied to vectors. The idea is to map elements in $\mbR_+^p$ to $\mbR^p$ via $t^{-1}$, perform standard linear operations in $\mbR^p$, and then map results back to $\mbR_+^p$ using $t$. Specifically, the transformed addition and scalar multiplication are defined as follows, respectively: for $\bx_1, \bx_2 \in \mbR_+^p$ and $a \in \mbR$,
\begin{equation} \label{e:tlo}
\bx_1 \oplus \bx_2 = t\{t^{-1}(\bx_1) +t^{-1}(\bx_2)\}, \quad\quad
a \circ x = t\{at^{-1}(x)\}.
\end{equation}
The additive identity is defined as $t(\bzero)$, and the additive inverse of $\bx \in \mbR_+^p$ is defined as $\ominus \bx = t\{ -t^{-1}(\bx)\}$. \cite{cooley:thibaud:2019} further define a scalar product on $\mbR_+^p$, which induces the norm as follows: for $\bx_1$, $\bx_2$, $\bx \in \mbR_+^p$, 
\begin{equation} \label{e:in}
\lip \bx_1, \bx_2 \rip_+ = t^{-1}(\bx_1)^{\top} t^{-1}(\bx_2), \quad\quad \|\bx\|_+^2 = \|t^{-1}(\bx)\|_2^2.
\end{equation}
In Supplementary material of~\cite{cooley:thibaud:2019}, it is shown that  $\mbR_+^p$ equipped with the operations in~\eqref{e:tlo} form a vector space, and when further equipped with the scalar product in~\eqref{e:in}, forms an inner product space. We develop our methodology  using~\eqref{e:tlo} and~\eqref{e:in} in Sections~\ref{s:TR} and~\ref{s:PTC}.


Following~\cite{cooley:thibaud:2019}, we apply the transformed-linear operations with $t(y) = \log \{1 + \exp(y)\}$ to nonnegative regularly varying random vectors. This transformation satisfies $\lim_{y \to \infty} t(y)/y = \lim_{x \to \infty} t^{-1}(x)/x =1$, ensuring that large values remain unchanged and thus preserving the extremal behavior. With the specified transformation, the marginals are additionally required to satisfy the following {\it lower tail condition} to fully preserve regular variation: 
\begin{equation} \label{e:lowt}
sP\lb X_i \le \exp(-cb(s))\rb \to 0, \quad \quad s \to \infty,
\end{equation}
where $c>0$ and $i \in [p]$. It precludes any of the marginals from placing nonzero mass at zero, ensuring that the upper tail remains unaffected under the transformed scalar multiplication for negative multipliers. This condition is met by standard regularly varying distributions such as the Fr{\' e}chet and Pareto distributions. Other transforms may be used with an appropriately adjusted lower tail condition in place of $t$. 

With the {\it lower tail condition}~\eqref{e:lowt}, the following lemma holds for the specified $t$, which is stated as Lemma A.1 in~\cite{cooley:thibaud:2019}. 

\begin{lemma}~\label{l:tinv}
Suppose that~\eqref{e:lowt} holds. Then, if $\bX \in RV_+^p(\ag)$ with exponent measure $\nu_{\bX}$ on $\mbE^p$, i.e., satisfying~\eqref{e:def1}, then $t^{-1}(\bX)$ is regularly varying with $-\ag$ in $\mbR^p$, and its limit measure $\nu_{t^{-1}(\bX)}$ on $[-\infty, \infty]^p/\{\bzero\}$ satisfies $\nu_{t^{-1}(\bX)}(\cdot) = \nu_{\bX}(\cdot \cap \mbE^p)$.
\end{lemma}

This result implies that the tail dependence structure of $\bX$ in the limit remains invariant under $t^{-1}$, i.e., $H_{t^{-1}(\bX)}= H_{\bX}$ on $\Theta_+^{p-1}$. This property is key for relating the TPDM of $\bX$ to the {\it tail regression} and the {\it partial tail correlation}, both of which can be described using $t^{-1}(\bX)$.

\section{Tail Regression} \label{s:TR}

This section introduces our approach to regression for regularly varying random vectors in $\mathbb{R}_+^p$. The key idea is to transform the vectors to $\mbR^p$ while preserving regular variation and to perform regression for extremes in $\mbR^p$. Thus, the TPDM needs to be represented using the transformed vectors, as it plays an essential role in the regression. This representation is provided in Section~\ref{ss:TPDM_TL}, followed by the {\it tail regression} in Section~\ref{ss:TR}.

\subsection{Representation of TPDMs} \label{ss:TPDM_TL}

As shown in Lemma~\ref{l:tinv}, the tail dependence structure of $\bX$ in the limit is preserved under $t^{-1}$. This allows the TPDM of $\bX$ to be expressed in terms of $t^{-1}(\bX)$. The following proposition shows that the $(i,j)$th element of $\Sg_{\bX}$ is the limit of the expected cross-moments of normalized $t^{-1}(X_i)$ and $t^{-1}(X_j)$, given that $t^{-1}(\bX)$ is large.

\begin{proposition}~\label{p:lcov}
Suppose that $\bX = [X_1, \ldots, X_p]^{\top} \in RV_+^p(\ag)$ with angular measure $H_{\bX}$ defined on $\Theta_+^{p-1}$ and satisfies~\eqref{e:lowt}. Then, 
the $(i,j)th$ element of $\Sg_{\bX}$, as defined in~\eqref{e:TPDM}, is given by
\[
\sg_{X, ij} = m\lim_{s\to \infty} E \lb \frac{t^{-1}(X_i)}{\|t^{-1}(\bX)\|_2} \frac{t^{-1}(X_j)}{\|t^{-1}(\bX)\|_2} \Bigg| \|t^{-1}(\bX)\|_2 > s \rb, \quad i,j \in [p],
\]
where $m=H_{\bX}(\Theta_+^{p-1})$, i.e., the total mass of $H_{\bX}$, as defined in~\eqref{e:m}.
\end{proposition}

\begin{proof}



To show the convergence, we apply Lemmas~\ref{l:tinv} and~\ref{l:Htinv}, which establish the limiting behavior of $t^{-1}(\bX)$. Let $(R^{\prime}, \bW^{\prime}) = (\|t^{-1}(\bX)\|_2, t^{-1}(\bX)/\|t^{-1}(\bX)\|_2)$. Then, it follows that
\[
mE \lb \frac{t^{-1}(\bX)_i}{\|t^{-1}(\bX)\|_2} \frac{t^{-1}(\bX)_j}{\|t^{-1}(\bX)\|_2} \Bigg| \|t^{-1}(\bX)\|_2 > s \rb =\int_{\Theta^{p-1}} w_i w_j \ mP [\bW^{\prime} \in d\bw | R^{\prime} > s].
\]
Consider $f: \Theta^{p-1} \to \mbR$ defined by $f(\bw) = w_i w_j$. The function $f$ is continuous with compact support. Thus, by Lemma~\ref{l:Htinv} and the definition of vague convergence, it follows that
\[
\lim_{s\to \infty} \int_{\Theta^{p-1}} w_i w_j \ mP \lb \bW^{\prime} \in d\bw | R^{\prime} >s\rb = \int_{\Theta^{p-1}} w_i w_j H_{t^{-1}(\bX)}(d\bw). 
\]
Since $t^{-1}(\bX)$ has no mass outside the nonnegative orthant in the limit and $H_{t^{-1}(\bX)} = H_{\bX}$ on $\Theta_+^{p-1}$ by Lemma~\ref{l:tinv}, we have that 
\[
\int_{\Theta^{p-1}} w_i w_j H_{t^{-1}(\bX)}(d\bw) = \int_{\Theta_+^{p-1}} w_i w_j H_{\bX}(d\bw) =\sg_{X, ij}.
\]

\end{proof}

Proposition~\ref{p:lcov} can be extended to represent the entire TPDM $\Sg_{\bX}$ of a regularly varying random vector $\bX$ as follows: 
\begin{equation} \label{e:lcov}
\Sg_{\bX} = m\lim_{s \to \infty} E \lb \frac{t^{-1}(\bX)}{\|t^{-1}(\bX)\|_2} \frac{t^{-1}(\bX)^{\top}}{\|t^{-1}(\bX)\|_2} \Bigg| \|t^{-1}(\bX)\|_2 > s \rb.  
\end{equation}
This representation corresponds to an extreme analogue of the covariance matrix, i.e., $E[\bY \bY^{\top}]$, where $\bY \in \mbR^{p}$ is a random vector.

\subsection{Multivariate Tail Regression} \label{ss:TR}

We consider the problem of regressing $X_p$ on $X_1, \ldots, X_{p-1}$, assuming that all variables are jointly regularly varying within the nonnegative orthant $\mbR_+^{p}$. To enable regression for extremes in $\mbR_+^{p}$, we use the transformed-linear operations outlined in Section~\ref{ss:TL}. This essentially involves transforming the variables to $\mbR^p$ while preserving their extremal behavior, performing linear regression focusing on extremes in $\mbR^p$, and then mapping the linear fit back to $\mbR_+^p$. It provides a transformed-linear fit for $X_p$ based on $X_1, \ldots, X_{p-1}$ at the extreme level. A remarkable feature of this approach is that the resulting regression coefficients can be computed directly from the TPDM of $\bX$ and take the same form as standard linear regression coefficients derived from a covariance matrix.

Suppose that $\bX=[\bX_{p-1}^{\top}, X_p]^{\top} \in RV_+^p(\ag)$, where $\bX_{p-1}=[X_1, \ldots, X_{p-1}]^{\top}$. Then, its TPDM can be expressed in the block matrix form
\begin{equation} \label{e:TPDM_X}
\Sg_{\bX}=
\left[ {\begin{array}{cc}
\Sg_{1:p-1, 1:p-1} & \Sg_{1:p-1, p} \\
\Sg_{p, 1:p-1} & \Sg_{p, p} \\
\end{array} } \right].    
\end{equation}
Using the transformed-linear operations, we define the magnitude of the regression error as
\[
\|X_p \ominus \bb^{\top} \circ \bX_{p-1}\|_+, \quad \bb \in \mbR^{p-1}.
\]
Our goal is to find the coefficient vector $\bb$ that minimizes the normalized regression error conditional on large values of $\bX$. Applying~\eqref{e:tlo} and~\eqref{e:in}, this error can be written as
\[
\|t^{-1}(X_p) - \bb^{\top} t^{-1}(\bX_{p-1})\|_2,    
\]
implying that the regression problem is equivalent to performing a linear regression of $t^{-1}(X_p)$ on $t^{-1}(\bX_{p-1})$ in $\mbR^p$ at extreme levels. The following theorem identifies the coefficient vector that provides the best fit in this {\it tail regression}. The resulting coefficients capture the extremal relationship between $X_p$ and $\bX_{p-1}$ and can thus be viewed as an extreme analogue of classical regression coefficients.

\begin{theorem}~\label{t:best}
Suppose that $\bX=[\bX_{p-1}^{\top}, X_p]^{\top} \in RV_+^p( \ag)$, where $\bX_{p-1}=[X_1, \ldots, X_{p-1}]^{\top}$, and satisfies~\eqref{e:lowt}. Assume that $\Sg_{\bX}$ in~\eqref{e:TPDM_X} is of full rank. Let $\bb^{\ast}$ be the value of $\bb \in \mbR^{p-1}$ that minimizes
\[
m\lim_{s \to \infty} E \lb  \frac{ \|X_p \ominus \bb^{\top} \circ \bX_{p-1} \|_+^2}{\|\bX\|_+^2} \Bigg| \|\bX\|_+ > s \rb,
\]
where $m$ is defined in~\eqref{e:m}. Then, $\bb^{\ast}$ is uniquely determined as $\Sg_{1:p-1,1:p-1}^{-1}\Sg_{1:p-1,p}$.
\end{theorem}


\begin{proof}

With the transformed-linear operations given in~\eqref{e:tlo} and~\eqref{e:in}, we have that
\begin{align*}
&E \lb  \frac{ \|X_p \ominus \bb^{\top} \circ \bX_{p-1} \|_+^2}{\|\bX\|_+^2} \Bigg| \|t^{-1}(\bX)\|_2> s \rb \\
&=E \lb  \frac{ \|t^{-1}(X_p) - \bb^{\top} t^{-1}(\bX_{p-1}) \|_2^2}{\|t^{-1}(\bX)\|_2^2} \Bigg| \|t^{-1}(\bX)\|_2> s \rb\\
&=E \lb  \frac{ t^{-1}(X_p)t^{-1}(X_p)}{\|t^{-1}(\bX)\|_2^2}  \Bigg| \|t^{-1}(\bX)\|_2> s \rb 
- E \lb \frac{t^{-1}(X_p)t^{-1}(\bX_{p-1})^{\top} }{\|t^{-1}(\bX)\|_2^2}  \Bigg| \|t^{-1}(\bX)\|_2> s \rb \bb \\
&\quad - \bb^{\top} E \lb \frac{t^{-1}(\bX_{p-1})t^{-1}(X_p)}{\|t^{-1}(\bX)\|_2^2}  \Bigg| \|t^{-1}(\bX)\|_2> s \rb 
+ \bb^{\top} E \lb \frac{t^{-1}(\bX_{p-1}) t^{-1}(\bX_{p-1})^{\top} }{\|t^{-1}(\bX)\|_2^2}  \Bigg| \|t^{-1}(\bX)\|_2> s \rb \bb
\end{align*}
By Proposition~\ref{p:lcov} and equivalently by~\eqref{e:lcov}, we have that
\begin{equation} \label{e:hess}
m\lim_{s \to \infty} E \lb  \frac{ \|X_p \ominus \bb^{\top} \circ \bX_{p-1} \|_+^2}{\|\bX\|_+^2} \Bigg| \|\bX\|_+ > s \rb
= \Sg_{p, p}
- \Sg_{p, 1:p-1} \bb - \bb^{\top} \Sg_{1:p-1, p}
+ \bb^{\top} \Sg_{1:p-1,1:p-1} \bb.   
\end{equation}
The Hessian matrix of the right-hand side of~\eqref{e:hess} is 2$\Sg_{1:p-1,1:p-1}$. Since $\Sg_{1:p-1,1:p-1}$ is invertible by assumption, the Hessian is positive definite. Therefore,~\eqref{e:hess} attains its unique minimum at $\bb^{\ast} = \Sg_{1:p-1,1:p-1}^{-1}\Sg_{1:p-1,p}$.

\end{proof}

This result indicates that the coefficient can be derived using the TPDM of $\bX$ and takes the same form as its non-extreme counterpart in~\eqref{e:best-nonex}. We emphasize that our result is obtained under the assumption that $\bX \in RV_+^p(\ag)$, without the need to construct a vector space of nonnegative random variables via transformed-linear combinations of independent regularly varying random variables as done in~\cite{lee2021transformed}. This makes our results applicable to a much broader class of models.


By applying the transformed scalar multiplication in~\eqref{e:tlo}, we obtain the best transformed-linear fit for $X_p$ from the {\it tail regression}, given by 
\[
\bb^{\ast \top} \circ \bX_{p-1}= t\{\Sg_{p, 1:p-1}\Sg_{1:p-1,1:p-1}^{-1} t^{-1}(\bX_{p-1})\} \in \mbR_+.
\]
The following remark provides a geometric interpretation of $\bb^{\ast \top} \circ \bX_{p-1}$.

\begin{remark} \label{r:best}

\noindent Let $\cV=\{\bb^{\top} \circ \bX_{p-1}: \bb \in \mbR^{p-1} \}$ denote the set of all transformed-linear combinations of $\bX_{p-1}$. Then, $\bb^{\ast \top} \circ \bX_{p-1}$ is the best approximation of $X_p$ within $\cV$ in the sense described in Theorem~\ref{t:best}, and can thus be seen as analogous to an orthogonal projection at the extreme level. To see this, observe that
\begin{equation} \label{e:orth}
m\lim_{s \to \infty} E \lb  \frac{ \lip \bb^{\top} \circ \bX_{p-1}, X_p \ominus \bb^{\ast \top} \circ \bX_{p-1} \rip_+ }{\|\bX\|_+^2} \Bigg| \|\bX\|_+ > s \rb
= 0, \quad\quad \forall \bb \in \mbR^{p-1}.      
\end{equation}
This implies that the residual $X_p \ominus \bb^{\ast \top} \circ \bX_{p-1}$ is orthogonal to all elements of $\cV$ in the tail. Consequently, a Pythagorean-type property holds:
\begin{align*}
&\lim_{s \to \infty} E \lb  \frac{ \|X_p\|_+^2}{\|\bX\|_+^2} \Bigg| \|\bX\|_+ > s \rb \\
&=\lim_{s \to \infty} E \lb  \frac{ \|\bb^{\ast \top} \circ \bX_{p-1}\|_+^2}{\|\bX\|_+^2} \Bigg| \|\bX\|_+ > s \rb + \lim_{s \to \infty} E \lb  \frac{ \| X_p \ominus \bb^{\ast \top} \circ \bX_{p-1} \|_+^2 }{\|\bX\|_+^2} \Bigg| \|\bX\|_+ > s \rb,   
\end{align*}
which supports the geometric interpretation.


\end{remark}

\section{Partial Tail Correlation} \label{s:PTC}

In Section~\ref{ss:PTC}, we introduce the concept of {\it partial tail correlation}, which is constructed using the residuals from the {\it tail regression} in Section~\ref{s:TR}. Section~\ref{ss:PTCTLC} explores its relationship with the corresponding coefficients from the regression.

\subsection{Partial Tail Correlation between Pairs of Variables} \label{ss:PTC}

We develop the notion of {\it partial tail correlation} for pairs of elements of a vector $\bX \in RV_+^p(\ag)$. It quantifies the tail dependence between two components in $\bX$ after accounting for the effect of the others.

To begin, partition $[p]$ into disjoint sets $K=\{i,j\}\in [p]$, $i\neq j$, and $L = [p]\setminus K$. With this partition, elements of $\bX$ can be rearranged as $[\bX_K^{\top}, \bX_L^{\top}]^{\top}$, where $\bX_K=[X_i, X_j]^{\top}$ and $\bX_L = \bX_{\setminus (i,j)}$. 
The TPDM of $\bX$ can then be expressed in the block matrix form
\begin{equation} \label{e:TPDM_KL}
\Sg_{\bX} = \begin{bmatrix}
\Sg_{KK} & \Sg_{KL} \\
\Sg_{LK} & \Sg_{LL}
\end{bmatrix}.    
\end{equation}

The partial tail correlation between $X_i$ and $X_j$ given $\bX_L$ is defined as the correlation of the tail residuals from the regression of $X_i$ and $X_j$ on $\bX_L$ at the extreme level. By Theorem~\ref{t:best}, these residuals are given by
\begin{equation} \label{e:RE}
\bX_K \ominus \Sg_{KL}\Sg_{LL}^{-1} \circ \bX_L \in \mbR_+^2, 
\end{equation}
which represents the remaining errors in $\bX_K$ after removing its tail dependence on $\bX_L$. Based on~\eqref{e:lcov}, we define the $2 \times 2$ tail conditional covariance matrix of $\bX_{K}$ given $\bX_{L}$ as
\begin{equation} \label{e:K|L}  
\Sg_{K|L} :=m\lim_{s \to \infty} E \lb  \frac{ \lip (\bX_K \ominus \Sg_{KL}\Sg_{LL}^{-1} \circ \bX_L)^{\top}, (\bX_K \ominus \Sg_{KL}\Sg_{LL}^{-1} \circ \bX_L)^{\top} \rip_+}{\|\bX\|_+^2} \Bigg| \|\bX\|_+ > s \rb, 
\end{equation}
where $m$ is defined in~\eqref{e:m}. Analogous to its non-extreme counterpart, the {\it partial tail correlation} is defined using $\Sg_{K|L}$.

\begin{definition} \label{d:ptc}
Given that $\bX = [\bX_K^{\top}, \bX_L^{\top}]^{\top} \in RV_+^{p}(\ag)$, where $\bX_K=[X_i, X_j]^{\top}$, $i \neq j\in [p]$,  and $\bX_L = \bX_{\setminus (i,j)}$, the partial tail correlation between $X_i$ and $X_j$ given $\bX_L$ is defined as
\[
\rho_{ij \mid L} = \frac{[\Sg_{K|L}]_{12}}{\sqrt{[\Sg_{K|L}]_{11}[\Sg_{K|L}]_{22}}},
\]
where $\Sg_{K|L}$ is defined as~\eqref{e:K|L}. 
\end{definition}

By applying the transformed-linear operations in~\eqref{e:tlo} and~\eqref{e:in} to~\eqref{e:K|L}, it can be shown that $\rho_{ij \mid L}$ corresponds to the correlation between the residuals from the regression of $t^{-1}(X_i)$ and $t^{-1}(X_j)$ on $t^{-1}(\bX_L)$ in $\mathbb{R}^p$ at the extreme level. Since $t^{-1}$ preserves the extremal behavior of $\bX$ as shown in Lemma~\ref{l:tinv}, the measure $\rho_{ij \mid L}$ represents the strength and direction of the extremal relationship between $X_i$ and $X_j$ that remains after accounting for the influence of $\bX_L$. Its value lies in the range $[-1,1]$, where nonzero values indicate an extremal association given $\bX_L$, with the sign denoting its direction. A zero partial tail correlation implies that $X_i$ and $X_j$ are partially uncorrelated given $\bX_L$ in the tail.

We conclude this section by presenting a useful finding on the tail conditional covariance matrix $\Sg_{K|L}$, followed by two remarks based on this result.
The following proposition demonstrates that the TPDM of $\bX$ determines $\Sg_{K|L}$. This property is analogous to its non-extreme counterpart in~\eqref{e:cov_K|L_1}.

\begin{proposition} \label{p:K|L}
Suppose that $\bX=[\bX_K^{\top}, \bX_L^{\top}]^{\top} \in RV_+^p(\ag)$, where $\bX_K=[X_i, X_j]^{\top}$, $i \neq j\in [p]$, and $\bX_L = \bX_{\setminus (i,j)}$, and $\bX$ satisfies~\eqref{e:lowt}. Assume that  $\Sg_{\bX}$ in~\eqref{e:TPDM_KL} is of full rank. Then, 
\[
\Sg_{K|L} =\Sg_{KK} - \Sg_{KL}\Sg_{LL}^{-1}\Sg_{LK},
\]
where $\Sg_{K|L}$ is defined as~\eqref{e:K|L}. 
\end{proposition}

\begin{proof}

Using the transformed-linear operations given in~\eqref{e:tlo} and~\eqref{e:in}, we have that
\begin{align*}
&\Sg_{K|L} \\
&=m\lim_{s \to \infty} E \lb  \frac{ \{t^{-1}(\bX_K) - \Sg_{KL}\Sg_{LL}^{-1} t^{-1}(\bX_L)\} \{t^{-1}(\bX_K) - \Sg_{KL}\Sg_{LL}^{-1} t^{-1}(\bX_L)\}^{\top}}{\|t^{-1}(\bX)\|_2^2} \Bigg| \|t^{-1}(\bX)\|_2  > s \rb \\
&=m\lim_{s \to \infty} \Bigg\{ E \lb  \frac{ t^{-1}(\bX_K)t^{-1}(\bX_K)^{\top}}{\|t^{-1}(\bX)\|_2^2}  \Bigg| \|t^{-1}(\bX)\|_2 > s \rb 
- E \lb \frac{t^{-1}(\bX_K)t^{-1}(\bX_L)^{\top} }{\|t^{-1}(\bX)\|_2^2}  \Bigg| \|t^{-1}(\bX)\|_2 > s \rb \Sg_{LL}^{-1}\Sg_{LK} \\
&\quad - \Sg_{KL}\Sg_{LL}^{-1} E \lb \frac{t^{-1}(\bX_L)t^{-1}(\bX_K)^{\top}}{\|t^{-1}(\bX)\|_2^2}  \Bigg| \|t^{-1}(\bX)\|_2 > s \rb \\
&\quad + \Sg_{KL}\Sg_{LL}^{-1} E \lb \frac{t^{-1}(\bX_L) t^{-1}(\bX_L)^{\top} }{\|t^{-1}(\bX)\|_2^2}  \Bigg| \|t^{-1}(\bX)\|_2 > s \rb \Sg_{LL}^{-1}\Sg_{LK} \Bigg\}.
\end{align*}
Then, by~\eqref{e:lcov}, it simplifies to $\Sg_{K|L} = \Sg_{KK} - \Sg_{KL}\Sg_{LL}^{-1}\Sg_{LK}$.

\end{proof}

\begin{remark} \label{r:ptc}

\noindent (i) Proposition~\ref{p:K|L} suggests that $\Sg_{K|L}$ can be expressed as a quadratic form in terms of the TPDM $\Sg_{\bX}$. Consider the block matrix
\begin{equation} \label{e:A}
A := \begin{bmatrix}
A_1 & A_2 
\end{bmatrix} = \begin{bmatrix}
I_{2\times2} & -\Sg_{KL}\Sg_{LL}^{-1} 
\end{bmatrix}^{\top}
\in \mbR^{p \times 2},
\end{equation}
where $A_1, A_2 \in \mbR^{p \times1}$ are the first and second column vectors of $A$, respectively. Then, it follows from~\eqref{e:TPDM_KL} that
\begin{equation} \label{e:aK|L}
\Sg_{K|L} = \Sg_{KK} - \Sg_{KL}\Sg_{LL}^{-1}\Sg_{LK} = A^{\top} \begin{bmatrix}
\Sg_{KK} & \Sg_{KL} \\
\Sg_{LK} & \Sg_{LL}
\end{bmatrix}A = A^{\top} \Sg_{\bX} A.
\end{equation}
This quadratic expression for $\Sg_{K|L}$ is particularly useful in Section~\ref{s:E}, where we introduce an estimator for $\rho_{ij \mid L}$ and derive its asymptotic properties.

\noindent (ii) It follows directly from Proposition~\ref{p:K|L} and Schur's formula for block-matrix inversion that under the assumptions in Proposition~\ref{p:K|L},
\[
\rho_{ij \mid L}=0 \quad\quad {\rm iff} \quad\quad [\Sg_{\bX}^{-1}]_{ij}=0.
\]
This extends the classical link between zero partial correlation and the inverse covariance matrix to the extreme setting.

\end{remark}

\subsection{Partial Tail Correlation and Tail Regression Coefficients} \label{ss:PTCTLC}

Suppose that $\bX_K=\{X_1,X_p\}$ and $\bX_L=\{X_2, \ldots, X_{p-1}\}$. Then, $\rho_{1p \mid L}$ quantifies the strength of the tail dependence between $X_1$ and $X_p$ after removing the effect of $\bX_L$. This suggests that the {\it partial tail correlation} is closely related to the regression coefficients $\bb^{\ast}$ from the {\it tail regression}, presented in Theorem~\ref{t:best}, as its first element measures the effect of $X_1$ on $X_p$ after controlling for the influence of $\bX_L$ in extremes. We formalize this relationship in the following theorem. While we continue to assume $\bX_K=\{X_1,X_p\}$ in the theorem for simplicity, the argument follows similarly when $\bX_K =\{X_i,X_j\}$ with $i \neq j  \in [p]$.

\begin{theorem} \label{t:PTC_TLC}
Suppose that $\bX=[\bX_K^{\top}, \bX_L^{\top}]^{\top} \in RV_+^p(\ag)$, where $\bX_K=[X_1, X_p]^{\top}$ and $\bX_L = \bX_{\setminus (1,p)}$, and $\bX$ satisfies~\eqref{e:lowt}. Assume that $\Sg_{\bX}$ in~\eqref{e:TPDM_KL} is of full rank. Then, $\rho_{1p \mid L} = 0$ if and only if $b_1^{\ast} = 0$, where $\bb^{\ast} = [b_1^{\ast}, \ldots, b_{p-1}^{\ast}]^{\top}$ denotes the regression coefficient from the {\it tail regression} of $X_p$ on $\bX_{p-1}= \{X_1, \ldots, X_{p-1}\}$, as specified in Theorem~\ref{t:best}.

\end{theorem}

\begin{proof}

It is equivalent to show that $[\Sg_{K|L}]_{12} = 0$ if and only if $b_1^{\ast} = 0$. We will first work on $[\Sg_{K|L}]_{12}$ and $b_1^{\ast}$ separately, and then establish their relationship.

Let $\bc^{\ast}, \bd^{\ast} \in \mbR^{p-2}$ be the coefficients resulting from the {\it tail regression} of $X_1$ on $\bX_L$ and $X_p$ on $\bX_L$, respectively. Then, it follows from~\eqref{e:K|L} that
\[
[\Sg_{K|L}]_{12} =m\lim_{s \to \infty} E \lb  \frac{ \lip X_1 \ominus \bc^{\ast \top} \circ \bX_L, X_p \ominus \bd^{\ast \top} \circ \bX_L \rip_+}{\|\bX\|_+^2} \Bigg| \|\bX\|_+ > s \rb.
\]
Let $\cV_L = \{\bb^{\top} \circ \bX_{L}: \bb \in \mbR^{p-1} \}$. Then, by~\eqref{e:orth} in Remark~\ref{r:best}, $X_1 \ominus \bc^{\ast \top} \circ \bX_L$ and $\bd^{\ast \top} \circ \bX_L$ are orthogonal at the extreme level. This implies that
\begin{equation} \label{e:SgKL12}
[\Sg_{K|L}]_{12}=m\lim_{s \to \infty} E \lb  \frac{ \langle X_1 \ominus \bc^{\ast \top} \circ \bX_L, X_p \rangle_+}{\|\bX\|_+^2} \Bigg| \|\bX\|_+ > s \rb.
\end{equation}

We now investigate $b_1^{\ast}$. To begin, consider the set spanned by $X_1 \ominus \bc^{\ast \top} \circ \bX_L$ and $\bX_L$. By Theorem~\ref{t:best} and Remark~\ref{r:best}, there exists a unique $\ba^{\ast} = [a_1^{\ast}, \ba_{2:p-1}^{\ast \top}]^{\top} \in \mbR^{p-1}$, where $\ba_{2:p-1}^{\ast} = [a_2^{\ast}, \ldots, a_{p-1}^{\ast}]^{\top}$, that provides the best approximation of $X_p$ within the set at the extreme level. The fitted $X_p$ is given by
\[
a_1^{\ast} \circ (X_1 \ominus \bc^{\ast \top} \circ \bX_L) \oplus \ba_{2:p-1}^{\ast \top} \circ \bX_L = a_1^{\ast} \circ X_1 \oplus (\ba_{2:p-1}^{\ast \top} \ominus a_1^{\ast} \circ \bc^{\ast \top}) \circ \bX_L. 
\]
Note that $X_1 \ominus \bc^{\ast \top} \circ \bX_L$ and $\bX_L$ span the same set as $\bX_{p-1}$ because the vectors in each set can be written as linear combinations of the others. Therefore, by the uniqueness of the regression coefficient established in Theorem~\ref{t:best}, we have $a_1^{\ast} = b_1^{\ast}$. The residual from the regression of $X_p$ is then given by
\[
X_p \ominus \{b_1^{\ast} \circ (X_1 \ominus \bc^{\ast \top} \circ \bX_L) \oplus \ba_{2:p-1}^{\ast \top} \circ \bX_L\}.
\]
Since $X_1 \ominus \bc^{\ast \top} \circ \bX_L$ belongs to the set spanned by $X_1 \ominus \bc^{\ast \top} \circ \bX_L$ and $\bX_L$, and the residual must be orthogonal to the set at the extreme level, we have that 
\[
m\lim_{s \to \infty} E \lb  \frac{ \langle X_p \ominus \{b_1^{\ast} \circ (X_1 \ominus \bc^{\ast \top} \circ \bX_L) \oplus \ba_{2:p-1}^{\ast \top} \circ \bX_L\}, X_1 \ominus \bc^{\ast \top} \circ \bX_L \rangle_+}{\|\bX\|_+^2} \Bigg| \|\bX\|_+ > s \rb =0.
\]
Also, since $\ba_{2:p-1}^{\ast \top} \circ \bX_L \in \cV_L$, $X_1 \ominus \bc^{\ast \top} \circ \bX_L$ and $\ba_{2:p-1}^{\ast \top} \circ \bX_L$ are orthogonal at the extreme level. This implies that 
\begin{equation}\label{e:SgKL12_1}
m\lim_{s \to \infty} E \lb  \frac{ \langle X_p \ominus \{b_1^{\ast} \circ (X_1 \ominus \bc^{\ast \top} \circ \bX_L)\}, X_1 \ominus \bc^{\ast \top} \circ \bX_L \rangle_+}{\|\bX\|_+^2} \Bigg| \|\bX\|_+ > s \rb=0.
\end{equation}

We turn to the relationship between $[\Sg_{K|L}]_{12}$ and $b_1^{\ast}$. It follows from~\eqref{e:SgKL12} and~\eqref{e:SgKL12_1} that
\begin{align*}
[\Sg_{K|L}]_{12} & =m\lim_{s \to \infty} E \lb  \frac{ \langle X_p, X_1 \ominus \bc^{\ast \top} \circ \bX_L \rangle_+}{\|\bX\|_+^2} \Bigg| \|\bX\|_+ > s \rb\\
& = m\lim_{s \to \infty} E \lb  \frac{ \langle b_1^{\ast} \circ (X_1 \ominus \bc^{\ast \top} \circ \bX_L), X_1 \ominus \bc^{\ast \top} \circ \bX_L \rangle_+}{\|\bX\|_+^2} \Bigg| \|\bX\|_+ > s \rb\\
& = b_1^{\ast} m\lim_{s \to \infty} E \lb  \frac{ \| X_1 \ominus \bc^{\ast \top} \circ \bX_L\|_+^2}{\|\bX\|_+^2} \Bigg| \|\bX\|_+ > s \rb.
\end{align*}
By substituting $\bc^{\ast} = \Sg_{LL}^{-1}\Sg_{L1}$, we obtain that $[\Sg_{K|L}]_{12} = b_1^{\ast}\lp \Sg_{11} - \Sg_{1L}\Sg_{LL}^{-1}\Sg_{L1}\rp$. Since $\det(\Sg_{\bX}) = \det(\Sg_{LL})\det(\Sg_{11} -\Sg_{1L}\Sg_{LL}^{-1}\Sg_{L1})$ by the Schur complement and $\Sg_{\bX}$ is assumed to be full rank, $\Sg_{11} -\Sg_{1L}\Sg_{LL}^{-1}\Sg_{L1} \neq 0$, which completes the proof.

\end{proof}

\section{Estimation of Partial Tail Correlation} \label{s:E}

We develop a statistical inference procedure for the {\it partial tail correlation}, which facilitates identifying sparse relationships among the components of $\bX$ in the tail. We introduce an estimator of $\rho_{ij \mid L}$ in Section~\ref{ss:EE} and establish its consistency in Section~\ref{ss:C}. Section~\ref{ss:HT} presents a hypothesis test to assess whether the tail dependence between two components in $\bX$ remains significant after accounting for the effects of others. 

\subsection{Estimator for $\rho_{ij \mid L}$} \label{ss:EE}

The estimation of the TPDM $\Sg_{\bX}$ is essential for computing the partial tail correlation $\rho_{ij \mid L}$. Recall from Definition~\ref{d:ptc} that $\rho_{ij \mid L}$ is defined as the standardized off-diagonal element of the tail conditional covariance matrix $\Sg_{K|L}$. According to Proposition~\ref{p:K|L} (or equivalently,~\eqref{e:aK|L}), $\Sg_{K|L}$ can be written as $\Sg_{K|L} = A^{\top} \Sg_{\bX}A$, where $A$ is constructed using submatrices of $\Sg_{\bX}$ as specified in~\eqref{e:A}. This formulation implies that, once $\Sg_{\bX}$ is estimated, $\rho_{ij \mid L}$ can be readily computed. Thus, we begin by presenting the estimator for $\Sg_{\bX}$ proposed by~\cite{cooley:thibaud:2019}.

Suppose that $\bX_l = [\bX_{lK}^{\top}, \bX_{lL}^{\top}]^{\top}$, for $1 \le l \le n$, are i.i.d. copies of $\bX = [\bX_K^{\top}, \bX_L^{\top}]^{\top}  \in RV_+^{p}(\ag)$, where $\bX_K=[X_i, X_j]^{\top}$, $i \neq j\in [p]$, and $\bX_L = \bX_{\setminus (i,j)}$. Let $(R_l, \bW_l) = (\|\bX_l\|, \bX_l / \|\bX_l\|)$ denote the radial and angular components of $\bX_l$. Since we are only interested in estimating $\Sg_{\bX}$, we do not use the transformed-linear operations. Any norm can be used to obtain $\{(R_l, \bW_l),1 \le l \le n \}$.

We first select the $k$ largest observations based on their norm $R_l$, and estimate the TPDM $\Sg_{\bX}$ using the following estimator:
\begin{equation} \label{e:Sgnk}
\widehat{\Sg}_{\bX}:=\widehat{\Sg}_{\bX}(n,k)
=\frac{m}{k} \sum_{l=1}^{n} \bW_l \bW_l^{\top} I_{R_{l} \ge R_{(k)}},
\end{equation}
where $R_{(k)}$ is the $k$th largest order statistics with the convention $R_{(1)} = \max\{R_{1}, \ldots, R_{n}\}$.  This estimator thus follows a standard peaks-over-threshold method in that it focuses on the extreme part of a sample and captures the structure of tail dependence in $\bX$. It generalizes the estimator for the {\it extremal dependence measure} introduced by~\cite{larsson:resnick:2012} to a $p$-dimensional regularly varying $\bX$.

Define the sample counterpart of $A$ in~\eqref{e:A} as
\begin{equation} \label{e:Ahat}
\widehat{A} := \begin{bmatrix}
\widehat{A}_1 & \widehat{A}_2 
\end{bmatrix} = \begin{bmatrix}
I_{2\times2} & -\widehat{\Sg}_{KL}\widehat{\Sg}_{LL}^{-1} 
\end{bmatrix}^{\top}
\in \mbR^{p \times 2},
\end{equation}
where $\widehat{A}_1, \widehat{A}_2 \in \mbR^{p \times1}$ are the first and second column vectors of $\widehat{A}$, respectively, and $\widehat{\Sg}_{KL}$ and $\widehat{\Sg}_{LL}$ are submatrices within $\widehat{\Sg}_{\bX}$ in~\eqref{e:Sgnk}. Based on Proposition~\ref{p:K|L} (or equivalently,~\eqref{e:aK|L}), we define the estimator for $\Sg_{K|L}$ as
\begin{equation} \label{e:est_K|L}
\widehat{\Sg}_{K|L}(n,k) :=\widehat{A}^{\top} \widehat{\Sg}_{\bX}\widehat{A}.    
\end{equation}
Finally, following Definition 4.1, the estimator for $\rho_{ij \mid L}$ is defined as
\begin{equation} \label{e:est_rho}
\hat{\rho}_{ij \mid L}(n,k) :=  \frac{[\widehat{\Sg}_{K|L}(n,k)]_{12}} {\sqrt{[\widehat{\Sg}_{K|L}(n,k)]_{11}}\sqrt{[\widehat{\Sg}_{K|L}(n,k)]_{22}} }. 
\end{equation}

As evident from its construction, the asymptotic properties of $\hat{\rho}_{ij \mid L}(n,k)$ largely depend on that of $\widehat{\Sg}_{\bX}$. Proposition~\ref{p:Xcon} below establishes that $\widehat{\Sg}_{\bX}$ consistently captures the tail dependence structure of $\bX$. To our knowledge, consistency for this specific form of $\widehat{\Sg}_{\bX}$ has not been previously established, so we provide the result here for completeness. This serves as a foundation for deriving asymptotic properties of $\hat{\rho}_{ij \mid L}(n,k)$ in Sections~\ref{ss:C} and~\ref{ss:HT}. We assume that the number of large observations $k$ increases with $n$ such that $k/n \to 0$, which is a standard condition in heavy-tailed analysis for deriving asymptotic properties. We assume throughout the paper that this condition holds.

\begin{proposition} \label{p:Xcon}

Suppose that $\bX_l = [X_{l1}, \ldots, X_{lp}]^{\top}$, $1 \le l \le n$, are i.i.d. copies of $\bX = [X_1, \ldots, X_p]^{\top}  \in RV_+^{p}(\ag)$. Then, we have that
\[
\widehat{\Sg}_{\bX} \convP \Sg_{\bX},
\]
where $\widehat{\Sg}_{\bX}$ and $\Sg_{\bX}$ are defined in~\eqref{e:Sgnk} and~\eqref{e:TPDM}, respectively. 
\end{proposition}

\begin{proof}

To prove the consistency of $\widehat{\Sg}_{\bX}$ for $\Sg_{\bX}$, we derive several weak convergence results for tail empirical measures via a sequence of continuous mappings.

By replacing $b(\cdot)$ with $b(m\cdot)$ in~\eqref{e:def2}, Theorem 5.3 (ii) of Resnick (2007) implies that
\[
\frac{m}{k}  \sum_{l=1}^{n} \ep_{(R_l/b(n/k), \bW_l)} \Rightarrow \nu_{\ag} \times H_{\bX}, \quad {\rm in}\ M_+((0, \infty] \times \Theta_+^{p-1}).    
\]
Additionally, by Lemma~\ref{l:R} (iii), we have that $R_{(k)}/b(n/k) \convP 1$. From these two convergence results, we obtain the joint weak convergence
\begin{equation} \label{e:Xcon1}
\lp \frac{m}{k} \sum_{l=1}^n  \ep_{(R_l/b(n/k), \bW_l)}, \frac{R_{(k)}}{b(n/k)}\rp \Rightarrow (\nu_{\ag} \times H_{\bX}, 1).   
\end{equation}
Consider the operator $G:M_+((0, \infty] \times \Theta_+^{p-1}) \times [0, \infty) \to M_+((0, \infty] \times \Theta_+^{p-1})$ defined by $G(U, x)(A \times B) = U(xA \times B)$. Then, $G$ is continuous at $(\nu_{\ag} \times H_{\bX},1)$ by Lemma 6 of~\cite{kim:kokoszka:2023b}. By applying $G$ to~\eqref{e:Xcon1}, it follows that
\begin{equation} \label{e:Xcon2}
\frac{m}{k} \sum_{l=1}^n  \ep_{(R_l/R_{(k)}, \bW_l)}\Rightarrow \nu_{\ag} \times H_{\bX}, \quad \quad {\rm in}\  M_+((0, \infty] \times \Theta_+^{p-1}).    
\end{equation}
Now, apply the continuous map $U \mapsto U([1, \infty] \times \cdot)$, from $M_+((0, \infty] \times \Theta_+^{p-1})$ to $M_+(\Theta_+^{p-1})$, to~\eqref{e:Xcon2}. It then follows that 
\begin{equation} \label{e:Xcon3}
\frac{m}{k} \sum_{l=1}^n  \ep_{\bW_l} I_{R_l \ge R_{(k)}} \Rightarrow H_{\bX},\quad\quad {\rm in}\ M_+(\Theta_+^{p-1}).    
\end{equation}
We then apply the continuous map $S \mapsto \int_{\Theta_+^{p-1}} \bw \bw^{\top} S(d\bw)$, from $M_+(\Theta_+^{p-1})$ to $\mathbb{R}^{p\times p}$, to~\eqref{e:Xcon3}, which implies that
\[
\widehat{\Sg}_{\bX}=\int_{\Theta_+^{p-1}} \bw \bw^{\top} \frac{m}{k} \sum_{l=1}^n  \ep_{\bW_l \in d\bw} I_{R_l \ge R_{(k)}}  \convP \int_{\Theta_+^{p-1}} \bw \bw^{\top} H_{\bX} (d\bw)=\Sg_{\bX}.
\]

\end{proof}

\subsection{Consistency of $\hat{\rho}_{ij \mid L}(n,k)$} \label{ss:C}

In this section, we show that $\hat{\rho}_{ij \mid L}(n,k)$ is consistent for $\rho_{ij \mid L}$. This follows from asymptotic properties of $\widehat{\Sg}_{K|L}(n,k)$ in~\eqref{e:est_K|L}, which we establish first. We begin by considering the following random matrix
\begin{equation} \label{e:est_K|L_p}
\widetilde{\Sg}_{K|L}(n,k) :=A^{\top}\widehat{\Sg}_{\bX}A,  
\end{equation}
where $A$ in~\eqref{e:A} is composed of submatrices of $\Sg_{\bX}$ in~\eqref{e:TPDM_KL}. The matrix $\widetilde{\Sg}_{K|L}(n,k)$ is not a statistic, as it involves $A$, which is a parameter. We work with $\widetilde{\Sg}_{K|L}(n,k)$ because it can be shown to be consistent for $\Sg_{K|L}$. We then prove that the difference between $\widehat{\Sg}_{K|L}(n,k)$ and $\widetilde{\Sg}_{K|L}(n,k)$ becomes asymptotically negligible, leading to the consistency of $\widehat{\Sg}_{K|L}(n,k)$ for $\Sg_{K|L}$. We work under the following assumptions.

\begin{assumption} \label{a:iid}
The random vectors $\bX_l = [\bX_{lK}^{\top}, \bX_{lL}^{\top}]^{\top}$, $1 \le l \le n$, are i.i.d. copies of $\bX = [\bX_K^{\top}, \bX_L^{\top}]^{\top}$, where $\bX_K = [X_i, X_j]^{\top}$ for $i \ne j \in [p]$, and $\bX_L = \bX_{\setminus (i,j)}$. We assume that $\bX \in RV_+^p(\alpha)$ and that its TPDM $\Sg_{\bX}$ is full rank.
\end{assumption}

The following proposition shows the consistency of $\widetilde{\Sg}_{K|L}(n,k)$ for the tail conditional covariance matrix $\Sg_{K|L}$ under the assumption.

\begin{proposition} \label{p:K|Lcon}
Under Assumption~\ref{a:iid}, we have that 
\[
\widetilde{\Sg}_{K|L}(n,k) \convP \Sg_{K|L},
\]
where $\widetilde{\Sg}_{K|L}(n,k)$ and $\Sg_{K|L}$ are defined in~\eqref{e:est_K|L_p} and~\eqref{e:K|L}, respectively. 
\end{proposition}

\begin{proof}

Consider the continuous mapping $\varphi:\mbR^{p\times p} \to \mbR^{2\times 2}$ defined by $\varphi(\Sg) = A^{\top}\Sg A$, for $\Sg \in \mbR^{p\times p}$. From Proposition~\ref{p:Xcon}, it follows that applying $\varphi$ to $\widehat{\Sg}_{\bX} \convP \Sg_{\bX}$ leads to
\[
\widetilde{\Sg}_{K|L}(n,k)=A^{\top}\widehat{\Sg}_{\bX}A \convP A^{\top}\Sg_{\bX}A = \Sg_{K|L},
\]
where the last equality follows from~\eqref{e:aK|L}.

\end{proof}

By proving that $\widehat{\Sg}_{K|L}(n,k)$ is asymptotically close to $\widetilde{\Sg}_{K|L}(n,k)$, we establish the consistency of  $\widehat{\Sg}_{K|L}(n,k)$ for $\Sg_{K|L}$. This result is presented in the following theorem.

\begin{theorem} \label{t:K|Lcon}
Under Assumption~\ref{a:iid}, we have that 
\[
\widehat{\Sg}_{K|L}(n,k) \convP \Sg_{K|L},
\]
where $\widehat{\Sg}_{K|L}(n,k)$ and $\Sg_{K|L}$ are defined in~\eqref{e:est_K|L} and~\eqref{e:K|L}, respectively. 
\end{theorem}

\begin{proof}

Recall the definitions of $A$ and $\widehat{A}$ from~\eqref{e:A} and~\eqref{e:Ahat}. Observe that
\begin{align*}
&\widehat{\Sg}_{KL}\widehat{\Sg}_{LL}^{-1} - {\Sg}_{KL}{\Sg}_{LL}^{-1} \\
& = (\widehat{\Sg}_{KL}- {\Sg}_{KL})(\widehat{\Sg}_{LL}^{-1}- {\Sg}_{LL}^{-1})  +(\widehat{\Sg}_{KL}- {\Sg}_{KL}){\Sg}_{LL}^{-1} +{\Sg}_{KL}(\widehat{\Sg}_{LL}^{-1}- {\Sg}_{LL}^{-1}) \convP 0.
\end{align*}
The convergence follows from Proposition~\ref{p:Xcon} and Lemma~\ref{l:Sgeq}, and thus we conclude that $\widehat{A} - A \convP 0$. Now, observe that
\begin{align*}
\widehat{\Sg}_{K|L}(n,k) - \widetilde{\Sg}_{K|L}(n,k) &= \widehat{A}^{\top} \widehat{\Sg}_{\bX}\widehat{A} - A^{\top} \widehat{\Sg}_{\bX} A\\
&= (\widehat{A} - A)^{\top} \widehat{\Sg}_{\bX} (\widehat{A} - A)  + (\widehat{A} - A)^{\top} \widehat{\Sg}_{\bX} A + A^{\top} \widehat{\Sg}_{\bX} (\widehat{A} - A).
\end{align*}
Since $\widehat{\Sg}_{\bX} \convP \Sg_{\bX}$ by Proposition~\ref{p:Xcon} and $\widehat{A} - A \convP 0$, it follows that $\widehat{\Sg}_{K|L}(n,k) - \widetilde{\Sg}_{K|L}(n,k) \convP 0$. Combining this with Proposition~\ref{p:K|Lcon} completes the proof.

\end{proof}

The consistency of the estimator for the partial tail correlation $\rho_{ij \mid L}$ readily follows from Theorem~\ref{t:K|Lcon}, and thus the proof is omitted.

\begin{corollary} \label{c:PTCcon}
Under Assumption~\ref{a:iid}, we have that 
\[
\hat{\rho}_{ij \mid L}(n,k) \convP \rho_{ij \mid L},
\]
where $\hat{\rho}_{ij \mid L}(n,k)$ and $\rho_{ij \mid L}$ are defined in~\eqref{e:est_rho} and Definition~\ref{d:ptc}, respectively. 
\end{corollary}

\subsection{Hypothesis test for $\rho_{ij \mid L}$} \label{ss:HT}

For statistical inference, we develop a hypothesis testing procedure to evaluate the significance of the extremal relationship between $X_i$ and $X_j$ given $\bX_L$.

According to  Definition~\ref{d:ptc}, testing whether the partial tail correlation $\rho_{ij \mid L} =0$ is equivalent to assessing the following hypothesis
\begin{equation} \label{e:HT}
H_0: [\Sg_{K|L}]_{12} =0,
\end{equation}
where $[\Sg_{K|L}]_{12}$ represents the off-diagonal of the tail conditional covariance matrix $\Sg_{K|L}$ in~\eqref{e:K|L}. We thus use $[\widehat{\Sg}_{K|L}(n,k)]_{12}$ in~\eqref{e:est_K|L}, as a statistic, and establish its asymptotic normality under the null hypothesis~\eqref{e:HT}.

Typically, deriving asymptotic normality in heavy-tailed analysis requires the concept of second--order regular variation, see e.g., \cite{haeusler:teugels:1985}, \cite{csorgo:deheuvels:mason:1985}, \citeauthor{resnick:starica:1997a} (\citeyear{resnick:starica:1997a}, \citeyear{resnick:starica:1997b}). 
It is considered a refinement of regular variation, which allows a more precise characterization of the convergence rate of tail probabilities to its limit measure. We use a multivariate extension of second-order regular variation, which is a sufficient condition for the one considered in \cite{larsson:resnick:2012}. This condition ensures that the radial and angular components of $\bX$ become independent at a sufficiently fast rate, as specified below.

\begin{assumption} \label{a:2nd}
The random vectors $\bX_l = [\bX_{lK}^{\top}, \bX_{lL}^{\top}]^{\top}$, $1 \le l \le n$, are i.i.d. copies of $\bX = [\bX_K^{\top}, \bX_L^{\top}]^{\top}$, where $\bX_K = [X_i, X_j]^{\top}$ for $i \ne j \in [p]$, and $\bX_L = \bX_{\setminus (i,j)}$. We assume that there exist a sequence $b(s)\to\infty$ and an angular measure $H_{\bX}$ on $\Theta_+^{p-1}$ such that the radial and angular components of $\bX$, i.e.,$(R, \bW)$, satisfy
\begin{equation}\label{e:2nd}
\sqrt{k}\lbr \frac{n}{k} P \lb \lp \frac{R}{b(n/k)}, \bW  \rp \in \cdot \rb - \nu_{\ag}\times  \frac{1}{m}H_{\bX}\rbr \convv 0, \quad\quad {\rm in}\ M_+((0,\infty] \times \Theta_+^{p-1}). 
\end{equation}
The TPDM $\Sg_{\bX}$ in~\eqref{e:TPDM_KL} is assumed to be of full rank.
\end{assumption}

Under the second-order regular variation condition, the following theorem establishes the asymptotic normality of $[\widehat{\Sg}_{K|L}(n,k)]_{12}$ under the null hypothesis~\eqref{e:HT}. The main idea of the proof is based on convergence of empirical processes, and the full proof is provided in Appendix~\ref{s:app2} due to its length.

\begin{theorem}~\label{t:AN}
Under Assumption~\ref{a:2nd} and null hypothesis~\eqref{e:HT}, we have that
\[
\sqrt{k}[\widehat{\Sg}_{K|L}(n,k)]_{12} \Rightarrow N(0, \tau^2),
\]
where $\widehat{\Sg}_{K|L}(n,k)$ is defined in~\eqref{e:est_K|L}, and $\tau^2= m^2 \var[A_1^{\top}\widetilde{\bW} \widetilde{\bW}^{\top}A_2]$. Here,  $\widetilde{\bW}$ is a random vector on $\Theta_+^{p-1}$ with $P[\widetilde{\bW} \in \cdot] = m^{-1} H_{\bX}$, $A_1$ and $A_2$ are defined in~\eqref{e:A}, and $m$ is defined in~\eqref{e:m}.
\end{theorem}

The asymptotic normality of $\hat{\rho}_{ij \mid L}$ when $\rho_{ij \mid L}=0$ follows directly from combining Theorems~\ref{t:K|Lcon} and~\ref{t:AN}, hence the proof is omitted.

\begin{corollary}~\label{c:AN}
Under Assumption~\ref{a:2nd} and null hypothesis~\eqref{e:HT}, we have that
\[
\sqrt{k}\hat{\rho}_{ij \mid L}(n,k) \Rightarrow N\lp 0, \frac{\tau^2}{[\Sg_{K|L}]_{11}[\Sg_{K|L}]_{22}} \rp,
\]
where $\hat{\rho}_{ij \mid L}$ is defined in~\eqref{e:est_rho} and $\Sg_{K|L}$ is given in~\eqref{e:K|L}.
\end{corollary}

Although a hypothesis test for $\rho_{ij \mid L}=0$ can be constructed based on Corollary~\ref{c:AN}, we propose a test for $[\Sg_{K|L}]_{12} =0$ in~\eqref{e:HT}, as it requires estimating only $\tau^2$ for the variance, resulting in reduced estimation error. From~\eqref{e:HT}, we obtain that
\[
\tau^2= m^2 \var[A_1^{\top}\widetilde{\bW} \widetilde{\bW}^{\top}A_2] = m^2 E[(A_1^{\top}\widetilde{\bW} \widetilde{\bW}^{\top}A_2)^2]. 
\]
Therefore, $\tau^2$ can be estimated by
\begin{equation}\label{e:tau2}
\hat{\tau}^2 := \frac{m^2}{k} \sum_{l=1}^{n} (\widehat{A}_1^{\top}\bW_l \bW_l^{\top} \widehat{A}_2)^2 I_{R_{l} \ge R_{(k)}},    
\end{equation}
and the test rejects the null hypothesis~\eqref{e:HT} if
\begin{equation}\label{e:test}
z_{n,k}:=\frac{\sqrt{k}}{\hat{\tau}}|[\widehat{\Sg}_{K|L}(n,k)]_{12}|> z_q,
\end{equation}
where $z_q$ is the upper
quantile of the standard normal distribution  denoted by $\Phi(z_q) = 1- q$, $q \in [0,1]$. 

The total mass $m$ must be computed to obtain $\hat{\tau}^2$ in~\eqref{e:tau2} and $\widehat{\Sg}_{\bX}$ in~\eqref{e:Sgnk}. It
can be replaced by the dimension $p$ when the data are suitably pre-processed, as shown in~\eqref{e:mp}. Otherwise, $m$ can be estimated using the empirical estimator $\frac{1}{k} \sum_{l=1}^n I_{R_l \ge (n/k)^{1/\ag}}$. This is motivated by the asymptotic relation $\frac{n}{k}P[R \ge b(n/k)] \to m$, which follows from~\eqref{e:def2} with $b(\cdot)$ replaced by $b(m\cdot)$, and $b(t) \sim t^{1/\ag}$. 

We conclude this section by proving the test's consistency under the alternative $[\Sg_{K|L}]_{12} \neq 0$, which justifies its ability to detect deviations from the null.
\begin{theorem}~\label{t:PO}
Under Assumption~\ref{a:2nd} with the alternative $[\Sg_{K|L}]_{12} \neq 0$, we have that
\[
\sqrt{k}|[\widehat{\Sg}_{K|L}(n,k)]_{12}| \convP \infty.
\]
\end{theorem}

\begin{proof}
From~\eqref{e:Ahat} and~\eqref{e:est_K|L}, it follows that 
\begin{align*}
&\sqrt{k}[\widehat{\Sg}_{K|L}(n,k)]_{12} \\
&= \sqrt{k}(\widehat{A}_1 - A_1)^{\top} \widehat{\Sg}_{\bX}\widehat{A}_2 + A_1^{\top} \widehat{\Sg}_{\bX} \sqrt{k}(\widehat{A}_2 - A_2) + A_1^{\top} \sqrt{k}(\widehat{\Sg}_{\bX} - \Sg_{\bX}) A_2 + \sqrt{k} A_1^{\top} \Sg_{\bX} A_2.
\end{align*}
All terms except the last converge to zero in probability by Proposition~\ref{p:kXcon} and Lemma~\ref{l:Sgeq}. The last term diverges because $A_1 \Sg_{\bX} A_2 = [\Sg_{K|L}]_{12} \neq 0$, which completes the proof.

\end{proof}

\section{Simulation study}
\label{sec:simulation}

We conduct simulation studies to evaluate the finite-sample performance of the proposed estimators and hypothesis test developed in Section~\ref{s:E}.

\subsection{Estimation}
\label{sec:estimation}
Following the estimation procedure in Section~\ref{ss:EE}, we assess the bias and variability of the proposed estimators: $\widehat{\Sigma}_{\bX}(n,k)$ in~\eqref{e:Sgnk}, $\widehat{\Sigma}_{K|L}(n,k)$ in~\eqref{e:est_K|L}, and $\hat{\rho}_{ij|L}(n,k)$ in~\eqref{e:est_rho} across varying $(n,k)$ under 1,000 repeated iterations.

Let $\bm{X}=[X_1,\ldots,X_p]^\top$ in $\reals_+^p$.
To induce positive extremal dependence among its components, we employ the matrix-based construction of Corollary 1 in \cite{cooley:thibaud:2019}.
Specifically, let $C=[c_{ij}]=(\bc_1,\ldots,\bc_q)\in \reals^{p\times q}$ be a matrix with $\max_{i=1,\ldots,p}c_{ij}>0$ for all $j=1,\ldots,q$.
We then construct $\bX$ with positive dependence via the transformed matrix multiplication
\begin{equation*}
    \bX=C\circ \bZ,
\end{equation*}
where $\bZ=(Z_1,\ldots,Z_q)^\top$ is a vector of i.i.d. $Z_j\in RV_+^1(\alpha)$ meeting the lower tail condition~\eqref{e:lowt}, and sharing a common scaling, $s\prob(Z_j>b(s) z)\rightarrow z^{-\alpha}$ as $s\rightarrow \infty$.
Then, $\bX\in RV_+^p(\alpha)$ and its angular measure is $H_{\bX}(\cdot)=\sum_{j=1}^q \|\bc_j^{(0)}\|^\alpha\delta_{\bc_j^{(0)}/\|\bc_j^{(0)}\|}(\cdot),$ where $\delta(\cdot)$ denotes the Dirac measure and $c_{ij}^{(0)}:=\max(c_{ij},0)$ is applied componentwise.
Setting $\alpha=2$ and defining the angular measure with respect to the $L_2$-norm yields a particularly simple and convenient TPDM form $\Sigma_{C\circ \bZ}=C^{(0)}C^{(0)^\top},$ where $C^{(0)}=[c_{ij}^{(0)}]$.

To generate multivariate samples from $\bX=C\circ \bZ$, we specifically define $Z_j=(1-U_j)^{-1/2}-\delta$ for $j\in[q]$, where $U_j$ is the uniform distribution on $[0,1]$.
This yields a shifted unit-scale Pareto distribution with $\alpha=2$, so that $Z_j\in RV_+^1(2)$.
The shift $\delta=0.9352$ is chosen such that $E[t^{-1}(Z_j)]=0$, centering the preimages of the transformed data which we found reduces bias in TPDM estimation.
We consider a non-negative matrix $C\in\reals^{7\times 30}$ whose elements are a specific realization from a uniform distribution on $[0,5]$, and normalize each row to have unit norm.
Then, the specified TPDM is $\Sigma_{\bX}=CC^\top$.
We generate $n$ independent shifted Pareto random samples $\bZ_1,\ldots,\bZ_n$ and obtain $\bX_1,\ldots,\bX_n$ via the transformed matrix multiplication.
Under the $L_2$-norm, we decompose each observation into radial and angular components $\{(R_l,\bW_l),1\le l \le n\}$, and retain the threshold exceedances $\{(R_l,\bW_l):R_l\ge R_{(k)}\}$ for TPDM estimation, where $R_{(k)}$ is the $k$th largest order statistics.

\begin{figure}[ht]
\centering
\includegraphics[width=4cm]{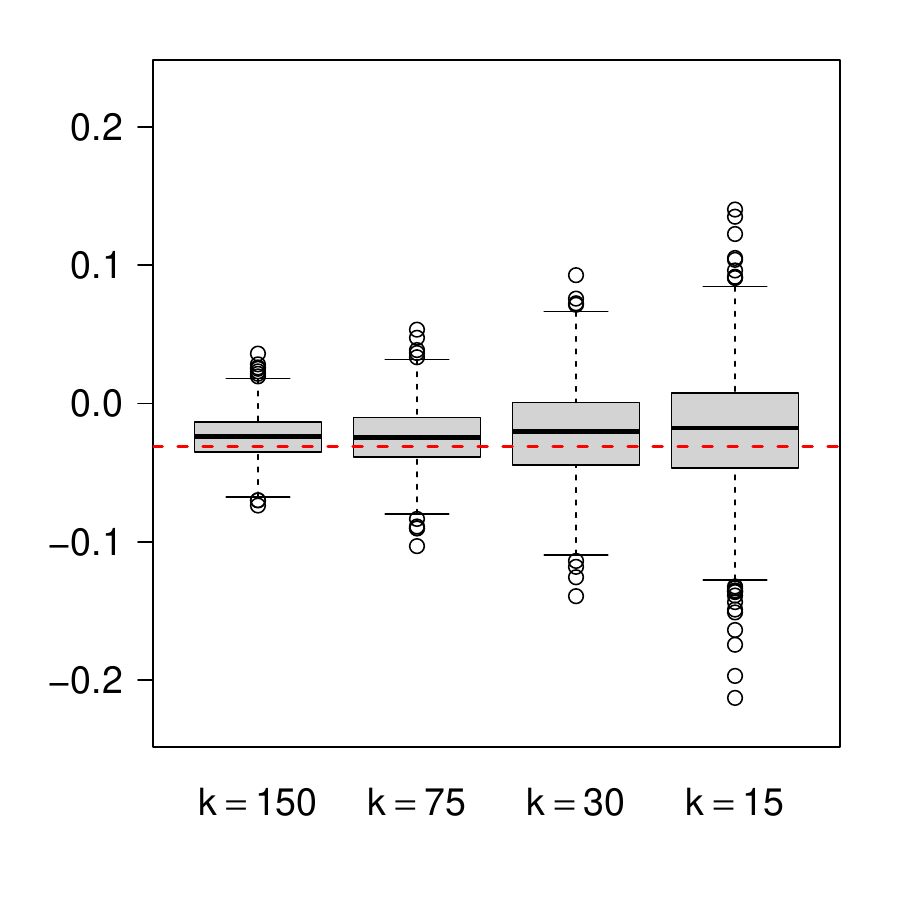}
\includegraphics[width=4cm]{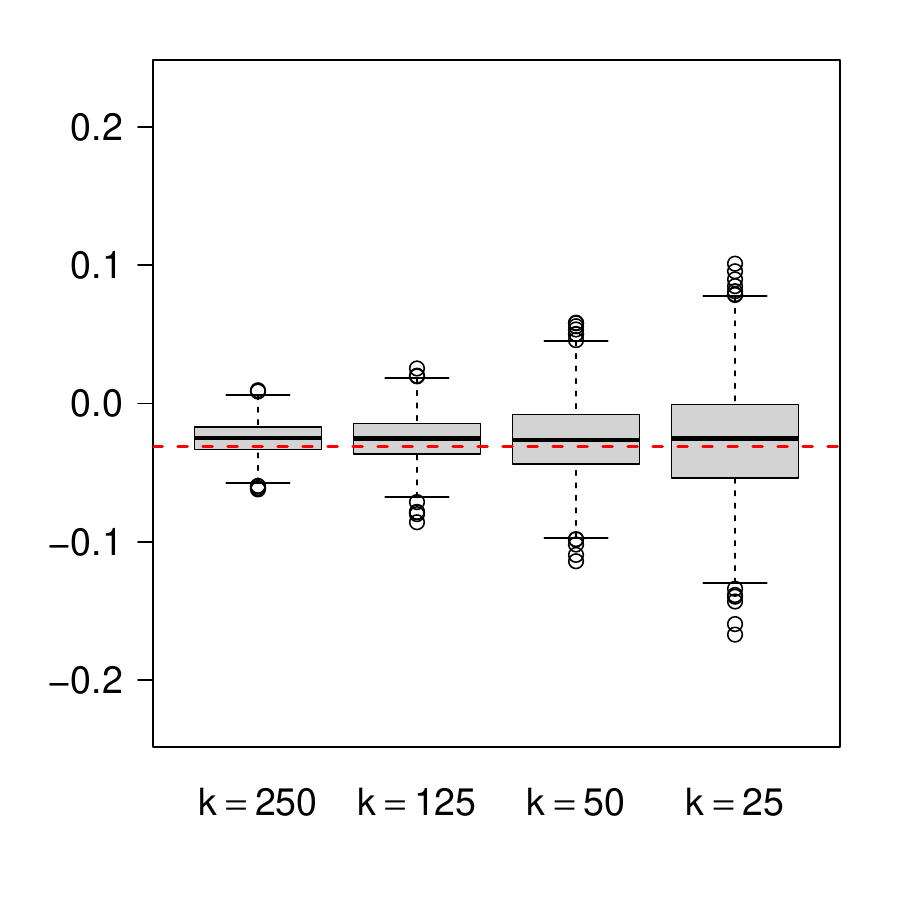}
\includegraphics[width=4cm]{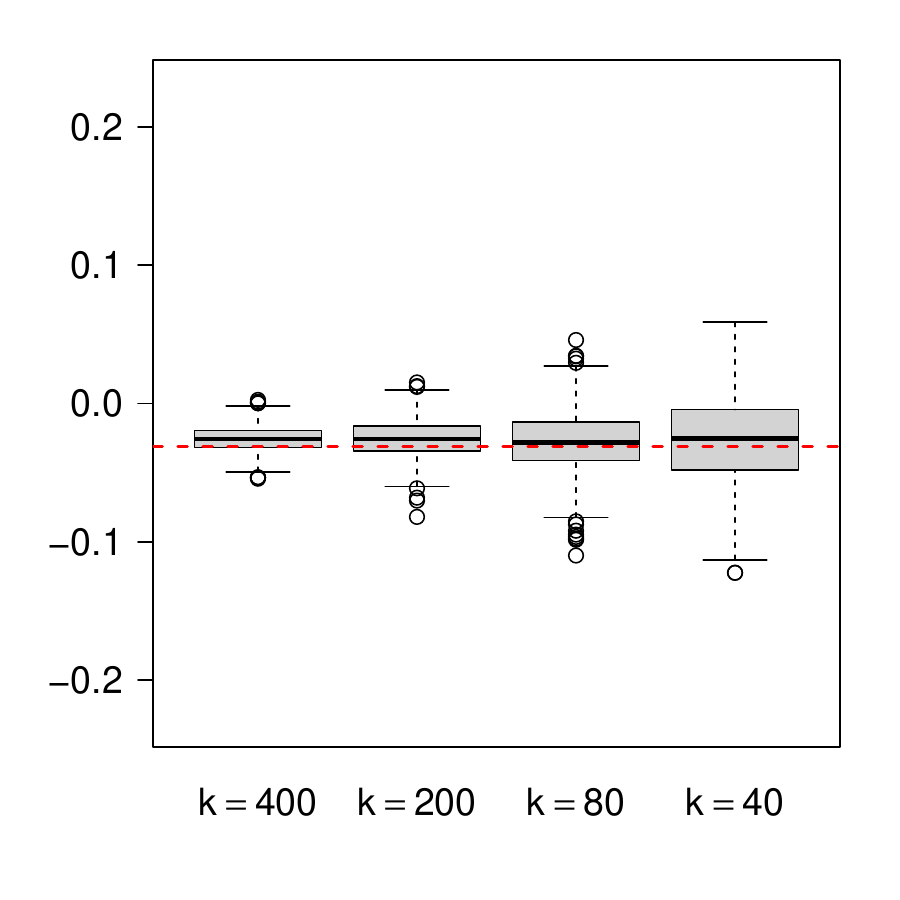}\\
\includegraphics[width=4cm]{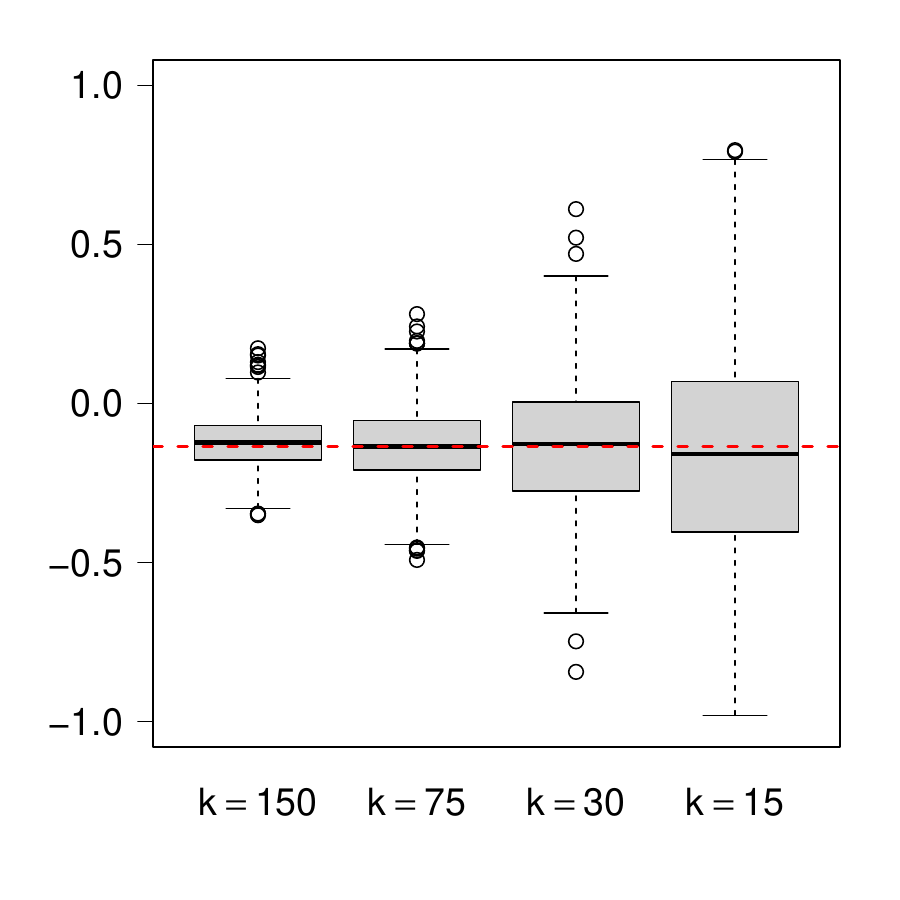}
\includegraphics[width=4cm]{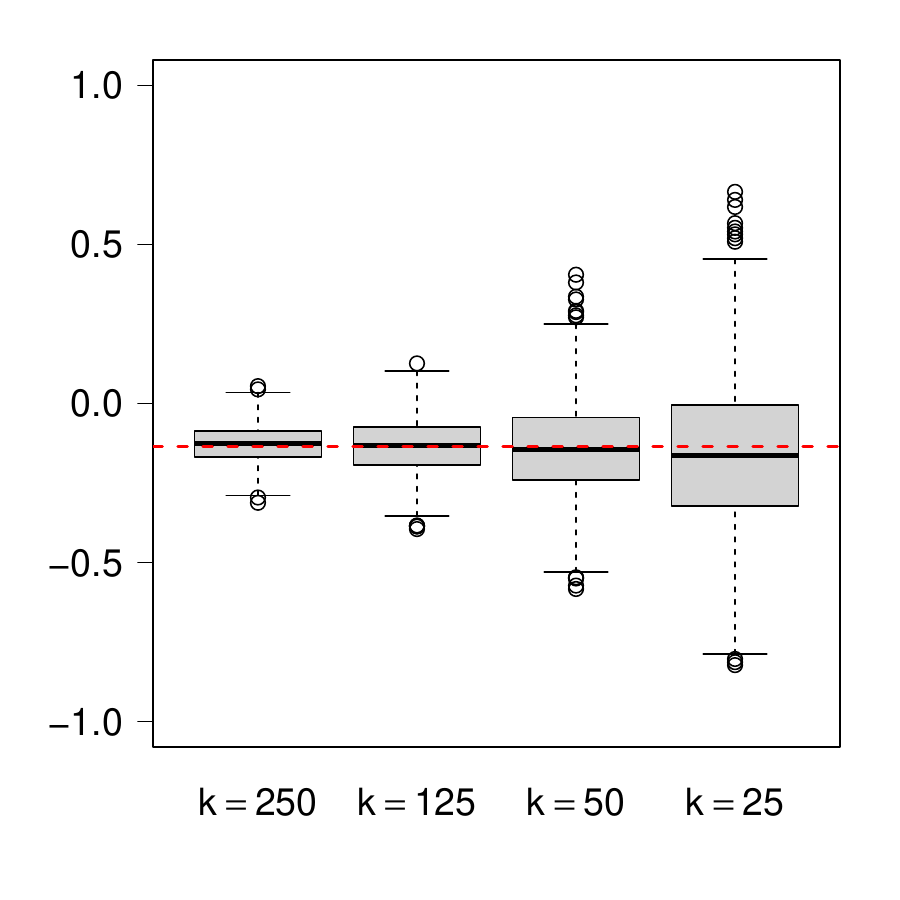}
\includegraphics[width=4cm]{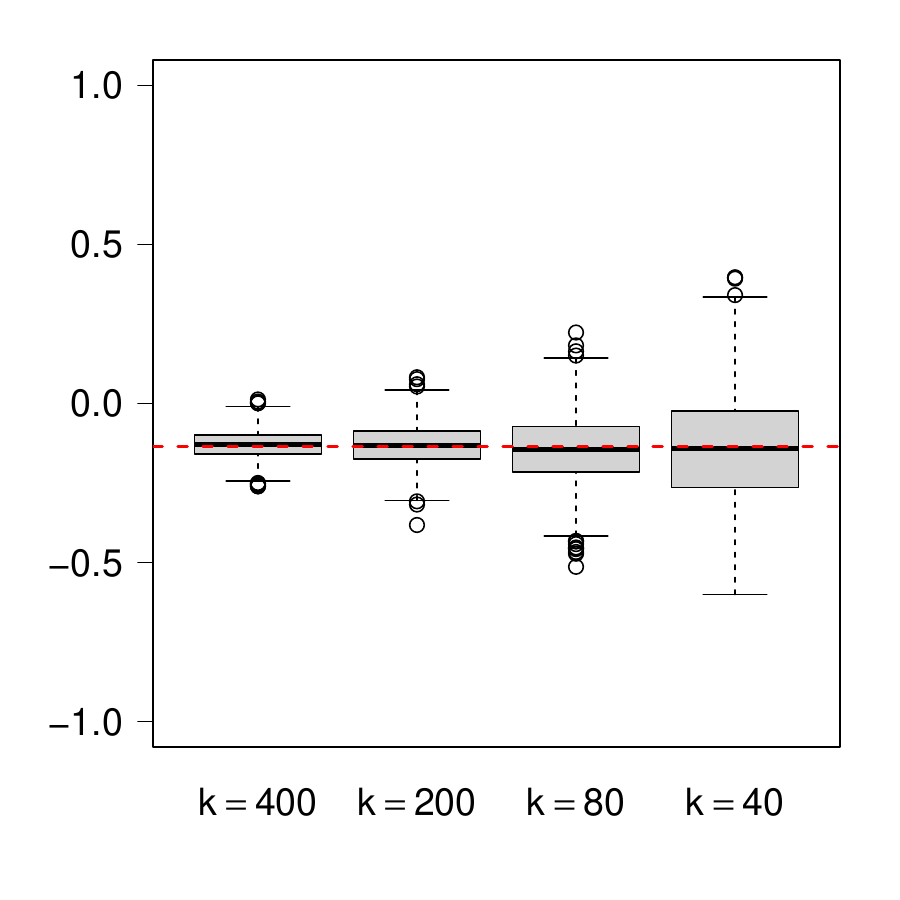}
\caption{\label{fig:condTPDM_est}
The top row shows box-plots of the off-diagonal element of the tail conditional covariance estimates $[\widehat{\Sg}_{K|L}(n,k)]_{12}$, under $1,000$ iterations, across sample sizes $n\in\cbr{1500,2500,4000}$ from left to right.
The bottom row shows box-plots of the partial tail correlation estimates $\hat{\rho}_{ij|L}$.
Red line segments indicate the specified values $[\Sg_{K|L}]_{12}=-0.03$ and $\rho_{12|L}(n,k)=-0.13$.
}
\end{figure}

Using these exceedances, we evaluate the bias and variability of the TPDM estimator, $\widehat{\Sigma}_{\bX}(n,k)$ over a range of $(n,k)$.
We consider three different sample sizes, $n\in\{1 500,2 500,4 000\}$, and four threshold ratios, $k/n\in\{0.1, 0.05, 0.02, 0.01\}.$
Box-plots of the off-diagonal TPDM estimates, provided in the Supplementary Material, show that increasing the effective sample size reduces variability while decreasing the threshold ratio $k/n$ reduces bias.
Our primary interest is in the asymptotic behavior of the off-diagonal element of $[\widehat{\Sg}_{K|L}(n,k)]_{12}$ in~\eqref{e:est_K|L} and $\hat{\rho}_{ij|L}(n,k)$ in~\eqref{e:est_rho}.
Figure~\ref{fig:condTPDM_est} presents box-plots of these estimates, showing reduced bias and variability for larger sample sizes across different $(n,k)$.

\subsection{Hypothesis test}
\label{sec:hypothesis}
We evaluate the empirical coverage rates and power of the test based on $z_{n,k}$ in~\eqref{e:test}.
We consider a finite version of the extremal AR(1) transformed-linear model to induce sparsity.
Specifically, let $C\in\reals^{15\times 15}$ be a lower triangular Toeplitz matrix with elements $C_{ij}=\phi^{i-j}I_{i\ge j}$, where $\phi\in(0,1)$ induces positive dependence among the components of $\bX$.
The relationship between the $X_j$'s is understood through the sequential generating equation
\begin{equation*}
    X_j=\phi\circ X_{j-1}\oplus Z_j,\quad X_0=0\,\,\, a.s.,
\end{equation*}
where the $Z_j$s are defined in Section~\ref{sec:estimation}.
The corresponding TPDM is $\Sg_{\bX}=CC^\top$.
The inverse TPDM has off-diagonal elements $\Theta_{ij}=-\phi$ if $|i-j|=1$ and $\Theta_{ij}=0$ if $|i-j|>1$.
This structure induces sparsity in the inverse TPDM, conveying that the partial tail correlation between $X_i$ and $X_j$ is zero whenever $|i-j|>1$.
As in Section~\ref{sec:estimation}, box-plots of the estimators $\widehat{\Sigma}_{\bX}(n,k)$ and $\widehat{\Sigma}_{K|L}(n,k)$, under $1000$ iterations using the Toeplitz matrix $C$, are included in the Supplementary Material.

We perform hypothesis tests for zero elements in the inverse TPDM.
Specifically, for $K=\{2,4\}$ and $L=[p]\setminus\{2,4\}$, we test the null hypothesis $H_0: [\Sg_{K|L}]_{12}=0$ under the AR(1) model with $\phi=0.7$.
For various combinations of $(n,k)$, we evaluate coverage rates using the asymptotic normality of $[\widehat{\Sg}_{K|L}(n,k)]_{12}$ and the test statistic $z_{n,k}$.
At the 5\% significance level, Table~\ref{tab:coverage} shows that the tests achieve appropriate empirical coverage across all $(n,k)$ combinations under 1,000 repeated simulations.

\begin{table}[ht]
    \centering
    \begin{tabular}{l c c c c}
    \toprule
    & \multicolumn{4}{c}{Threshold ratio $k/n$}\\
    \cmidrule(lr){2-5}
    & 0.1 & 0.05 & 0.02 & 0.01 \\
      \midrule
      $n=1500$  & 0.939 & 0.947 & 0.930 & 0.931 \\
      $n=2500$ & 0.944 & 0.943 & 0.938 & 0.922 \\
      $n=4000$ & 0.957 & 0.954 & 0.951 & 0.932\\
      \bottomrule
    \end{tabular}
\caption{\label{tab:coverage}
Coverage rates of confidence intervals based on the asymptotic normality of $[\widehat{\Sg}_{K|L}(n,k)]_{12}$ under the null hypothesis $H_0: [\Sg_{K|L}]_{12}=0$ for the AR(1) model with $\phi=0.7$ based on 1,000 iterations across $(n,k).$
}
\end{table}

\begin{figure}[ht]
\centering
\includegraphics[width=5cm]{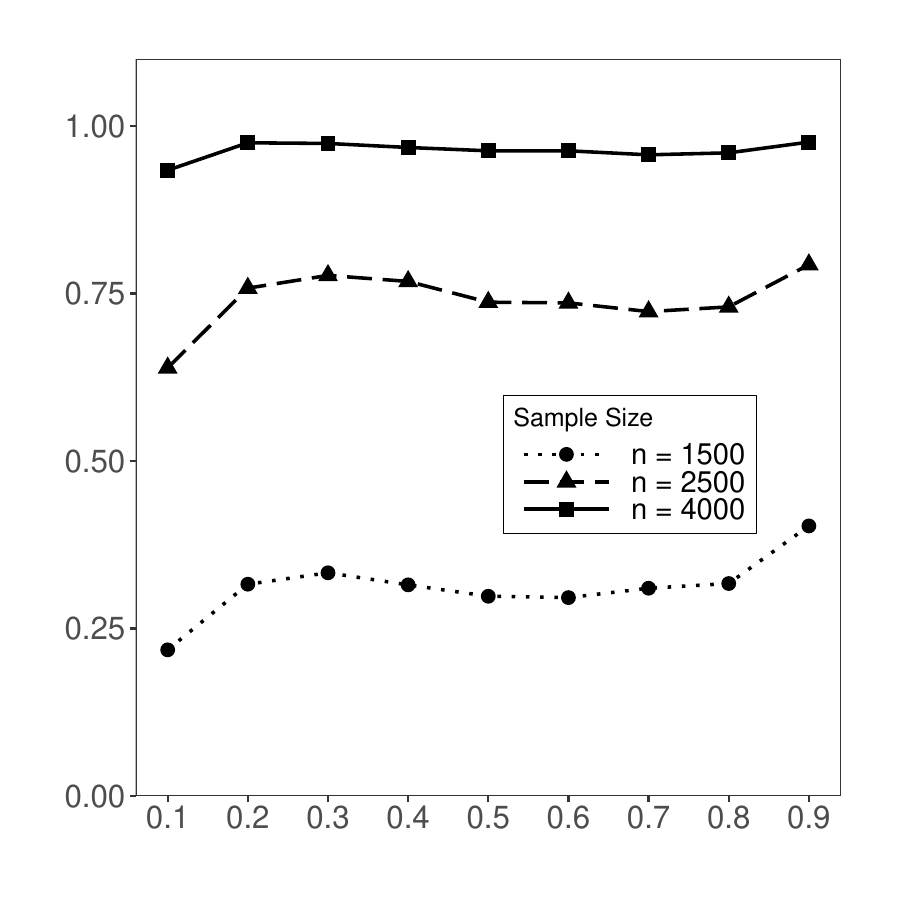}
\caption{\label{fig:power}
Empirical power curves plotted against $\phi\in\cbr{0.1,\ldots,0.9}$ for sample sizes $n\in\cbr{1500,2500,4000}$; significance level=5\%.
}
\end{figure}

We also conduct a power test using the pair $K=\cbr{1,2}$ and $L=[p]\setminus{\cbr{1,2}}$.
We calculate the rejection rate of $H_0:[\Sg_{K|L}]_{12}=0$ at the 5\% significance level across a sequence of $\phi\in\cbr{0.1,\ldots,0.9},$ for which the true off-diagonal element of $[\Sg_{K|L}]_{12}$ deviates from zero.
Figure~\ref{fig:power} displays power curves for sample sizes $n\in\cbr{1500,2500,4000}$ with a fixed ratio $k/n=0.03$.
Empirical power increases with sample size, consistent with Theorem~\ref{t:PO}, and also grows with $\phi$.
Even for small nonzero $\phi$, indicating weak partial tail correlation, the test achieves high power with large sample size.

\section{Danube river networks}
\label{sec:Danube}
\begin{figure}[ht]
\centering
\includegraphics[width=4.7cm]{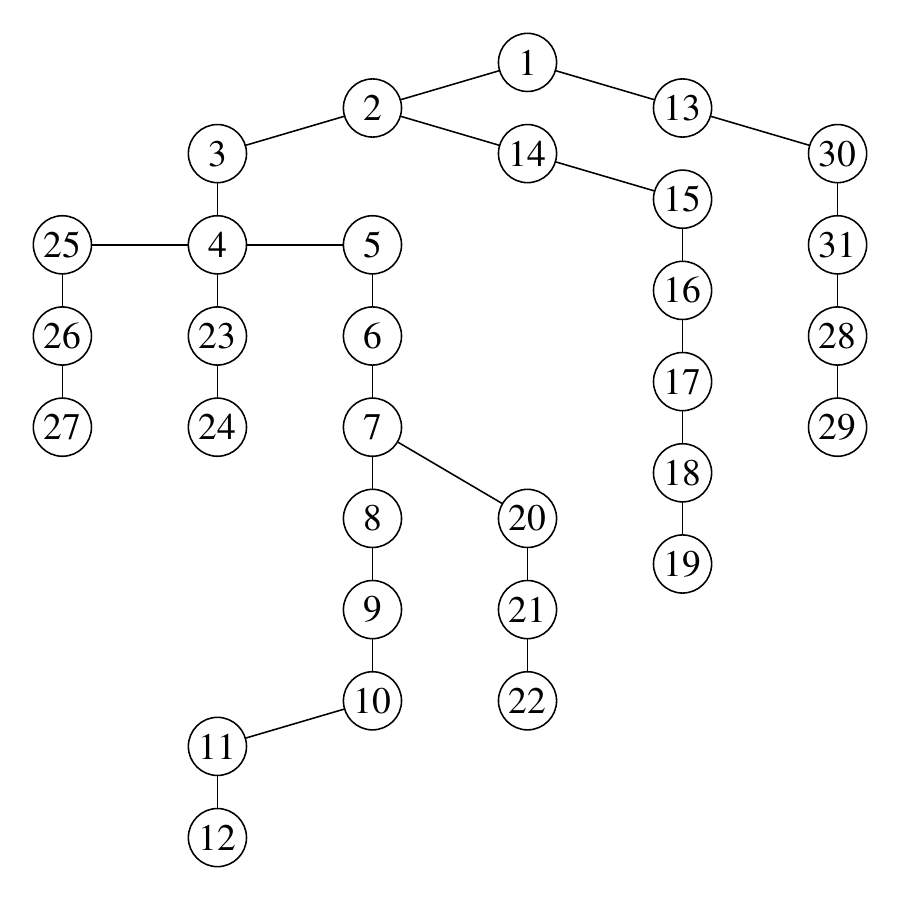}
\includegraphics[width=4.7cm]{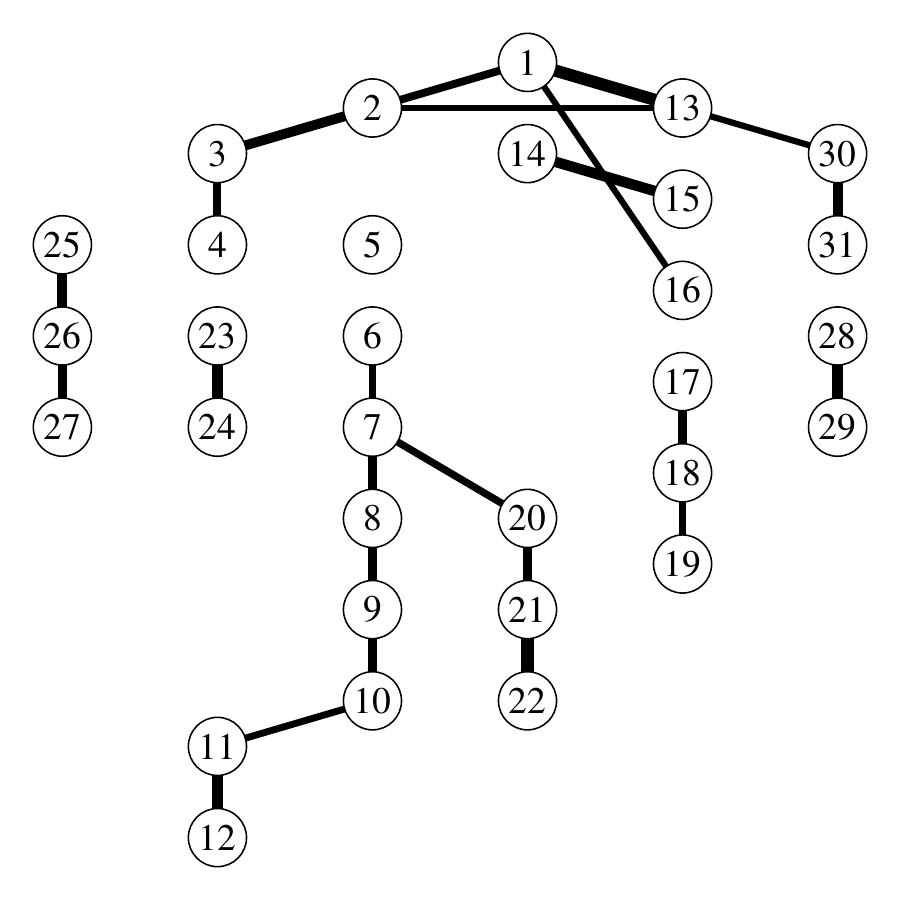}
\includegraphics[width=4.7cm]{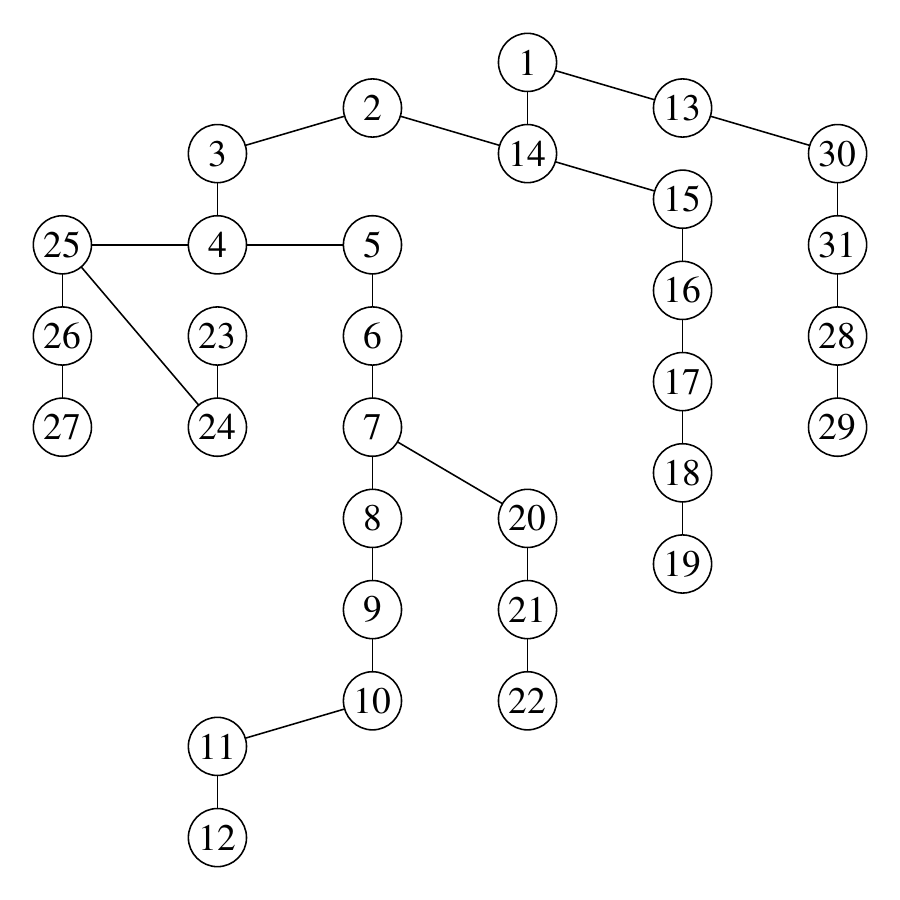}
\caption{\label{fig:danube}
Left: Graph of the physical river network in the upper Danube basin. 
Middle: Graph inferred from the hypothesis test, with edge thickness proportional to the magnitude of test statistic. 
Right: Extremal tree graph obtained from the fitted H{\"u}sler-Reiss model using a minimum spanning tree.
}
\end{figure}

We aim to test whether or not extreme river discharges between each pair of stations in the upper Danube basin exhibit significant partial tail correlation.
Our analysis uses average daily discharge records from 31 gauge stations for 1960-2010, obtained from the Bavarian Environmental Agency\footnote{http://www.gkd.bayern.de}.
The left panel of Figure~\ref{fig:danube} shows the physical river network, where the path $10\rightarrow\cdots\rightarrow 1$ represents the main channel and the remaining 21 stations are on tributaries.
This data set was first analyzed by \cite{asadi2015extremes}, who fit a spatial max-stable model to investigate extremal dependence across the network.
Following their preprocessing procedure, we focus on June, July, and August to reduce seasonality, reflecting both the concentration of major flood events in these summer months and that roughly half of the annual maxima occur during this period.
This results in $n=51\times 92=4,692$ daily discharges where all gauging stations have measurements.
No significant temporal trends are observed in the extreme stream-flows for this area, see \cite{asadi2015extremes}.

Extreme discharges tend to occur in clusters, implying the presence of temporal dependence.
In this work, rather than extracting clustered flood events as in \cite{asadi2015extremes}, we use the full sample of $n=4,692$ observations to retain adequate sample size.
We found that the declustered data set is not sufficient for obtaining stable inverse-matrix calculation required in our analysis.
Using the original series may lead to underestimation of uncertainty in parameter estimates and can be obscure dependencies between stations that are most evident in lagged measurements.
Incorporating a temporally dependent TPDM could potentially improve performance and represents an interesting direction for future work.

Let $\bX_l=[\bX_{lK}^\top,\bX_{lL}^\top]^\top$, $1\le l \le n$, be i.i.d. copies of $\bX=[\bX_K^\top,\bX_L^\top]^\top\in RV_+^p(\alpha),$ where $\bX_K=[X_i,X_j]^\top$, $i\ne j \in[p]$, and $\bX_L=\bX_{\setminus{(i,j)}}.$
As a standard preprocessing step to meet the tail equivalence assumption with $\alpha=2$, see e.g., Theorem 6.5 of \cite{resnick:2007}, 
we perform a marginal transformation via ranks.
Let $U_{l,j}=R_{l,j}/(n+1)$, where $R_{l,j}=\sum_{s=1}^n I_{(X_{s,j}\le X_{l,j})}$ is the rank of $X_{l,j}$ among $X_{1,j},\ldots,X_{n,j}$ for $j\in[p].$
We then define $Z_{l,j}=(1-U_{l,j})^{-1/2}-\delta$, obtaining a shifted Pareto distribution for each $j\in[p].$
Let $\bz_l$ denote the observed daily discharges on day $l$ in Pareto scale.
Under the $L_2$-norm and with $\alpha=2$, we define $(r_l,\bw_l)=(\|\bz_l\|_2,\bz_l/\|\bz_l\|_2)$.
For $k$ such that $k/n=0.1$, we obtain the TPDM estimate $\widehat{\Sg}_{\bZ}(n,k)=\frac{31}{k}\sum_{l=1}^n \bw_l\bw_l^\top I_{r_l\ge r_{(k)}},$
where $r_{(k)}$ is the $k$th largest value of $r_{(1)}\ge\cdots \ge r_{(n)}$.

From $\widehat{\Sg}_{\bZ}$, we compute the tail conditional covariance matrix estimate $\widehat{\Sg}_{K|L}(n,k)$ in~\eqref{e:est_K|L}, and then calculate the test statistic
$z_{n,k}$ in~\eqref{e:test}, where values near zero indicate week tail relationship between the station pair after accounting for all other stations.
We test the null hypothesis $H_0:[\Sg_{K|L}(n,k)]_{12}=0$ for each $i\ne j$.
To control the family-wise Type 1 error across multiple comparisons, we apply a Bonferroni adjustment, yielding a critical value $z_{crit}=3.87$.
We fail to reject the null hypothesis if $|z_{n,k}|<3.87$.
To visualize the results, we construct an {\em undirected} graph for the 31 stations where nodes $i$ and $j$ are connected if $\rho_{ij|L}$ is found to be significantly different from zero, as shown in the middle panel of Figure~\ref{fig:danube}.
The thickness of edge is proportional to the magnitude of test statistic, representing the strength of tail dependence after controlling for discharges at the remaining stations.
The resulting graph has 17 connected edges and reflects the overall structure of the physical river network.
In particular, significant partial tail correlation in discharge occurs mainly among stations within the same tributary, indicating clustered behavior.
This pattern is consistent with the expectation that, after accounting for discharge along the main channel, physically disconnected tributaries exhibit little additional extremal dependence.

For comparison, we consider the extremal tree modeling approach of \cite{engelke2022structure}.
This method represents extremal trees as sparse graphical structures, specifies bivariate H{\" u}sler-Reiss models along the tree, and estimate the underlying tree structure via a minimum spanning tree using the empirical extremal variogram as edge weights.
We implement this approach using the \texttt{graphicalExtremes} R package on the full dataset with standard Pareto margin $\alpha=2.$
The resulting extremal tree structure is fully connected due to its modeling assumption, as shown in the right panel of Figure~\ref{fig:danube}.
In contrast, our hypothesis testing procedure imposes no density or structural constraints, allowing for disconnected graphs and thus greater flexibility in capturing general extremal structures.

\section{Discussion}
\label{sec:discussion}
We have formulated a tail regression for multivariate extremes and introduced partial tail correlation within the framework of regular variation.
This methodology yields analogues of regression coefficients and partial correlations that closely parallel their classical non-extreme counterparts.
We further developed a hypothesis test for partial tail correlation, facilitating the identification of sparsity in extremal dependence structures and the construction of simplified models for extremes.
Compared with existing graphical models, our testing procedure imposes no density or structural constraints, allowing for disconnected graph and offering more flexibility in capturing general extremal structures.

Our framework with transformed linear operations allows for a focus on regular variation in the nonnegative orthant, which is often the direction of interest in extreme value analysis. 
The approach can, however, be extended to regular variation on the full real space using standard linear operations.
When contributions from other quadrants are considered, partial tail correlation should be interpreted with cautious, as the resulting dependence may reflect alignment along the same direction rather than orthogonality.

While our framework relies on assumptions of regular variation, several natural extensions remain for future work.
Incorporating temporal or spatial dependence into the TPDM could improve uncertainty quantification and better capture serial or spatial dependence among stations.
Establishing connections between generalized TPDMs, such as those in \cite{kiriliouk:chen:2022}, and our asymptotic results remain an open question.

\bigskip

\noindent{\bf \large Acknowledgments}\\
We thank Professor Dan Cooley at Colorado State University for helpful discussions during the development of this framework.
The framework and inference presented here extend and generalize J.L.'s Ph.D. research at Colorado State University, where the original concept of partial tail correlation was introduced in Chapter 4 of J.L.'s dissertation~\citep{lee2022phd}.
We also acknowledge independent and parallel work by \cite{gong2024partial} which investigate partial tail correlation for extremes from a model selection perspective rather than through hypothesis testing.
\bigskip

\noindent{\it Conflicts of interest:}  None declared.

\bigskip

\noindent{\bf \large Funding }\\
M.K. was partially supported by the United States National Science Foundation
grant DMS-2413516.
J.L. was partially supported by Engineering and Physical Sciences Research Council (EPSRC) grant EP/X010449/1.

\bigskip

\noindent{\bf \large Data availability}\\
The original Danube river data are available from the Bavarian Environmental Agency:\\ \texttt{http://www.gkd.bayern.de}.
The dataset analyzed in Section~\ref{sec:Danube} was obtained from the supplementary material of \cite{asadi2015extremes}.
The code used in the analysis is available at \texttt{https://github.com/JeongjinLee88/PTC}.




\clearpage

\appendix

\section{Preliminary Results} \label{s:app}

In this section, we put together several preliminary results needed for the proofs of the main results. While some of these results are known in the literature and are not particularly difficult to prove, presenting them here helps streamline the exposition and makes the paper self--contained.

The following lemma establishes a convergence result for $t^{-1}(\bX)$, which is used in the proof of Proposition~\ref{p:lcov}. Let $(R^{\prime}, \bW^{\prime}) = (\|t^{-1}(\bX)\|_2, t^{-1}(\bX)/\|t^{-1}(\bX)\|_2)$.

\begin{lemma}~\label{l:Htinv}
Suppose $\bX \in RV_+^p(\ag)$ and satisfies~\eqref{e:lowt}. Then, we have that
\[
mP \lb \bW^{\prime} \in \cdot | R^{\prime} > s\rb \convv H_{t^{-1}(\bX)}(\cdot),
\]
where $m$ is defined in~\eqref{e:def2}.
\end{lemma}

\begin{proof}

Since $t^{-1}(\bX)$ is regularly varying in $\mbR^p$ by Lemma~\ref{l:tinv}, there exists a sequence $c(s)\to \infty$ and an angular measure $H_{t^{-1}(\bX)}$ on $\Theta^{p-1}$ satisfying
\begin{equation} \label{e:def3}
sP\lb \lp \frac{R^{\prime}}{c(s)}, \bW^{\prime}  \rp \in \cdot \rb \convv \nu_{\ag} \times H_{t^{-1}(\bX)}, \quad \quad {\rm in}\  M_+((0, \infty] \times \Theta^{p-1}).    
\end{equation}
By applying the change of variables $t=c(s)$ to~\eqref{e:def3}, we have that
\[
c^{\leftarrow}(s) P\lb \lp \frac{R^{\prime}}{s}, \bW^{\prime}  \rp \in \cdot \rb \convv \nu_{\ag} \times H_{t^{-1}(\bX)}, \quad \quad {\rm in}\  M_+((0, \infty] \times \Theta^{p-1}). 
\]
Let $c(s)$ be chosen such that $c^{\leftarrow}(s)  = m/P[R^{\prime}>s]$. Then, it follows that
\[
\frac{m P\lb \lp R^{\prime}/s, \bW^{\prime}  \rp \in \cdot \rb}{P\lb R^{\prime}>s\rb}\convv \nu_{\ag} \times H_{t^{-1}(\bX)}, \quad \quad {\rm in}\  M_+((0, \infty] \times \Theta^{p-1}). 
\]
Since $(1,\infty] \times B$ is relatively compact for any measurable set $B \subset \Theta^{p-1}$, it follows that
\[
mP \lb \bW^{\prime} \in \cdot | R^{\prime} > s\rb \convv H_{t^{-1}(\bX)}(\cdot),
\]
which is valid because $m=H_{t^{-1}(\bX)}(\Theta^{p-1})$ by Lemma~\ref{l:tinv}.

\end{proof}

All results presented from here on serve as preliminaries for the main results in Section~\ref{s:E}. Lemma~\ref{l:R} includes properties of the radial component of $\bX$ when $\bX \in RV_+^p(\ag)$.

\begin{lemma} \label{l:R}

Suppose that $\bX \in RV_+^p(\ag)$, then

\noindent(i) Given any norm $\|\cdot\|$, $R=\|\bX\|$ is a nonnegative random variable whose distribution has a regularly varying tail with index $-\ag$,

\noindent(ii) $\frac{1}{k} \sum_{i=1}^n \ep_{R_i / b(n/k)} \Rightarrow \nu_{\ag}$, in $M_{+}(0,\infty]$,

\noindent(iii) $R_{(k)}/b(n/k) \convP 1$, in $[0,\infty)$.

\end{lemma}

\begin{proof}

Statement (i) follows from~\eqref{e:def2}. By (i), (ii) and (iii) follow directly from Theorem 4.1 and the proof of Theorem 4.2 of~\cite{resnick:2007}.

\end{proof}

Lemma~\ref{l:Sgeq} corresponds to equation (5.8.6) of~\cite{horn:jognson:2012}. It is needed to prove the convergence of inverse matrices in Theorems~\ref{t:K|Lcon} and~\ref{t:AN}, where the consistency of $\widehat{\Sg}_{K|L}$ and the asymptotic normality of $[\widehat{\Sg}_{K|L}]_{12}$ are established respectively.

\begin{lemma}~\label{l:Sgeq}
Let $\kg(\bSg) =\|\bSg^{-1}\|_F \|\bSg\|_F$, where $\|\cdot\|_F$ denotes the Frobenius norm. Then,
\[
\frac{\|\bSg^{-1} - \widehat{\bSg}^{-1}\|_F}{\|\bSg^{-1} \|_F} \le \frac{\kg(\bSg)}{1- \kg(\bSg) {\|\bSg - \widehat{\bSg} \|_F}/{\|\bSg\|_F}} \frac{\|\bSg - \widehat{\bSg} \|_F}{\|\bSg\|_F}.    
\]
\end{lemma}

The {\it De\'coupage de Le\'vy} is used in the proof of Theorem~\ref{t:AN}, so we state it in the following lemma, which is also stated on page 212 of~\cite{resnick:1987} and page 115 of~\cite{resnick:2007}.

\begin{lemma}~\label{l:Levy}
Suppose $\{X_n\}$ are i.i.d. random elements of a metric space $\mbS$ with Borel $\sg$-filed $\cS$. Fix a set $B \in \cS$ such that $P[X_1 \in B]>0$. Let $\tau_0^{\pm} = 0$, and $\tau_i^+ = \inf\{ j> \tau_{i-1}^+: X_j \in B\}$ for $i \ge 1$. The family $\{\tau_j^-, j \ge 0\}$ is defined similarly, with $B^c$ playing the role of $B$. Define the counting function $K_n = \sup\{i: \tau_i^+ \le n\}$. Then, $\{\tau_j^+, j \ge 0\}$, $\{\tau_j^-, j \ge 0\}$, $\{K_n\}$ are independent, and $\{\tau_j^+, j \ge 0\}$, $\{\tau_j^-, j \ge 0\}$ are i.i.d. with
\begin{align*}
&P[X_{\tau_1^+} \in A] = P[X_1 \in A| X_1 \in B], \quad A \subset B, \\
&P[X_{\tau_1^-} \in A] = P[X_1 \in A| X_1 \in B^c], \quad A \subset B^c.
\end{align*}

\end{lemma}

The following lemma states a result related to the second-order regular variation. It is used in the proof of Theorem~\ref{t:AN}.

\begin{lemma}\label{l:2nd}
Suppose that the radial and angular components of $\bX$, i.e.,$(R, \bW)$, satisfy~\eqref{e:2nd}. Then, we have that in $M_+((0,\infty] \times \Theta_+^{p-1})$,
\begin{equation}\label{e:2nd_1}
\sqrt{k} \lbr \frac{n}{k} P \lb \lp \frac{R}{b(n/k)}, \bW \rp \in \cdot \rb
- \frac{n}{k} P \lb \frac{R}{b(n/k)}  \in \cdot \rb \times \frac{1}{m}H_{\bX}(\cdot) \rbr \convv 0,
\end{equation}
where $m = H_{\bX}$ as defined in~\eqref{e:def2}. 
\end{lemma}

\begin{proof}

Take any relatively compact set $A \times S \subset (0,\infty] \times \Theta_+^{p-1}$, and observe that
\begin{align*}
&\sqrt{k} \lbr \frac{n}{k} P \lb \lp \frac{R}{b(n/k)}, \bW \rp \in A \times S \rb
- \frac{n}{k} P \lb \frac{R}{b(n/k)}  \in A \rb \times \frac{1}{m} H_{\bX}(S) \rbr \\
&=\sqrt{k} \lbr \frac{n}{k} P \lb \lp \frac{R}{b(n/k)}, \bW \rp \in A \times S \rb
- \nu_{\ag}(A) \frac{1}{m}H_{\bX}(S) \rbr\\
&\ \ \ \ + \sqrt{k}\lbr \nu_{\ag}(A) -  \frac{n}{k} P \lb \frac{R_1}{b(n/k)}  \in A \rb \rbr \frac{1}{m}H_{\bX}(S).
\end{align*}
This converges to 0 by~\eqref{e:2nd}, which completes the proof.

\end{proof}

Lemmas~\ref{l:N} and~\ref{l:Sgbar} provide preliminary results required for the proof of Proposition~\ref{p:kXcon}.



\begin{lemma}~\label{l:N}
Suppose that random vectors $\bX_l$, $1 \le l \le n$, are i.i.d. copies of $\bX$. We assume that there exist a sequence $b(s)\to\infty$ and an angular
measure $H_{\bX}$ on $\Theta_+^{p-1}$ such that the radial and angular components of $\bX$, i.e.,$(R, \bW)$, satisfy~\eqref{e:2nd}. Then, we have that for $t \ge 0$
\[
\frac{1}{k} \sum_{l=1}^n I_{R_l/b(n/k)\ge t^{-1/\ag}} - t = o_P(k^{-1/2}).
\]
\end{lemma}
\begin{proof}

From~\eqref{e:2nd}, Theorem 5.3 (ii) of Resnick (2007) implies that
\begin{equation} \label{e:gM}
\mu_n:=\sqrt{k} \lp \frac{m}{k}  \sum_{l=1}^{n} \ep_{R_l/b(n/k)} - \nu_{\ag}\rp \convP 0, \quad {\rm in}\ M_+(0, \infty].       
\end{equation}
Since any signed measures can be decomposed into positive and negative parts, we may assume without loss of generality that $\mu_n$ is positive. Consider the continuous map $g_M$ on $M_+(0,\infty]$, defined by $g_M(U) = U ([M,\infty])$, $M>0$. Applying $g_M$ with $M = t^{-1/\ag}$ to~\eqref{e:gM} completes the proof.


\end{proof}

\begin{lemma}~\label{l:Sgbar}
Under assumptions in Lemma~\ref{l:N},  we have that for $t \ge 0$
\[
\frac{m}{k} \sum_{l=1}^n W_{li}W_{lj}I_{R_l/b(n/k)\ge t^{-1/\ag}} - t\sg_{X, ij} = o_P(k^{-1/2}),
\]
where $\sg_{X, ij}$ is defined in~\eqref{e:TPDM}.
\end{lemma}
\begin{proof}

From~\eqref{e:2nd}, Theorem 5.3 (ii) of Resnick (2007) implies that
\begin{equation} \label{e:U}
\sqrt{k} \lp \frac{m}{k}  \sum_{l=1}^{n} \ep_{(R_l/b(n/k), \bW_l)} - \nu_{\ag} \times H_{\bX}\rp \convP 0, \quad {\rm in}\ M_+((0, \infty] \times \Theta_+^{p-1}).
\end{equation}
As in Lemma~\ref{l:N}, we once again assume without loss of generality that the sequence of measures on the left-hand side of~\eqref{e:U} is positive. First, apply the continuous map $U \mapsto U([t^{-1/\ag}, \infty] \times \cdot)$, from $M_+((0, \infty] \times \Theta_+^{p-1})$ to $M_+(\Theta_+^{p-1})$, to~\eqref{e:U}. It then follows that 
\begin{equation} \label{e:S}
\sqrt{k} \lp \frac{m}{k} \sum_{l=1}^n  \ep_{\bW_l} I_{R_l/b(n/k)\ge t^{-1/\ag}} -tH_{\bX} \rp \convP 0,\quad\quad {\rm in}\ M_+(\Theta_+^{p-1}).    
\end{equation}
Applying the continuous map $S \mapsto \int_{\Theta_+^{p-1}} w_i w_j S(d\bw)$, from $M_+(\Theta_+^{p-1})$ to $\mathbb{R}$, to~\eqref{e:S}, completes the proof.


\end{proof}

In the following proposition, we establish an asymptotic property of $\widehat{\Sg}_{\bX}$, defined in~\eqref{e:Sgnk}, under the second-order regular variation condition given in~\eqref{e:2nd}. This proposition plays a key role in the proof of Theorem~\ref{t:AN}.

\begin{proposition} \label{p:kXcon}
Under assumptions in Lemma~\ref{l:N}, we have that 
\[
\sqrt{k}(\widehat{\Sg}_{\bX} - \Sg_{\bX}) \convP 0,
\]
where $\widehat{\Sg}_{\bX}$ and $\Sg_{\bX}$ are defined in~\eqref{e:Sgnk} and~\eqref{e:TPDM}, respectively. 
\end{proposition}

\begin{proof}
It suffices to show that for any $i, j \in [p]$
\[
\sqrt{k}\lp \frac{m}{k} \sum_{l=1}^n W_{li}W_{lj}I_{R_l \ge R_{(k)}} - \sg_{X, ij} \rp \convP 0.
\]
Observe that
\begin{align*}
&\sqrt{k}\lp \frac{m}{k} \sum_{l=1}^n W_{li}W_{lj}I_{R_l \ge R_{(k)}} - \sg_{X, ij} \rp \\
&=\sqrt{k}\lp \frac{m}{k} \sum_{l=1}^n W_{li}W_{lj}I_{R_l \ge b(n/k)} - \sg_{X, ij} \rp+\sqrt{k}\lp \frac{m}{k} \sum_{l=1}^n W_{li}W_{lj}I_{R_l \ge R_{(k)}} - \frac{m}{k}\sum_{l=1}^n W_{li}W_{lj}I_{R_l \ge b(n/k)} \rp\\
&=: P_1(n) + Q_1(n).
\end{align*}
It follows by Lemma~\ref{l:Sgbar} with $t=1$ that $P_1(n) \convP 0$. It remains to show that $Q_1(n) \convP 0$. Since $|W_{li}W_{lj}| \le 1$, it follows that
\begin{align*}
\sqrt{k}\frac{m}{k}\Bigg|\sum_{l=1}^n W_{li}W_{lj}I_{R_l \ge R_{(k)}} - \sum_{l=1}^n W_{li}W_{lj}I_{R_l \ge b(n/k)} \Bigg| 
&\le \sqrt{k} \frac{m}{k} \sum_{l=1}^n |W_{li}W_{lj}| |I_{R_l \ge R_{(k)}} - I_{R_l \ge b(n/k)}|\\
& \le \sqrt{k} \frac{m}{k} \sum_{l=1}^n |I_{R_l \ge R_{(k)}} - I_{R_l \ge b(n/k)}|.
\end{align*}
Since $\sum_{l=1}^n I_{R_l \ge R_{(k)}}=k$, it implies that
\begin{align*}
&\sqrt{k} \frac{m}{k} \sum_{l=1}^n |I_{R_l \ge R_{(k)}} - I_{R_l \ge b(n/k)}|\\
&=\sqrt{k} \frac{m}{k} \lp  \sum_{l=1}^n I_{R_l \ge R_{(k)}} - \sum_{l=1}^n I_{R_l \ge b(n/k)} \rp I_{b(n/k) \ge R_{(k)}} + \sqrt{k} \frac{m}{k} \lp \sum_{l=1}^n I_{R_l \ge b(n/k)} - \sum_{l=1}^n I_{R_l \ge R_{(k)}} \rp I_{ b(n/k) < R_{(k)} }  \\
&\le m \Bigg|\sqrt{k} \lp \frac{1}{k} \sum_{l=1}^n I_{R_l \ge b(n/k)} -1 \rp \Bigg| = o_P(1),
\end{align*}
which follows from Lemma~\ref{l:N} with $t=1$.

\end{proof}

\newpage

\section{Proof of Theorem~\ref{t:AN}} \label{s:app2}

The structure of the proof is adapted from that used in the proof of Theorem 1 in \cite{larsson:resnick:2012}, where the asymptotic normality of an estimator for the {\it extremal dependence measure} is established. Our setting presents different challenges, as our estimator, designed to capture partial tail structure, differs in both its construction and the type of dependence it quantifies. This requires several new arguments including~\eqref{e:dSgKL} and distinct technical details, which are developed throughout various parts of the proof.

We begin by defining the following empirical process:
\[
\widehat{V}_n{(t)} = \frac{m}{\tau\sqrt{k}} \sum_{l=1}^n \widehat{A}_1^{\top}\bW_l \bW_l^{\top}\widehat{A}_2
I_{R_l/b({n}/{k}) \ge t^{-1/\ag}}, \ \ \ t \geq 0,
\]
where $\widehat{A}_1$, $\widehat{A}_2$ are defined in~\eqref{e:Ahat}. The main step in establishing the asymptotic normality of $\sqrt{k}[\widehat{\Sg}_{K|L}(n,k)]_{12}$ is to prove the weak convergence of $\widehat{V}_n$ to the standard Brownian motion $B$:
\begin{equation}\label{e:WntoW}
\widehat{V}_n \Rightarrow{} B,\ \ \ \ {\rm in}\ D[0,\infty),
\end{equation}
where $D[0,\infty)$ denotes the usual Skorokhod space.
Once this convergence is established, we can use the result $R_{(k)}/b(n/k) \convP 1$ from Lemma~\ref{l:R} (iii), to obtain the following joint convergence:
\[
\lp \widehat{V}_n(\cdot), \lp \frac{R_{(k)}}{b(n/k)} \rp ^{-\ag} \rp \Rightarrow{} (B(\cdot),1),\ \ \ \ {\rm in}\ D[0,\infty) \times [0,\infty).
\]
By applying the continuous map $(x(\cdot),c) \mapsto x(c)$ from $D[0, \infty) \times [0,\infty)$ to $[0, \infty)$, it follows that
\[
\sqrt{k}[\widehat{\Sg}_{K|L}(n,k)]_{12}=\frac{m}{\sqrt{k}} \sum_{l=1}^{n} \widehat{A}_1^{\top}\bW_l \bW_l^{\top}\widehat{A}_2 I_{R_{l} \ge R_{(k)}}  = \tau \widehat{V}_n\lp \lp \frac{R_{(k)}}{b(n/k)} \rp ^{-\ag} \rp \Rightarrow{} \tau B(1),
\]
which will complete the proof.

Observe that the empirical process $\widehat{V}_n(t)$ can be decomposed as follows:
\begin{align*}
\widehat{V}_n{(t)}&= \frac{m}{\tau\sqrt{k}} \sum_{l=1}^n A_1^{\top}\bW_l \bW_l^{\top}A_2
I_{R_l/b({n}/{k}) \ge t^{-1/\ag}} \\
& \quad + \frac{m}{\tau\sqrt{k}} \sum_{l=1}^n ( \widehat{A}_1^{\top}\bW_l \bW_l^{\top}\widehat{A}_2 - A_1^{\top}\bW_l \bW_l^{\top}A_2 )
I_{R_l/b({n}/{k}) \ge t^{-1/\ag}}.    
\end{align*}
Therefore, to establish~\eqref{e:WntoW}, it suffices to verify that
\begin{equation} \label{e:SgKLtoW}
V_n(t):=\frac{m}{\tau\sqrt{k}} \sum_{l=1}^n A_1^{\top}\bW_l \bW_l^{\top}A_2
I_{R_l/b({n}/{k}) \ge t^{-1/\ag}} \Rightarrow B(t),\ \ \ \ {\rm in}\ D[0,\infty),
\end{equation}
and for any $s \ge 0$,
\begin{equation} \label{e:dSgKL}
\sup_{0\le t \le s} \Bigg| \frac{m}{\tau\sqrt{k}} \sum_{l=1}^n ( \widehat{A}_1^{\top}\bW_l \bW_l^{\top}\widehat{A}_2 - A_1^{\top}\bW_l \bW_l^{\top}A_2 )
I_{R_l/b({n}/{k}) \ge t^{-1/\ag}} \Bigg| \convP 0.
\end{equation}

We first focus on~\eqref{e:SgKLtoW}, where we apply classical techniques based on the convergence of finite-dimensional distributions and tightness, see e.g., Theorem 13.5 of~\cite{billingsley:1999}. 

\medskip

\noindent{\it Convergence of finite-dimensional distributions:} This can be deduced from the following key argument: for $0 \le s < t$,
\begin{equation} \label{e:WtWs}
V_n(t)-V_n(s) = \frac{m}{\tau\sqrt{k}} \sum_{l=1}^n A_1^{\top}\bW_l \bW_l^{\top}A_2
I_{R_l/b(n/k) \in [t^{-1/\ag}, s^{-1/\ag})} \Rightarrow N(0, t-s).    
\end{equation}
Therefore, we prove~\eqref{e:WtWs} first, and then proceed to discuss the convergence of finite-dimensional distributions.

To show~\eqref{e:WtWs}, we first define the random number falling into the interval $[t^{-1/\ag}, s^{-1/\ag})$ by
\[
N_n := \sum_{l=1}^n I_{R_l/b(n/k) \in [t^{-1/\ag}, s^{-1/\ag})}.
\]
Let $\{l(r,n), r \ge 1\}$ be the random indices such that $R_l/b(n/k) \in [t^{-1/\ag}, s^{-1/\ag})$. Then, by Lemma~\ref{l:Levy}, $\{\bW_{l(r,n)}, r\ge 1 \}$ are i.i.d., and by~\eqref{e:2nd},
\begin{align}\label{e:PWlrn}
P\lb \bW_{l(1,n)} \in \cdot \rb &= P\lb \bW_1 \in \cdot | R_1/b(n/k) \in [t^{-1/\ag}, s^{-1/\ag})\rb  \\ \nonumber
&= \frac{\frac{n}{k}P\lb R_1/b(n/k) \in [t^{-1/\ag}, s^{-1/\ag}), \bW_1 \in \cdot \rb}{\frac{n}{k}P\lb R_1/b(n/k) \in [t^{-1/\ag}, s^{-1/\ag})\rb}\\  \nonumber
& \convv \frac{\nu_{\ag}[t^{-1/\ag}, s^{-1/\ag})  H_{\bX}(\cdot)/m}{\nu_{\ag}[t^{-1/\ag}, s^{-1/\ag}) } = \frac{1}{m}H_{\bX}(\cdot). 
\end{align}

Using $\{\bW_{l(r,n)}, r\ge 1 \}$ and $N_n$, we can express $V_n(t)-V_n(s)$ as the random sum of i.i.d. random variables and obtain the following decomposition:
\begin{align*}
&V_n(t)-V_n(s) \\
&= \frac{m}{\tau\sqrt{k}} \sum_{r=1}^{N_n} \lp A_1^{\top}\bW_{l(r,n)} \bW_{l(r,n)}^{\top} A_2 -  E[ A_1^{\top}\bW_{l(1,n)} \bW_{l(1,n)}^{\top} A_2 ]  \rp + \frac{mN_n}{\tau\sqrt{k}}  E[ A_1^{\top}\bW_{l(1,n)} \bW_{l(1,n)}^{\top} A_2 ] \\
&=: P(n) + Q(n)
\end{align*}
We will show that $P(n) \Rightarrow N(0, t-s)$ and $Q(n) \convP 0$. 

We first consider $P(n)$. By applying the functional central limit theorem for triangular arrays, it follows that
\[
\frac{1}{\sqrt{k}\sqrt{\var [A_1^{\top} \bW_{l(1,n)} \bW_{l(1,n)}^{\top} A_2 ]}} \sum_{r=1}^{[ku]} \lp A_1^{\top}\bW_{l(r,n)} \bW_{l(r,n)}^{\top}A_2 - E [ A_1^{\top}\bW_{l(1,n)} \bW_{l(1,n)}^{\top}A_2 ] \rp \Rightarrow B(u),
\]
in $D[0,\infty)$. Since $N_n/k \convP \nu_{\ag}[t^{-1/\ag}, s^{-1/\ag}) = t-s$ by Lemma~\ref{l:R} (ii), we have the joint convergence. Then, it follows by applying the continuous mapping $(x(\cdot),c) \mapsto x(c)$ that
\[
\frac{1}{\sqrt{k}\sqrt{\var[ A_1^{\top} \bW_{l(1,n)} \bW_{l(1,n)}^{\top} A_2]}} \sum_{r=1}^{N_n} \lp A_1^{\top}\bW_{l(r,n)} \bW_{l(r,n)}^{\top}A_2 - E[ A_1^{\top}\bW_{l(1,n)} \bW_{l(1,n)}^{\top}A_2 ] \rp \Rightarrow B(t-s),
\]
It remains to show that $\var[ A_1^{\top} \bW_{l(1,n)} \bW_{l(1,n)}^{\top} A_2] \to \tau^2/m^2$ to conclude $P(n) \Rightarrow N(0, t-s)$. Observe that
\begin{align*}
&\var [A_1^{\top} \bW_{l(1,n)} \bW_{l(1,n)}^{\top} A_2] \\ &= E [ (A_1^{\top} \bW_{l(1,n)} \bW_{l(1,n)}^{\top} A_2 )^2] - \lbr E [ A_1^{\top} \bW_{l(1,n)} \bW_{l(1,n)}^{\top} A_2] \rbr^2\\
&= \int_{\Theta_+^{p-1}} (A_1^{\top} \bw \bw^{\top} A_2)^2 P[\bW_{l(1,n)} \in d\bw] -  \lbr \int_{\Theta_+^{p-1}} A_1^{\top} \bw \bw^{\top} A_2 P[\bW_{l(1,n)} \in d\bw] \rbr^2.
\end{align*}
It then follows from~\eqref{e:PWlrn} that
\begin{align*}
&\var [A_1^{\top} \bW_{l(1,n)} \bW_{l(1,n)}^{\top} A_2] \\ 
&\to  \int_{\Theta_+^{p-1}} (A_1^{\top} \bw \bw^{\top} A_2)^2 \frac{1}{m}H_{\bX}(d\bw)  -  \lbr \int_{\Theta_+^{p-1}} A_1^{\top} \bw \bw^{\top} A_2 \frac{1}{m}H_{\bX}(d\bw) \rbr^2\\
&=  E[(A_1^{\top}\widetilde{\bW} \widetilde{\bW}^{\top}A_2)^2]  - \{ E[A_1^{\top}\widetilde{\bW} \widetilde{\bW}^{\top}A_2] \}^2= \var[A_1^{\top}\widetilde{\bW} \widetilde{\bW}^{\top}A_2] = \tau^2/m^2. 
\end{align*}

Turning to $Q(n)$, recall~\eqref{e:2nd_1} from Lemma~\ref{l:2nd} that
\[
\sqrt{k} \lbr \frac{n}{k} P \lb \lp \frac{R}{b(n/k)}, \bW \rp \in \cdot \rb
- \frac{n}{k} P \lb \frac{R}{b(n/k)}  \in \cdot \rb \times \frac{1}{m}H_{\bX}(\cdot) \rbr \convv 0.
\]
Consider the function $g: (0, \infty] \times \Theta_+^{p-1} \to \mbR$ defined by 
\[
g(r, \bw) = A_1^{\top}\bw \bw^{\top} A_2 I_{r \in [t^{-1/\ag}, s^{-1/\ag})},
\]
which is compactly supported and a.e. continuous on $(0, \infty] \times \Theta_+^{p-1}$. Therefore, applying $g$ to~\eqref{e:2nd_1} implies that
\begin{align*}
&\sqrt{k} \bigg\{ \frac{n}{k} E \lb A_1^{\top}\bW\bW^{\top}A_2 I_{R/b(n/k) \in [t^{-1/\ag}, s^{-1/\ag})}  \rb \\
&\quad\quad\quad\quad\quad\quad- \frac{n}{k} P \lb \frac{R}{b(n/k)}  \in [t^{-1/\ag}, s^{-1/\ag}) \rb \times E[A_1^{\top}\widetilde{\bW}\widetilde{\bW}^{\top}A_2] \bigg\} \to 0.    
\end{align*}
By the null hypothesis~\eqref{e:HT}, we have that
\begin{equation} \label{e:fHT}
[\Sg_{K|L}]_{12} = A_1^{\top} \Sg_{\bX}A_2 = mE[A_1^{\top}\widetilde{\bW}\widetilde{\bW}^{\top}A_2]=0.
\end{equation}
It implies that
\begin{equation} \label{e:ag}
\sqrt{k} \lbr \frac{n}{k} E \lb A_1^{\top}\bW\bW^{\top}A_2 I_{R/b(n/k) \in [t^{-1/\ag}, s^{-1/\ag})}  \rb \rbr \to 0.
\end{equation}
Then, it follows by~\eqref{e:2nd} and~\eqref{e:ag} that 
\begin{align*} 
&\sqrt{k}E[ A_1^{\top}\bW_{l(1,n)} \bW_{l(1,n)}^{\top}A_2 ] \\ \nonumber
&= \frac{1}{\frac{n}{k}P \lb R/b(n/k)  \in [t^{-1/\ag}, s^{-1/\ag}) \rb}  \sqrt{k} \lbr \frac{n}{k}E [ A_1^{\top}\bW_1\bW_1^{\top}A_2 I_{R_1/b(n/k) \in [t^{-1/\ag}, s^{-1/\ag})}   ] \rbr \\ \nonumber
& \to \frac{1}{t-s}\times 0 = 0. 
\end{align*}
Since $N_n/k\convP t-s$ by Lemma~\ref{l:R} (ii), it follows that 
\begin{align*}
Q(n) &= \frac{mN_n}{\tau k}  \lbr \sqrt{k}E[ A_1^{\top}\bW_{l(1,n)} \bW_{l(1,n)}^{\top}A_2 ] \rbr \convP 0.
\end{align*}
This completes the proof of~\eqref{e:WtWs}.

Now, to show the convergence of finite-dimensional distributions, take any $0 \le s <t$. It follows by~\eqref{e:WtWs} that $V_n(s) \Rightarrow N(0,s)$ and $V_n(t) - V_n(s) \Rightarrow N(0,t-s)$, and by Lemma~\ref{l:Levy} that $V_n(s)$ and $V_n(t) - V_n(s)$ are independent. Therefore, we obtain the joint weak convergence:
\[
(V_n(s), V_n(t) - V_n(s)) \Rightarrow (N(0,s), N(0,t-s)).
\]
By applying the continuous map $(x,y) \mapsto (x, x+y)$, we have that $(V_n(s), V_n(t)) \Rightarrow (N(0,s), N(0,t))$. This argument can be extended to any finite collection of time points.

\medskip

\noindent{\it Tightness:} By Theorem 13.5 of~\cite{billingsley:1999} (see also Proposition 2.1 of~\cite{resnick:starica:1997b}), it suffices to show that for any $0 \le r < s < t$,
\begin{equation} \label{e:tight}
\limsup_{n} E [|V_n(s) - V_n(r)|^2|V_n(t) - V_n(s)|^2] \le (t-r)^2.    
\end{equation}
Since $V_n(s) - V_n(r)$ and $V_n(t) - V_n(s)$ are independent by Lemma~\ref{l:Levy}, it follows that 
\begin{align*}
&E [|V_n(s) - V_n(r)|^2|V_n(t) - V_n(s)|^2] = E [|V_n(s) - V_n(r)|^2]E[|V_n(t) - V_n(s)|^2].
\end{align*}
We first focus on $E [|V_n(t) - V_n(s)|^2]$ and show that it converges to $t-s$. Since the $\bW_l$ are i.i.d. by assumption, we have that
\begin{align*}
&E[|V_n(t)-V_n(s)|^2] \\
&= \frac{m^2}{\tau^2k} E\Bigg[ \sum_{l=1}^n (A_1^{\top}\bW_l \bW_l^{\top}A_2)^2
I_{R_l/b(n/k) \in [t^{-1/\ag}, s^{-1/\ag})} \\
&\quad\quad\quad\quad+ \sum_{l\neq u} (A_1^{\top}\bW_l \bW_l^{\top}A_2
I_{R_l/b(n/k) \in [t^{-1/\ag}, s^{-1/\ag})})  (A_1^{\top}\bW_u \bW_u^{\top}A_2
I_{R_u/b(n/k) \in [t^{-1/\ag}, s^{-1/\ag})}) \Bigg]\\
&= \frac{m^2}{\tau^2} \Bigg\{ \frac{n}{k} E \lb (A_1^{\top}\bW_1 \bW_1^{\top}A_2)^2I_{R_1/b(n/k) \in [t^{-1/\ag}, s^{-1/\ag})} \rb \\
&\quad\quad\quad\quad+ \frac{n(n-1)}{k^2} \lp \sqrt{k}E \lb A_1^{\top}\bW_1 \bW_1^{\top}A_2
I_{R_1/b(n/k) \in [t^{-1/\ag}, s^{-1/\ag})} \rb \rp^2 \Bigg\}.
\end{align*}
The second term vanishes by~\eqref{e:ag}. The first term converges to $t-s$ since it follows by~\eqref{e:2nd},~\eqref{e:PWlrn}, and~\eqref{e:fHT} that
\begin{align*} 
&\frac{n}{k} E \lb (A_1^{\top}\bW_1 \bW_1^{\top}A_2)^2I_{R_1/b(n/k) \in [t^{-1/\ag}, s^{-1/\ag})} \rb \\
&=\frac{n}{k} P \lb R_1/b(n/k)  \in [t^{-1/\ag}, s^{-1/\ag}) \rb  E [ (A_1^{\top}\bW_{l(1,n)} \bW_{l(1,n)}^{\top}A_2)^2 ] \\ 
& \to (t-s)  E[(A_1^{\top}\widetilde{\bW} \widetilde{\bW}^{\top}A_2)^2]  = (t-s) \var[A_1^{\top}\widetilde{\bW} \widetilde{\bW}^{\top}A_2] = (t-s)\frac{\tau^2}{m^2}. 
\end{align*}
Similarly, it can be shown that $E[|V_n(s) - V_n(r)|^2] \to s-r$. Since $(s-r)(t-s) \le (t-r)^2$,~\eqref{e:tight} holds.

\medskip

Finally, we turn to~\eqref{e:dSgKL}. Define
\[
\overline{\Sg}_{\bX}(t) :=\overline{\Sg}_{\bX}(n,k;t) = \frac{m}{k} \sum_{l=1}^n  \bW_l \bW_l^{\top}  I_{R_l/b(n/k) \ge t^{-1/\ag}}.
\]
Then,~\eqref{e:dSgKL} is written as
\begin{align*}
&\sup_{0\le t \le s} \Bigg| \frac{m}{\tau\sqrt{k}} \sum_{l=1}^n ( \widehat{A}_1^{\top}\bW_l \bW_l^{\top}\widehat{A}_2 - A_1^{\top}\bW_l \bW_l^{\top}A_2 )I_{R_l/b({n}/{k}) \ge t^{-1/\ag}} \Bigg|    \\
&=\sup_{0\le t \le s} \Bigg|\frac{\sqrt{k}}{\tau} \lp \widehat{A}_1^{\top} \overline{\Sg}_{\bX}(t) \widehat{A}_2 - A_1^{\top}\overline{\Sg}_{\bX}(t) A_2\rp \Bigg|.
\end{align*}
With $\|\cdot\|_F$ denoting the Frobenious norm, it can be further decomposed as
\begin{align*}
&\frac{1}{\tau} \sup_{0\le t \le s} \big| \sqrt{k}(\widehat{A}_1 - A_1)^{\top} \overline{\Sg}_{\bX}(t)\widehat{A}_2  + A_1^{\top} \overline{\Sg}_{\bX}(t) \sqrt{k}(\widehat{A}_2 - A_2) \big|
\\
&\le \frac{1}{\tau}  \lbr \sqrt{k}\|\widehat{A}_1 - A_1\|_F^{\top} \sup_{0\le t \le s} \|\overline{\Sg}_{\bX}(t)\|_F\|\widehat{A}_2\|_F  + \|A_1\|_F^{\top} \sup_{0\le t \le s} \|\overline{\Sg}_{\bX}(t)\|_F \sqrt{k}\|\widehat{A}_2 - A_2\|_F \rbr \convP 0.
\end{align*}
The convergence follows from the fact that $\overline{\Sg}_{\bX}(t) \convP t\Sg_{\bX}$, as established in Lemma~\ref{l:Sgbar}. It also uses the fact that $\sqrt{k}(\widehat{A}_i - A_i) = o_P(1)$, $i \in  \{1,2\}$, and $\widehat{A}_2 \convP A_2$, which can be shown similarly to the proof of Theorem~\ref{t:K|Lcon}, using Proposition~\ref{p:kXcon} and Lemma~\ref{l:Sgeq}.

\clearpage

\bibliographystyle{chicago}
\renewcommand{\baselinestretch}{0.9}
\small
\bibliography{biblio}

@article{cooley:thibaud:2019,
    title={Decompositions of dependence for high-dimensional extremes},
    author={Cooley, D. and Thibaud, E.},
    journal={Biometrika},
    volume={106},
    number={3},
    pages={587--604},
    year={2019}}

@article{larsson:resnick:2012,
    title={Extremal dependence measure and extremogram: 
                        the regularly varying case},
    author={M. Larsson and S. I. Resnick},
    journal={Extremes},
    volume={15},
    number={2},
    pages={231--256},
    year={2012}}

@book{resnick:2007,
    Author = {S. I. Resnick},
    Publisher = {Springer},
    Title = {Heavy--{T}ail {P}henomena},
    Year = 2007}

@book{resnick:1987,
	Author = {S. I. Resnick},
	Publisher = {Springer},
	Title = {Extreme {V}alues, {R}egular {V}ariation, 
                    and {P}oint {P}rocesses},
	Year = {1987}}

@book{billingsley:1999,
	Address = {New York},
	Author = {P. Billingsley},
	Publisher = {Wiley},
	Title = {Convergence of {P}robability {M}easures; {S}econd {E}dition},
	Year = {1999}}

@article{csorgo:deheuvels:mason:1985,
  title={Kernel estimates of the tail index of a distribution},
  author={S. Cs{\"o}rg{\H o} and P. Deheuvels and D. Mason},
  journal={The Annals of Statistics},
  pages={1050--1077},
	Volume = {13},
  year={1985}}

@article{haeusler:teugels:1985,
  title={On asymptotic normality of Hill's estimator for the exponent of regular variation},
  author={E. Haeusler and J. L. Teugels},
  journal={The Annals of Statistics},
  volume={13},
  pages={743--756},
  year={1985},
  publisher={Institute of Mathematical Statistics}}

@book{horn:jognson:2012,
  title={Matrix analysis},
  author={R. A. Horn and C. R. Johnson},
  year={2012},
  publisher={Cambridge university press}}

@article{kim:kokoszka:2023b,
  title={Extremal dependence measure for functional data},
  author={M. Kim and P. Kokoszka},
  journal={Journal of Multivariate Analysis},
  volume={189},
  pages={104887},
  year={2022}}

@article{kiriliouk:chen:2022,
  title={Estimating probabilities of multivariate failure sets based on pairwise tail dependence coefficients},
  author={A. Kiriliouk and C. Zhou},
  journal={arXiv preprint arXiv:2210.12618},
  year={2022}}

@article{resnick:starica:1997a,
  title={Asymptotic behavior of {H}ill's estimator for autoregressive data},
  author={S. I. Resnick and C. St{\u a}ric{\u a}},
  journal={Communications in statistics. Stochastic models},
  volume={13},
  number={4},
  pages={703--721},
  year={1997},
  publisher={Taylor \& Francis}}

@article{resnick:starica:1997b,
  title={Smoothing the {H}ill estimator},
  author={S. I. Resnick and C. St{\u a}ric{\u a}},
  journal={Advances in Applied Probability},
  volume={29},
  number={1},
  pages={271--293},
  year={1997},
  publisher={Cambridge University Press}}

@book{dehaan:ferreira:2006,
	Author = {L. {de Haan} and A. Ferreira},
	Publisher = {Springer},
	Title = {Extreme {V}alue {T}heory: an {I}ntroduction},
	Year = {2006}}

@article{amendola2022conditional,
  title={Conditional independence in max-linear Bayesian networks},
  author={Am{\'e}ndola, Carlos and Kl{\"u}ppelberg, Claudia and Lauritzen, Steffen and Tran, Ngoc M},
  journal={The Annals of Applied Probability},
  volume={32},
  number={1},
  pages={1--45},
  year={2022},
  publisher={Institute of Mathematical Statistics}
}

@article{gissibl2021identifiability,
  title={Identifiability and estimation of recursive max-linear models},
  author={Gissibl, Nadine and Kl{\"u}ppelberg, Claudia and Lauritzen, Steffen},
  journal={Scandinavian Journal of Statistics},
  volume={48},
  number={1},
  pages={188--211},
  year={2021},
  publisher={Wiley Online Library}
}

@article{engelke2020graphical,
  title={Graphical models for extremes},
  author={Engelke, Sebastian and Hitz, Adrien S},
  journal={Journal of the Royal Statistical Society: Series B (Statistical Methodology)},
  volume={82},
  number={4},
  pages={871--932},
  year={2020},
  publisher={Wiley Online Library}
}

@article{husler1989,
  title={Maxima of normal random vectors: between independence and complete dependence},
  author={H{\"u}sler, J{\"u}rg and Reiss, Rolf-Dieter},
  journal={Statistics \& Probability Letters},
  volume={7},
  number={4},
  pages={283--286},
  year={1989},
  publisher={Elsevier}
}

@article{kluppelberg2021estimating,
  title={Estimating an extreme Bayesian network via scalings},
  author={Kl{\"u}ppelberg, Claudia and Krali, Mario},
  journal={Journal of Multivariate Analysis},
  volume={181},
  pages={104672},
  year={2021},
  publisher={Elsevier}
}

@article{tran2021estimating,
    author = {Tran, Ngoc Mai and Buck, Johannes and Klüppelberg, Claudia},
    title = "{Estimating a directed tree for extremes}",
    journal = {Journal of the Royal Statistical Society Series B: Statistical Methodology},
    volume = {86},
    number = {3},
    pages = {771-792},
    year = {2024},
    month = {02},
    issn = {1369-7412},
    doi = {10.1093/jrsssb/qkad165},
    url = {https://doi.org/10.1093/jrsssb/qkad165},
    eprint = {https://academic.oup.com/jrsssb/article-pdf/86/3/771/58504442/qkad165.pdf},
}

@article{gissibl2018max,
  title={Max-linear models on directed acyclic graphs},
  author={Gissibl, Nadine and Kl{\"u}ppelberg, Claudia},
  journal={Bernoulli},
  volume={24},
  number={4A},
  pages={2693--2720},
  year={2018},
  publisher={Bernoulli Society for Mathematical Statistics and Probability}
}

@book{whittaker:1990,
  title={Graphical Models in Applied Multivariate Statistics},
  author={J. Whittaker},
  year={1990},
  publisher={New York: Wiley}}

@book{lauritzen:1996,
  title={Graphical Models},
  author={S. L. Lauritzen},
  year={1996},
  publisher={Oxford University Press}}

@book{anderson:2003,
  title={An introduction to multivariate statistical analysis},
  author={T. W. Anderson},
  year={2003},
  publisher={New York: Wiley}}

@article{lee2021transformed,
  title={Transformed-linear prediction for extremes},
  author={Lee, Jeongjin and Cooley, Daniel},
  journal={arXiv preprint arXiv:2111.03754},
  year={2021}
}

@article{asadi2015extremes,
  title={Extremes on river networks},
  author={Asadi, Peiman and Davison, Anthony C and Engelke, Sebastian},
  journal={The Annals of Applied Statistics},
  volume={9},
  number={4},
  pages={2023--2050},
  year={2015},
  publisher={Institute of Mathematical Statistics}
}

@article{engelke2022structure,
  title={Structure learning for extremal tree models},
  author={Engelke, Sebastian and Volgushev, Stanislav},
  journal={Journal of the Royal Statistical Society: Series B (Statistical Methodology)},
  volume={84},
  number={5},
  pages={2055--2087},
  year={2022},
  publisher={Wiley Online Library}
}

@phdthesis{lee2022phd,
  title={Linear prediction and partial tail correlation for extremes},
  school={Colorado State University},
  author={Jeongjin Lee},
  year={2022}
}

@article{gong2024partial,
  title={Partial tail-correlation coefficient applied to extremal-network learning},
  author={Gong, Yan and Zhong, Peng and Opitz, Thomas and Huser, Rapha{\"e}l},
  journal={Technometrics},
  volume={66},
  number={3},
  pages={331--346},
  year={2024},
  publisher={Taylor \& Francis}
}

@article{van2016systematic,
  title={Systematic tail risk},
  author={Van Oordt, Maarten RC and Zhou, Chen},
  journal={Journal of Financial and Quantitative Analysis},
  volume={51},
  number={2},
  pages={685--705},
  year={2016},
  publisher={Cambridge University Press}}

@article{van2019estimating,
  title={Estimating systematic risk under extremely adverse market conditions},
  author={Van Oordt, Maarten RC and Zhou, Chen},
  journal={Journal of Financial Econometrics},
  volume={17},
  number={3},
  pages={432--461},
  year={2019},
  publisher={Oxford University Press}
}

\end{document}


\title{Supplementary Material}

\author{Mihyun Kim\textsuperscript{1}\thanks{Department of Statistics, West Virginia University, Morgantown, WV 26506, USA.}
\and Jeongjin Lee\textsuperscript{1}\thanks{School of Mathematical Sciences, Lancaster University, Fylde College, Lancaster, LA1 4YF, United Kingdom. Corresponding author. E-mail: j.lee58@lancaster.ac.uk}
}

\maketitle

\section{Additional numerical results}

\subsection{A matrix $C$ from a uniform distribution}

\begin{figure}[ht]
\centering
\includegraphics[width=3.5cm]{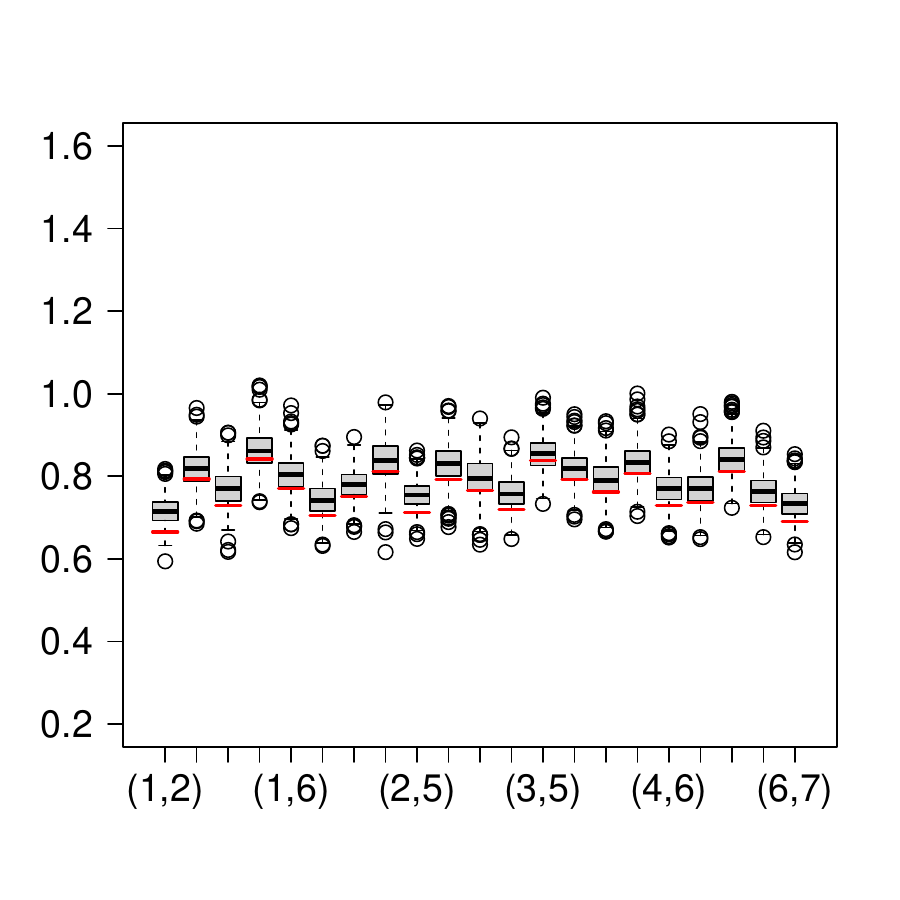}
\includegraphics[width=3.5cm]{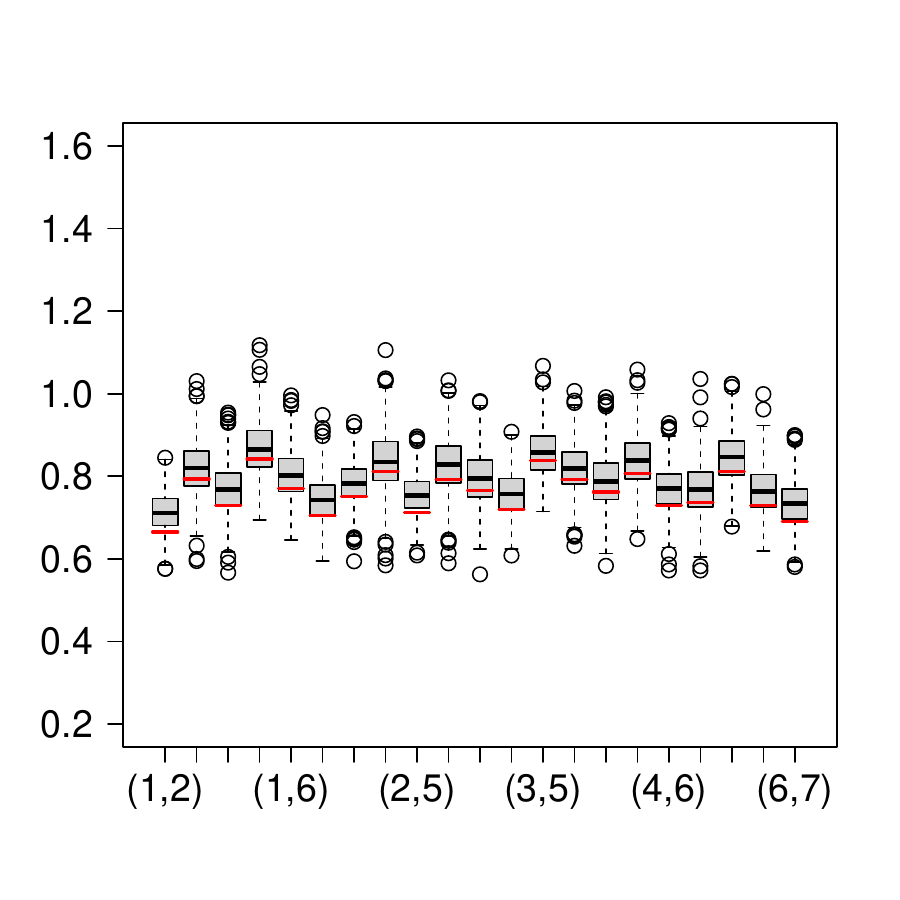}
\includegraphics[width=3.5cm]{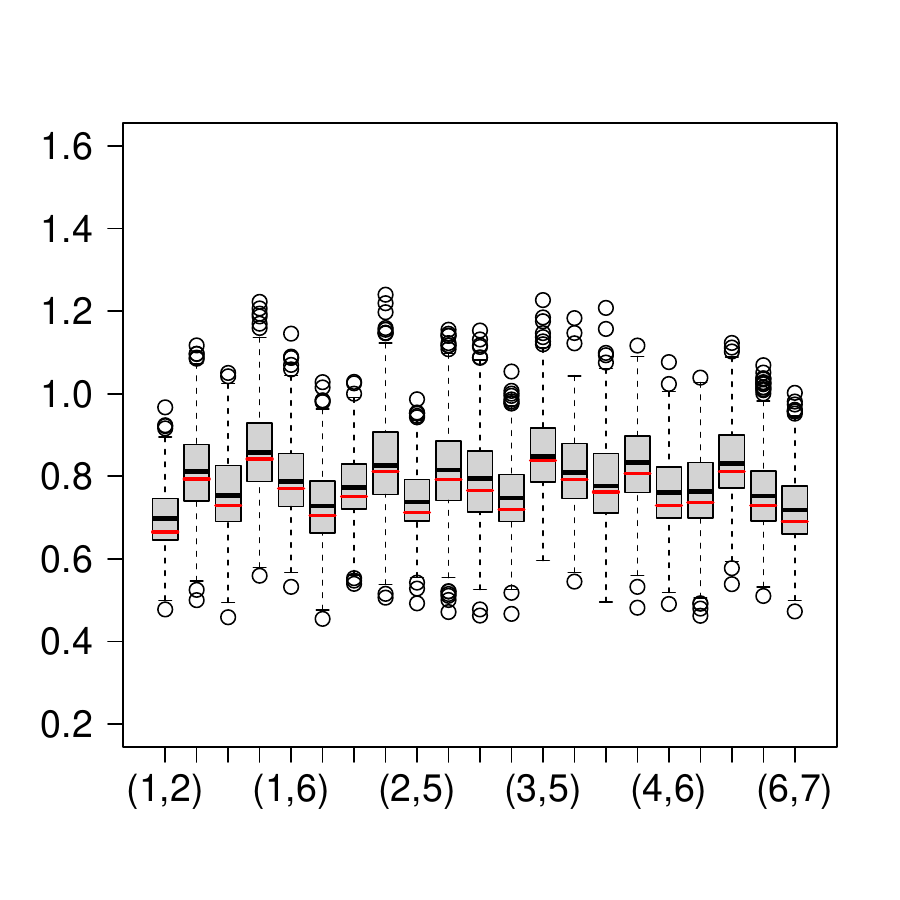}
\includegraphics[width=3.5cm]{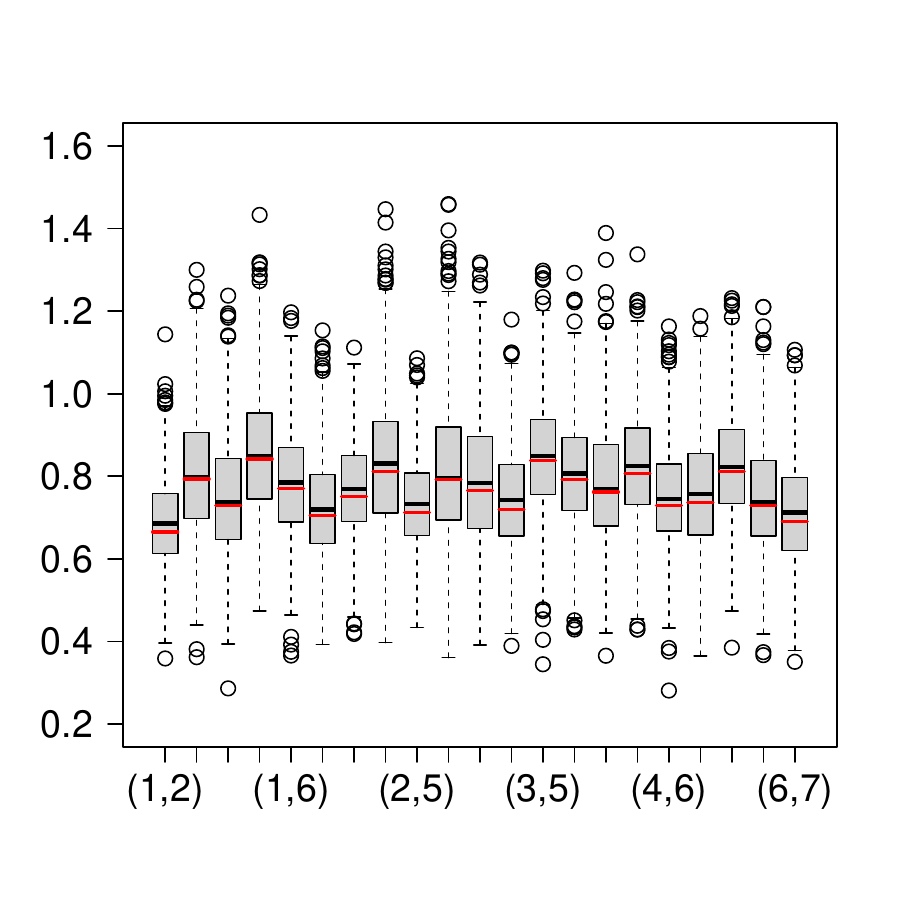}\\
\includegraphics[width=3.5cm]{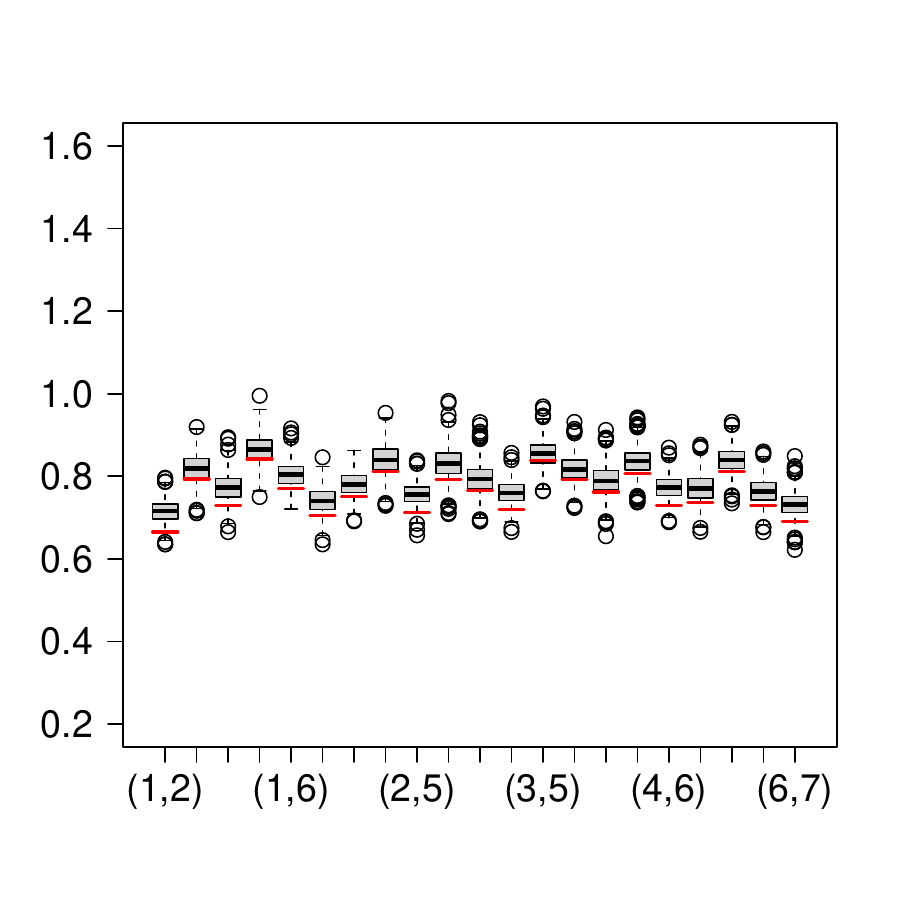}
\includegraphics[width=3.5cm]{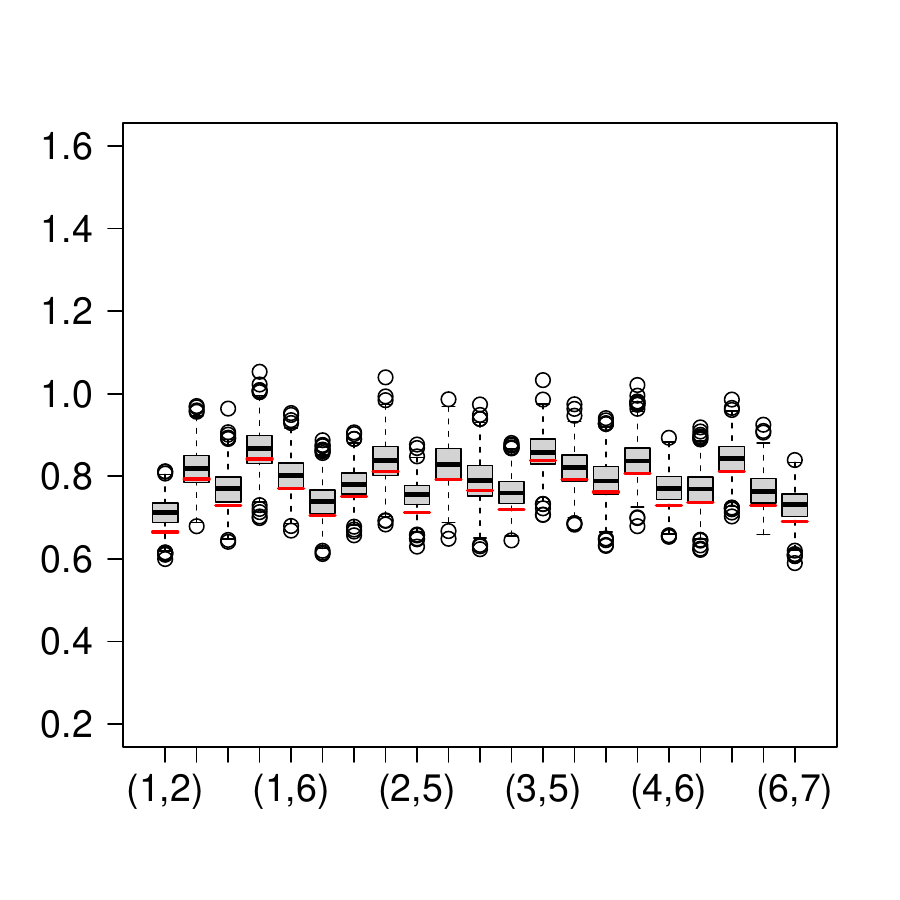}
\includegraphics[width=3.5cm]{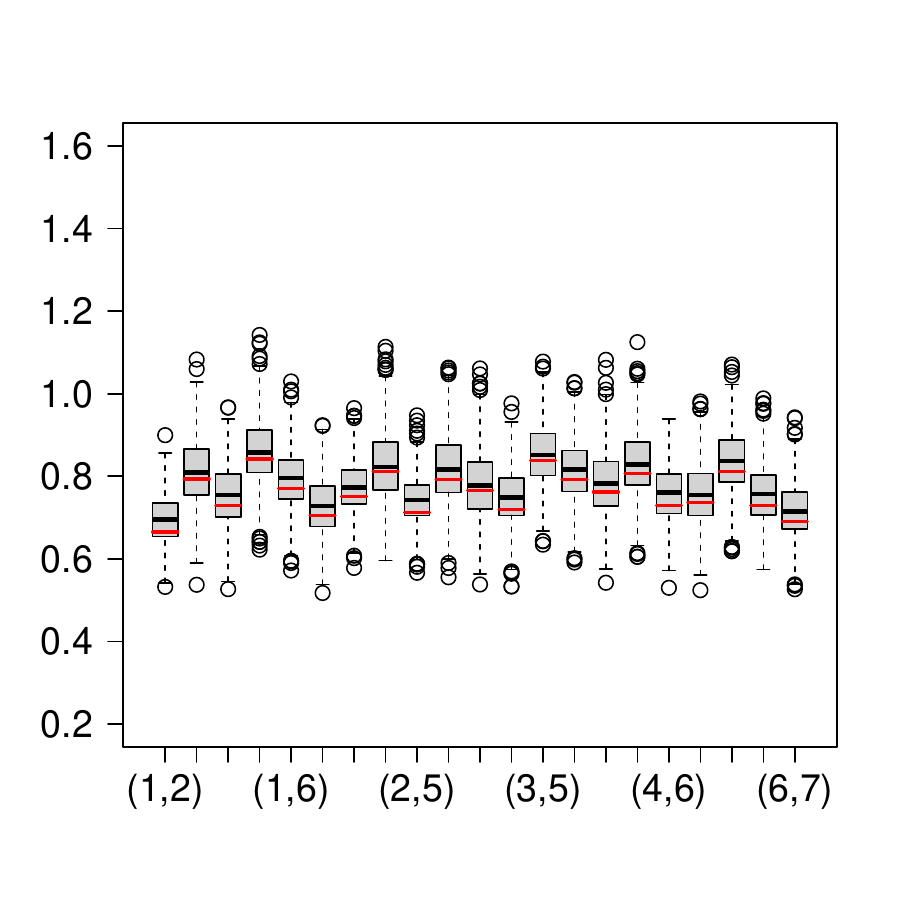}
\includegraphics[width=3.5cm]{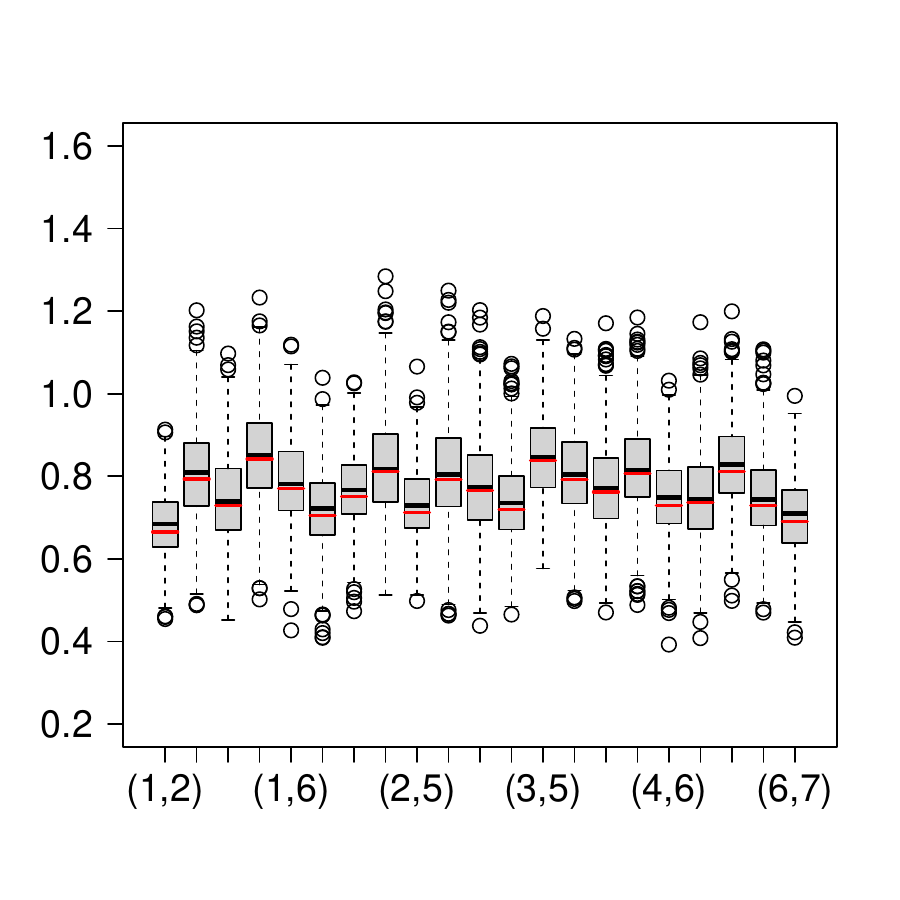}\\
\includegraphics[width=3.5cm]{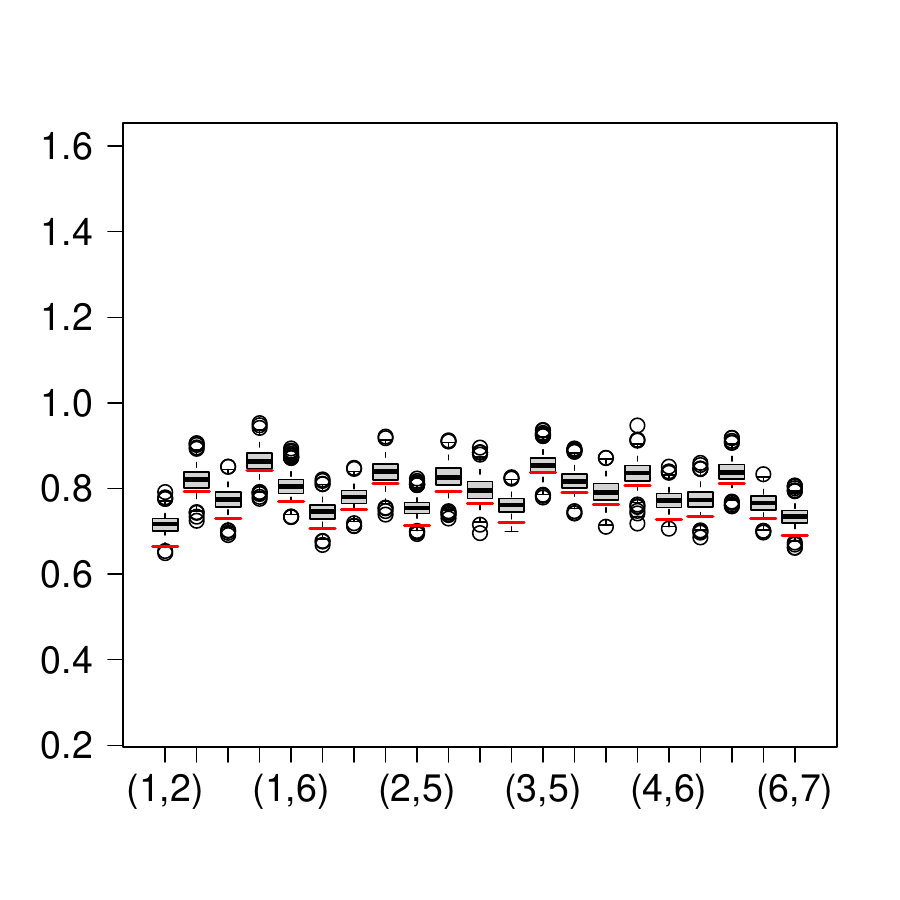}
\includegraphics[width=3.5cm]{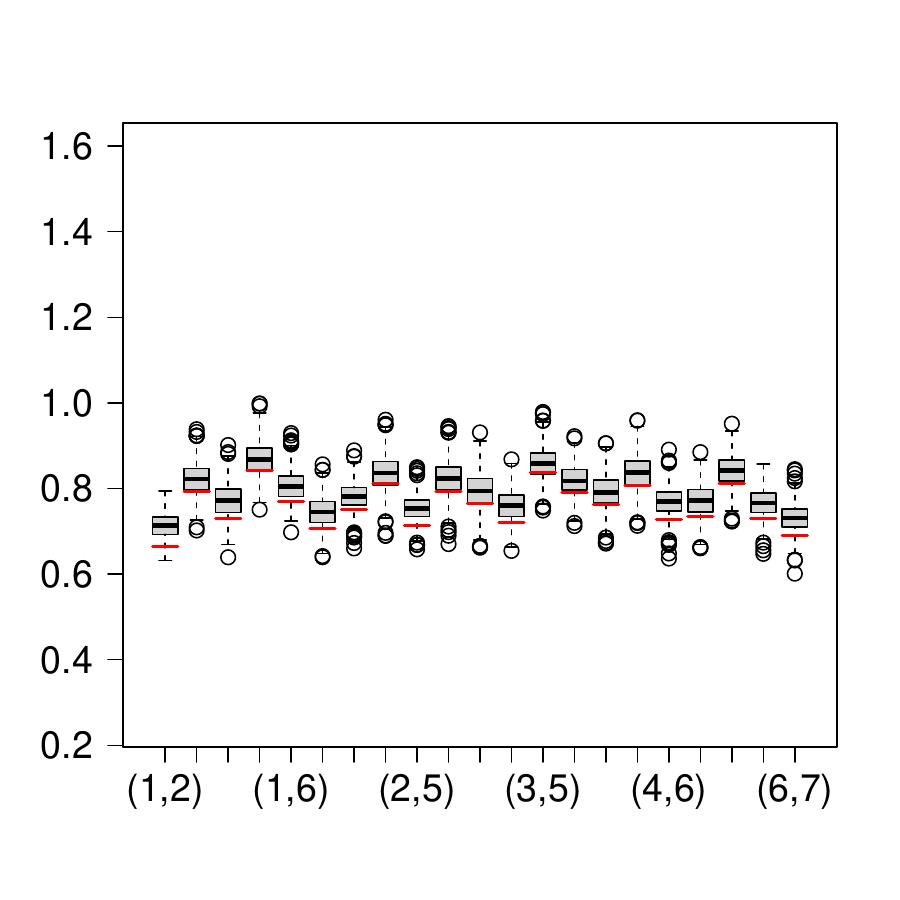}
\includegraphics[width=3.5cm]{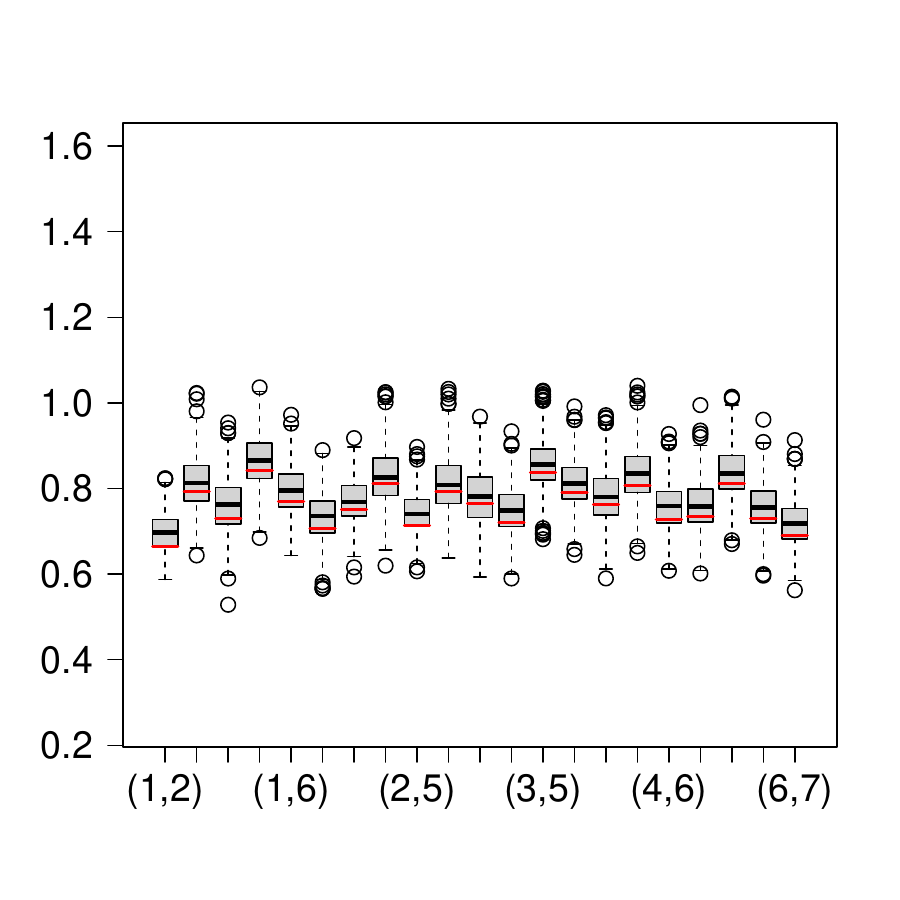}
\includegraphics[width=3.5cm]{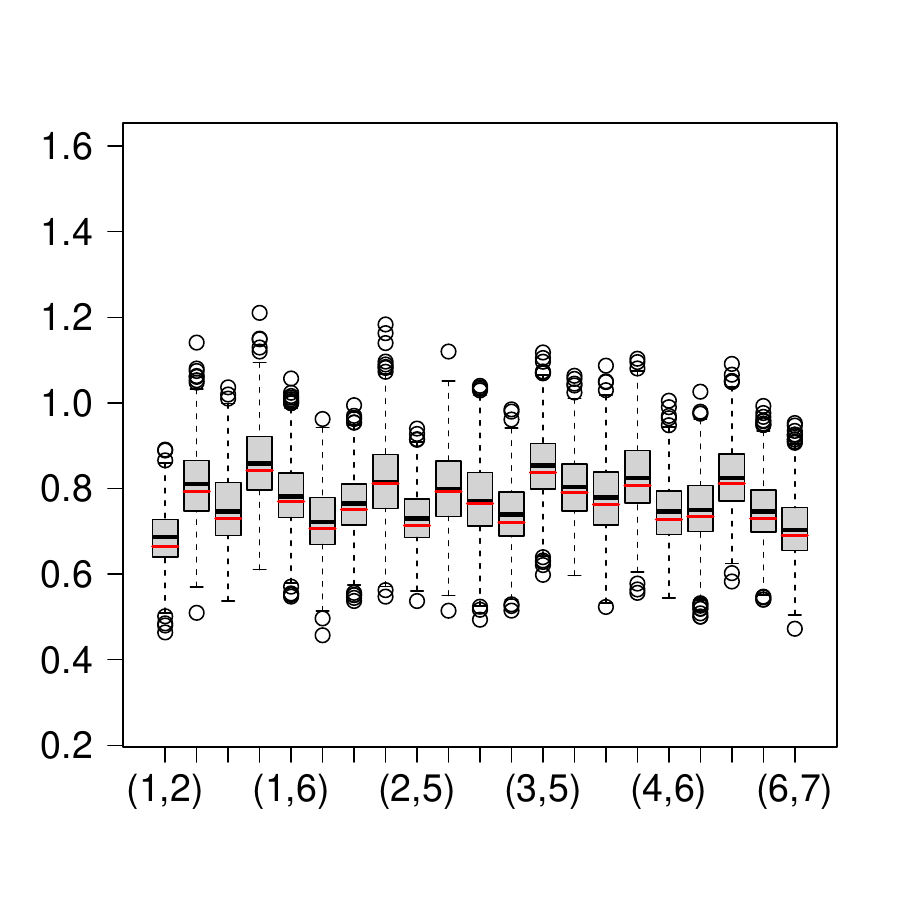}
\caption{\label{fig:TPDMest_unif}
Box-plots of the off-diagonal TPDM estimates across different $(n,k)$ under $1{,}000$ iterations.
Panels are arranged with sample sizes $n=1500, 2500, 4000$ from top to bottom and thresholds $k/n=0.1, 0.05, 0.02, 0.01$ from left to right.
Red lines indicate the specified off-diagonal TPDM values.
}
\end{figure}

We investigate the bias and variability of the TPDM estimator, $\widehat{\Sigma}_{\bX}(n,k)$ across varying $(n,k)$ under 1,000 iterations.
We consider three different sample sizes, $n\in\{1500,2500,4 000\}$ and four thresholds, $k/n\in\{0.1, 0.05, 0.02, 0.01\}.$
In the first setting, positive extremal dependence in $\bm{X}=(X_1,\ldots,X_7)^\top\in\reals_+^7$ is generated through a matrix multiplication $\bX=C\circ \bZ$ ,where the elements of the matrix $C\in\reals^{7\times 30}$ are taken from a specific realization from a uniform distribution on $[0,5].$
Figure~\ref{fig:TPDMest_unif} presents box-plots of the off-diagonal TPDM estimates with total mass $m=7$, showing that increasing the effective sample size reduces variability while decreasing the threshold ratio $k/n$ reduces bias.

\subsection{A matrix $C$ from an AR(1) model}
\begin{figure}[ht]
\centering
\includegraphics[width=4.5cm]{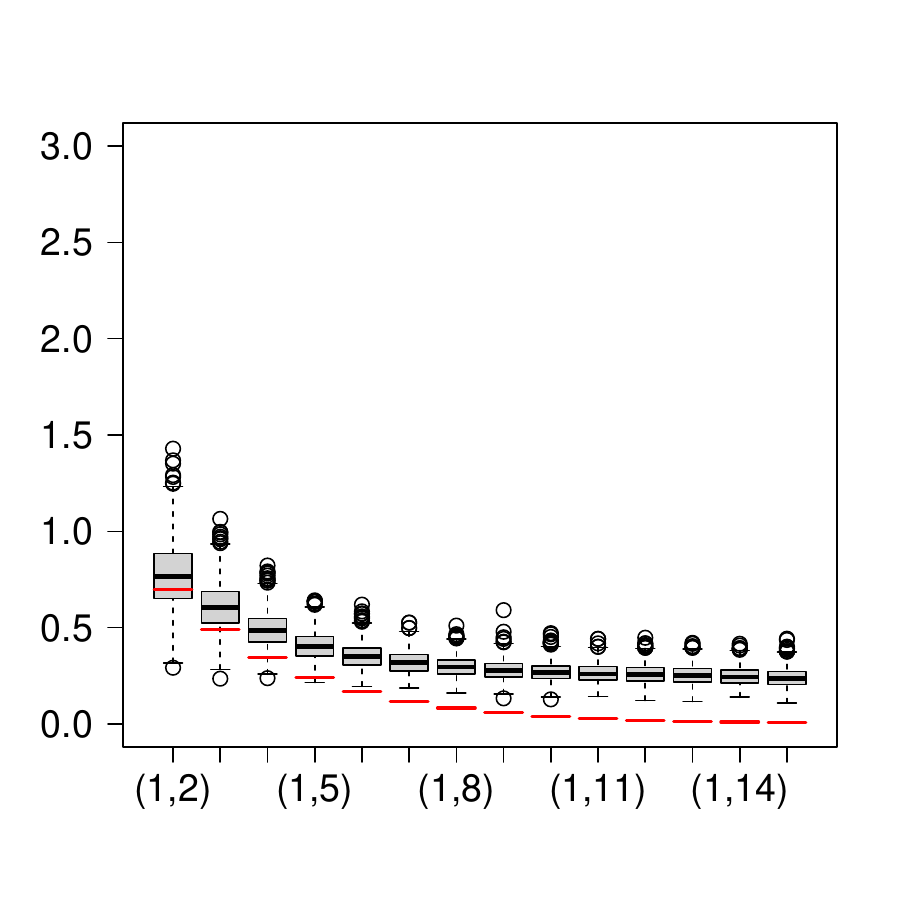}
\includegraphics[width=4.5cm]{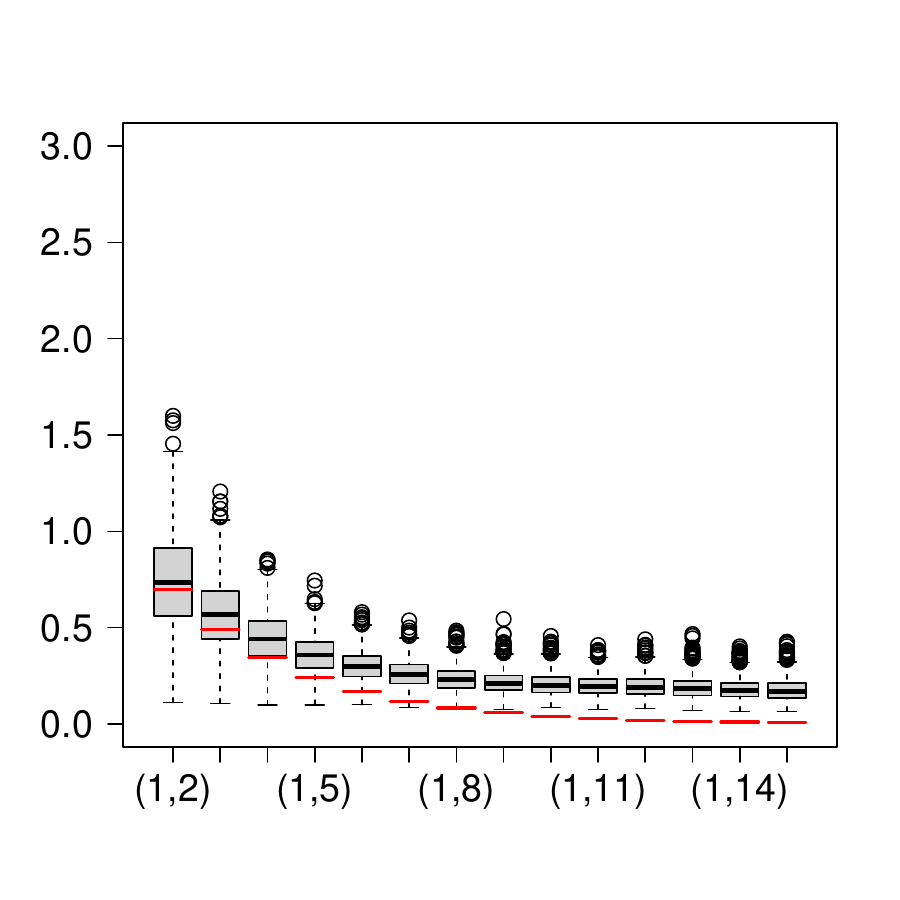}
\includegraphics[width=4.5cm]{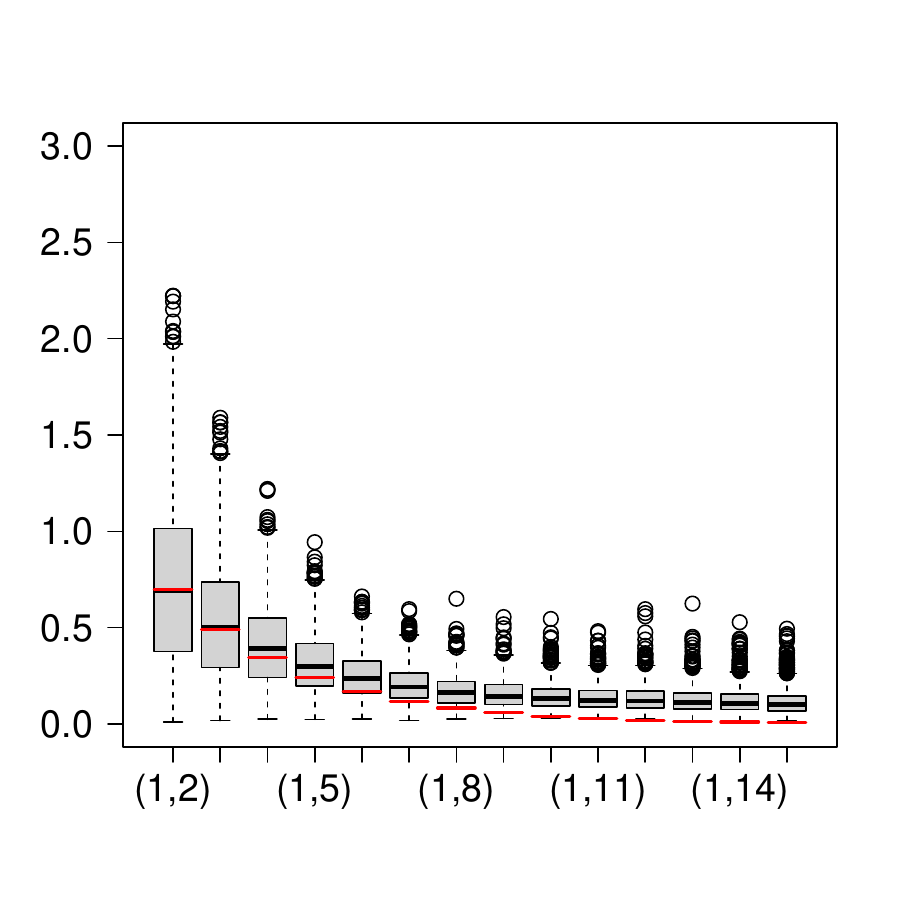}\\
\vspace{-0.3cm}
\includegraphics[width=4.5cm]{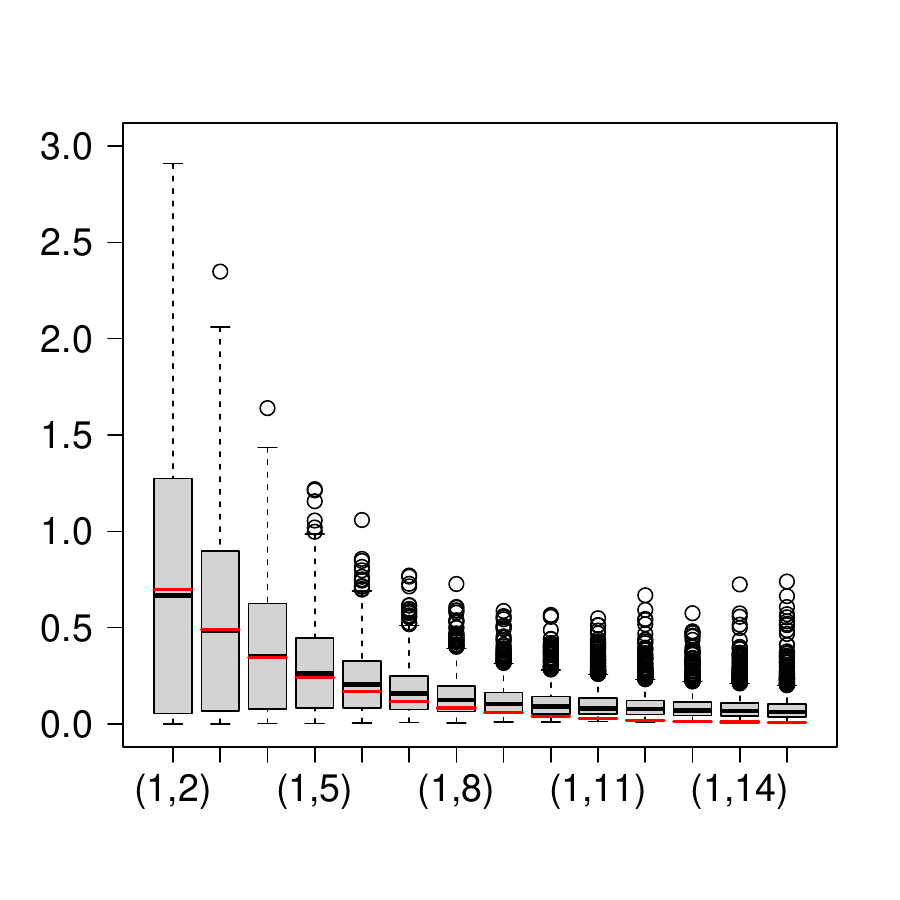}
\includegraphics[width=4.5cm]{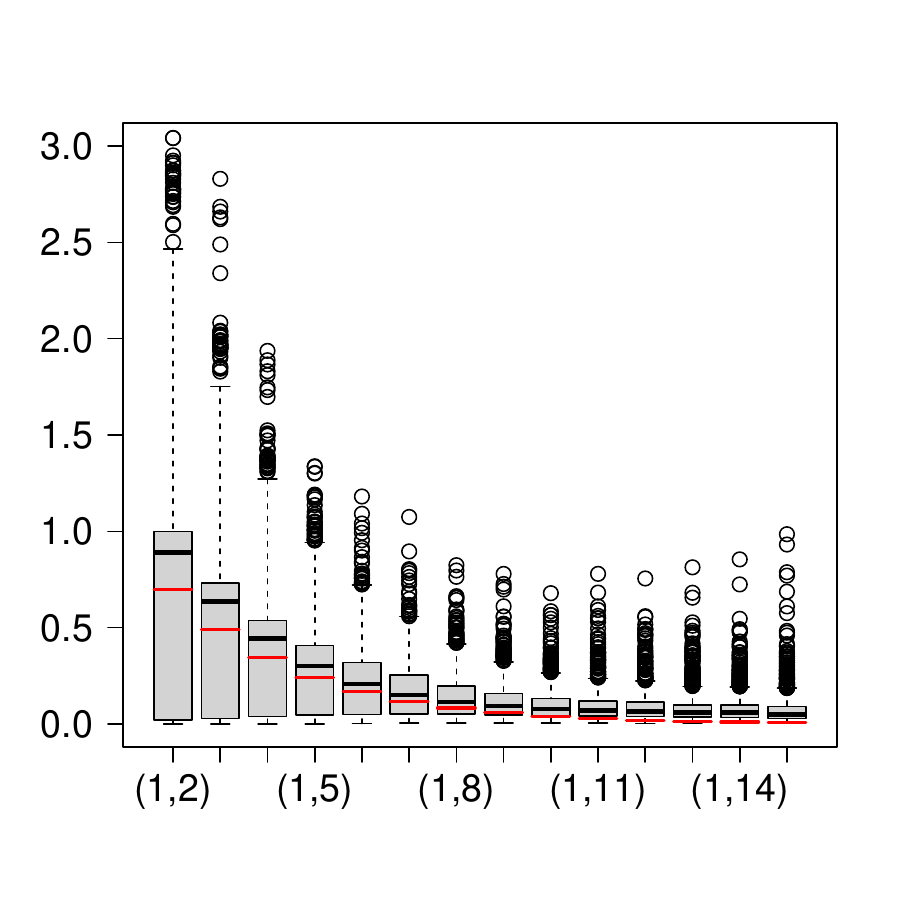}
\caption{\label{fig:TPDMest_AR_n1000}
Box-plots of the first-row off-diagonal TPDM estimates across different $(n,k)$ under $1,000$ iterations.
Panels are arranged with a sample size $n=1,500$ and thresholds $k/n\in\{0.1, 0.05, 0.02, 0.01, 0.007\}$ displayed from top to bottom and left to right.
Red lines indicate the specified off-diagonal TPDM values.
}
\end{figure}

\begin{figure}[ht]
\centering
\includegraphics[width=4.5cm]{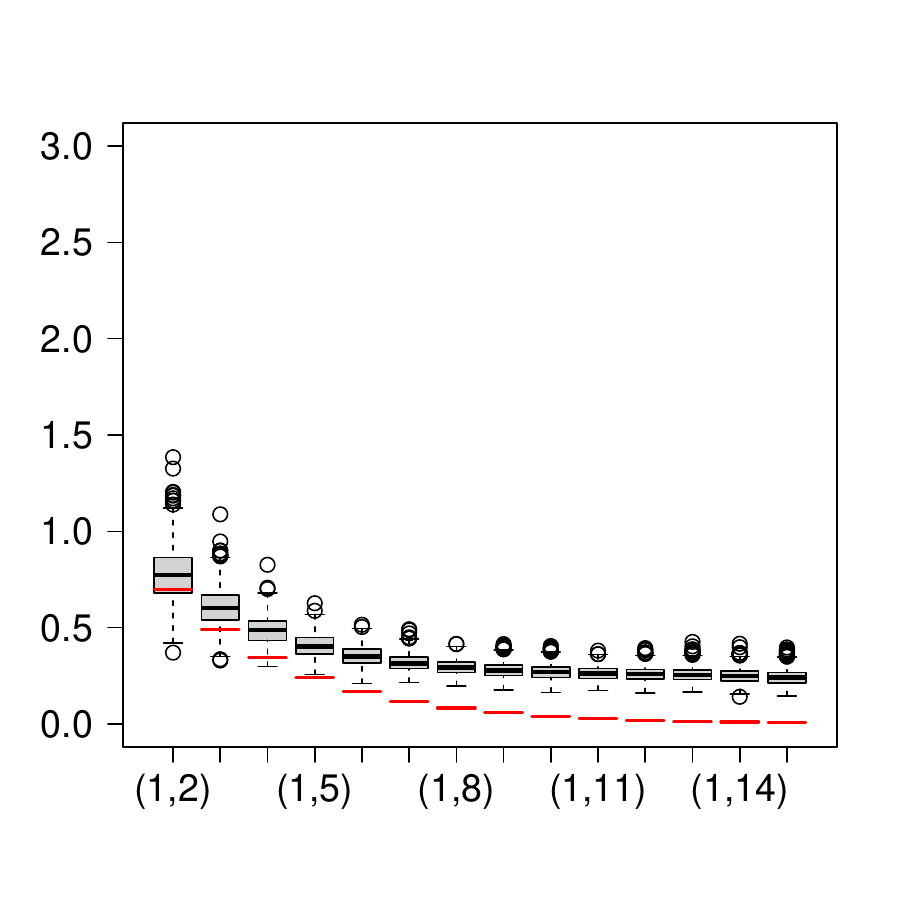}
\includegraphics[width=4.5cm]{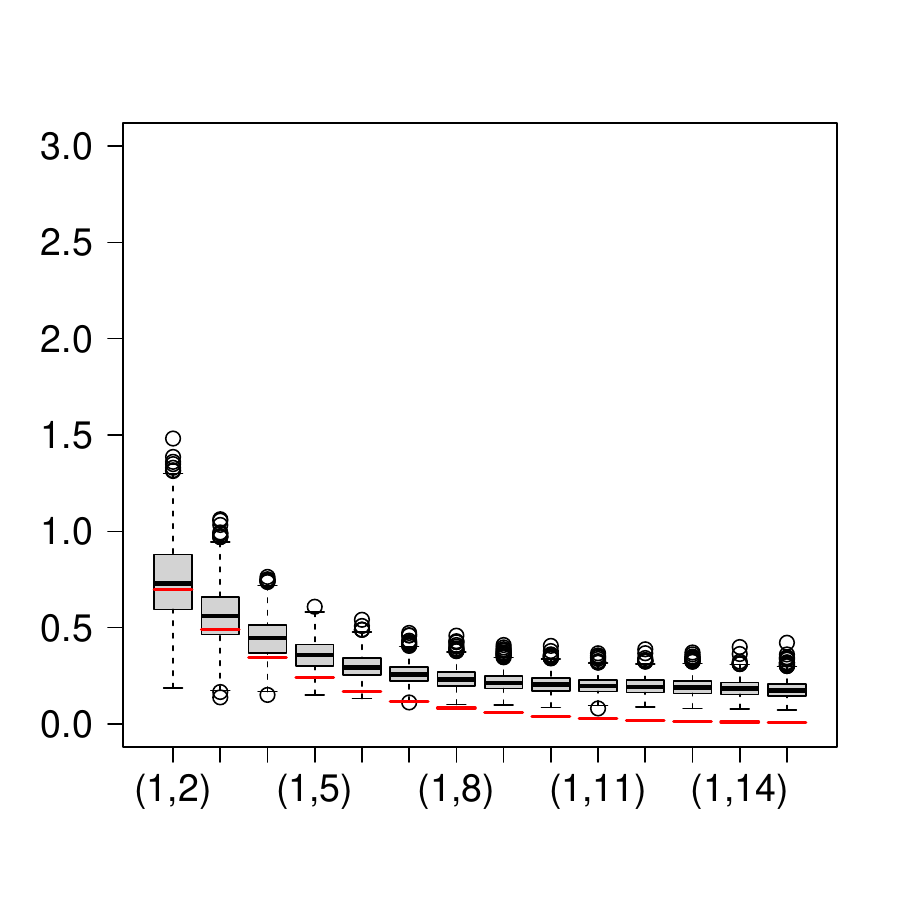}
\includegraphics[width=4.5cm]{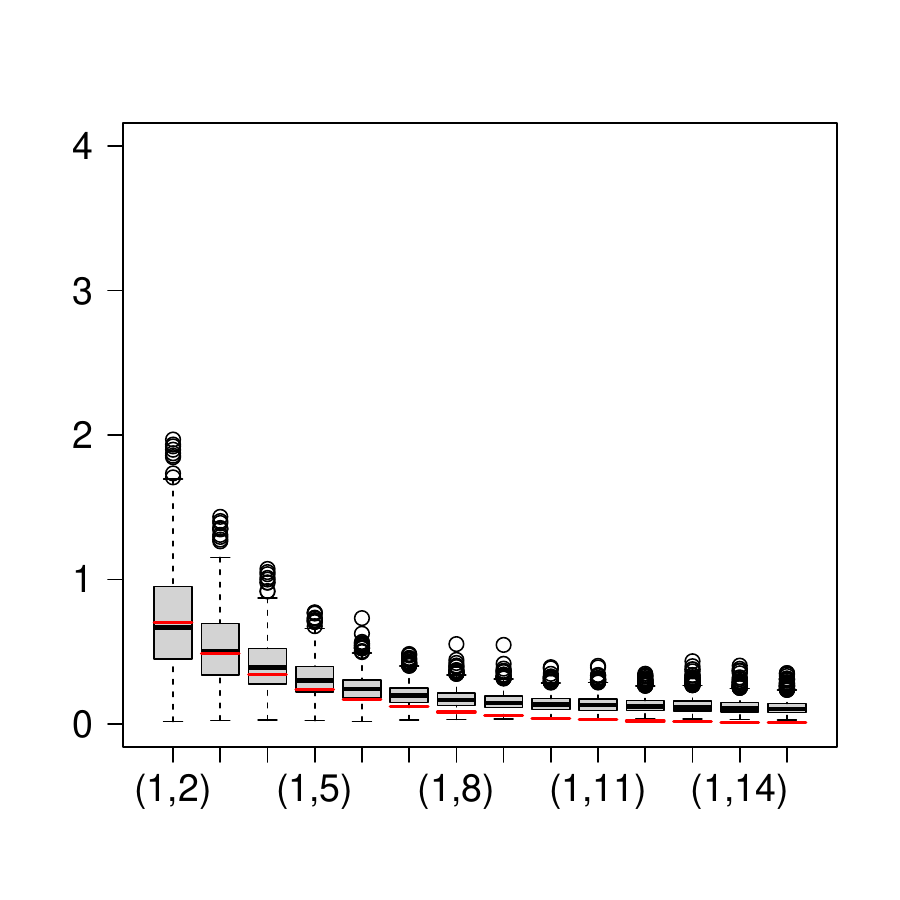}\\
\vspace{-0.3cm}
\includegraphics[width=4.5cm]{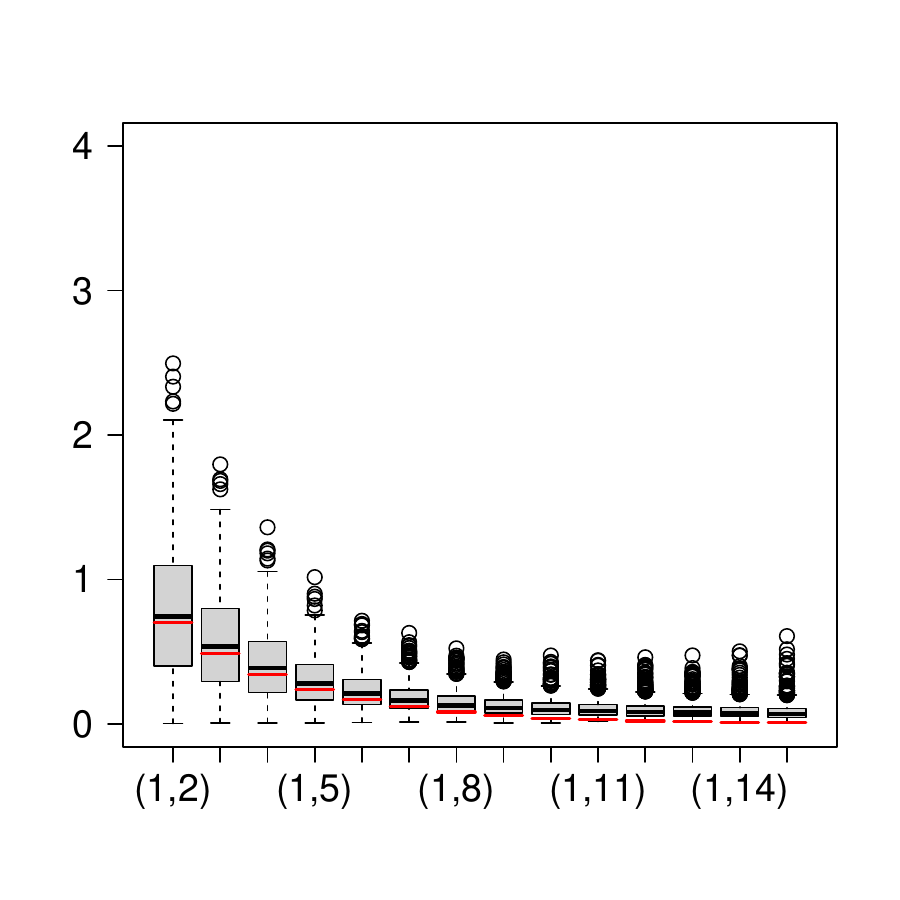}
\includegraphics[width=4.5cm]
{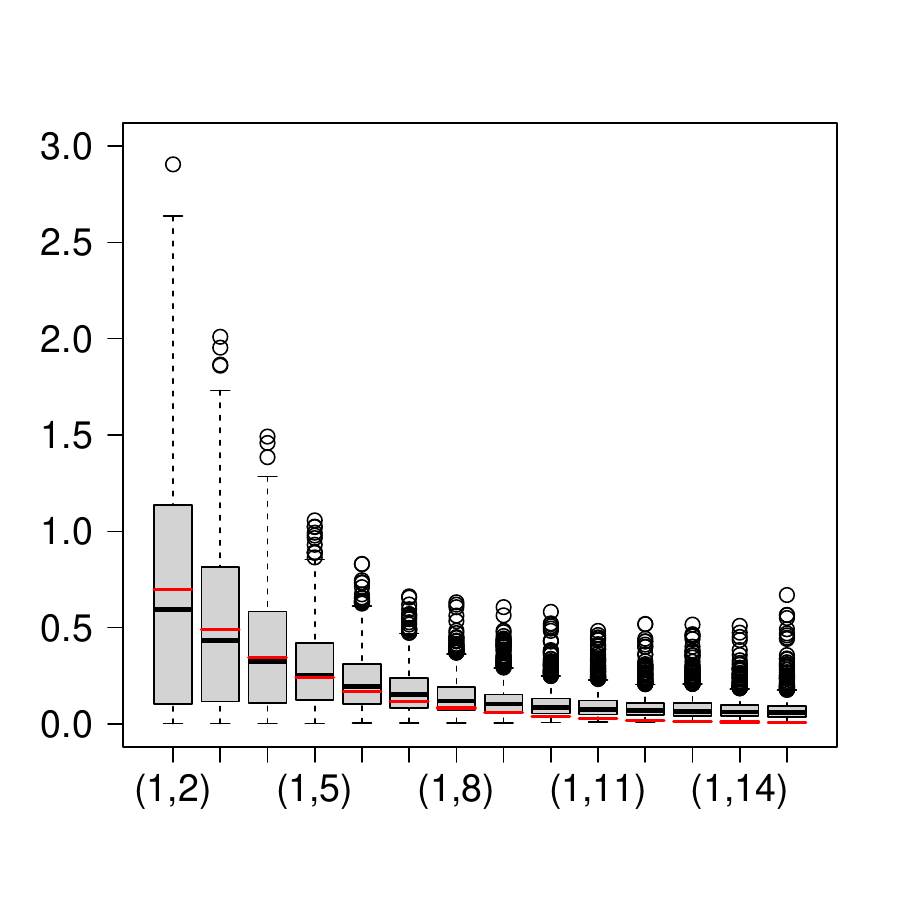}
\caption{\label{fig:TPDMest_AR_n2000}
The description is identical to that of Figure in~\ref{fig:TPDMest_AR_n1000}, with the different $n=2,500$.
}
\end{figure}

\begin{figure}[ht]
\centering
\includegraphics[width=4.5cm]
{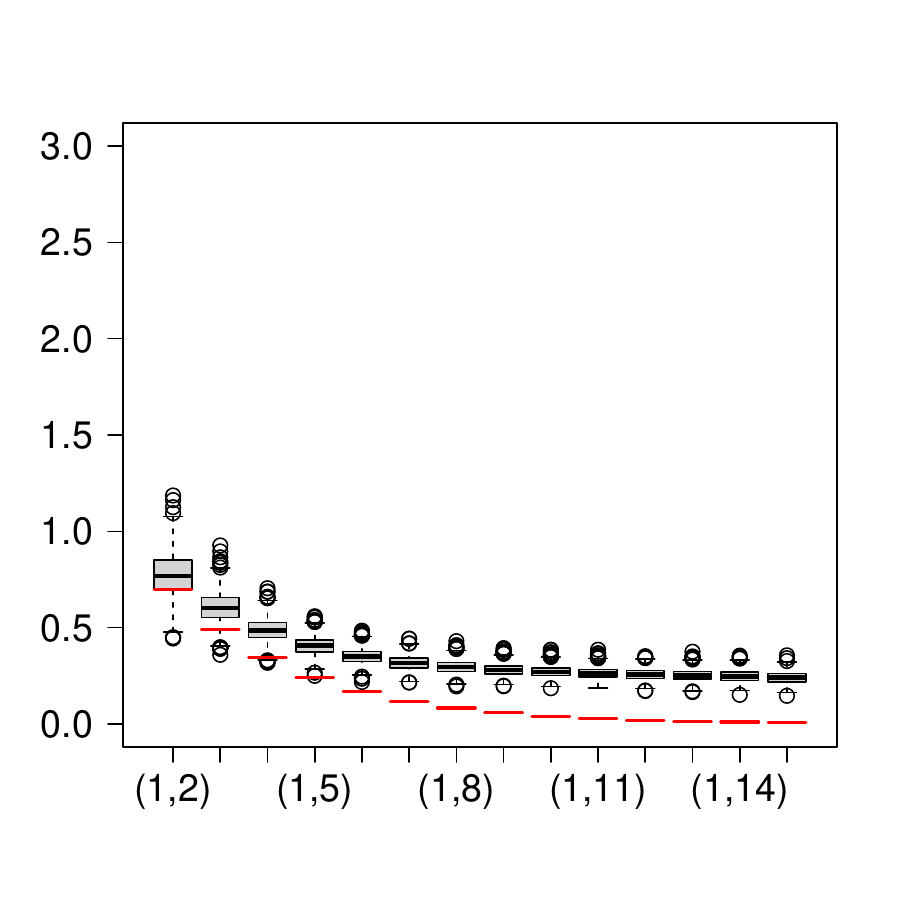}
\includegraphics[width=4.5cm]{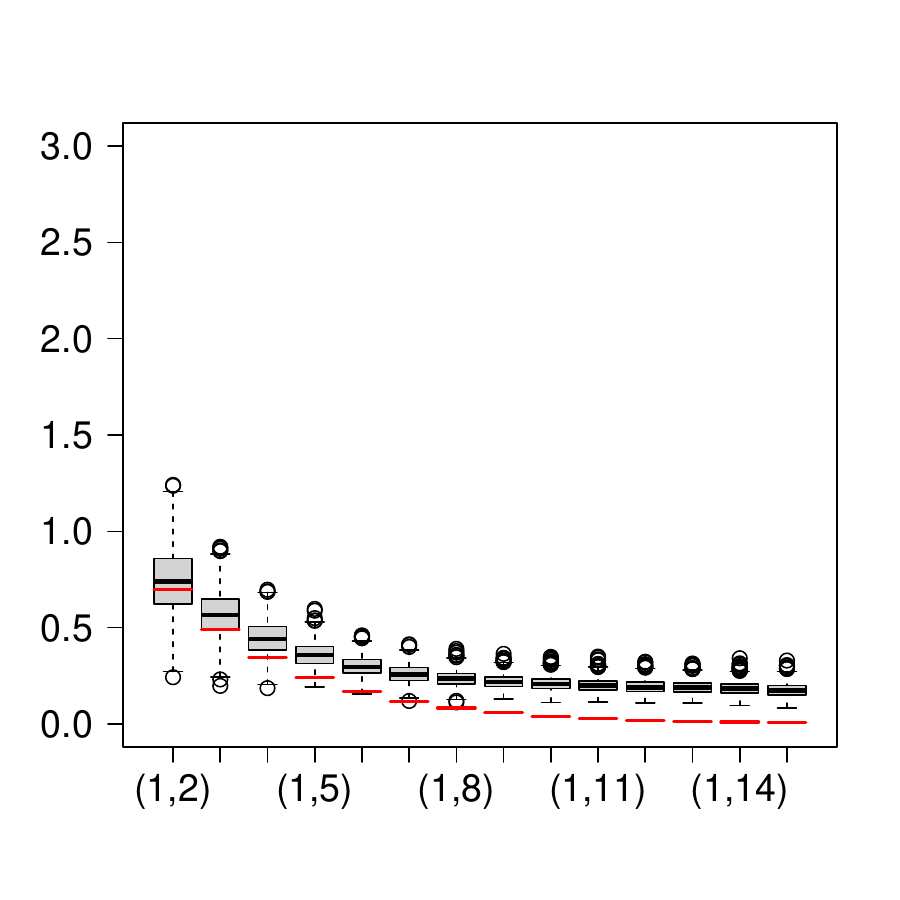}
\includegraphics[width=4.5cm]{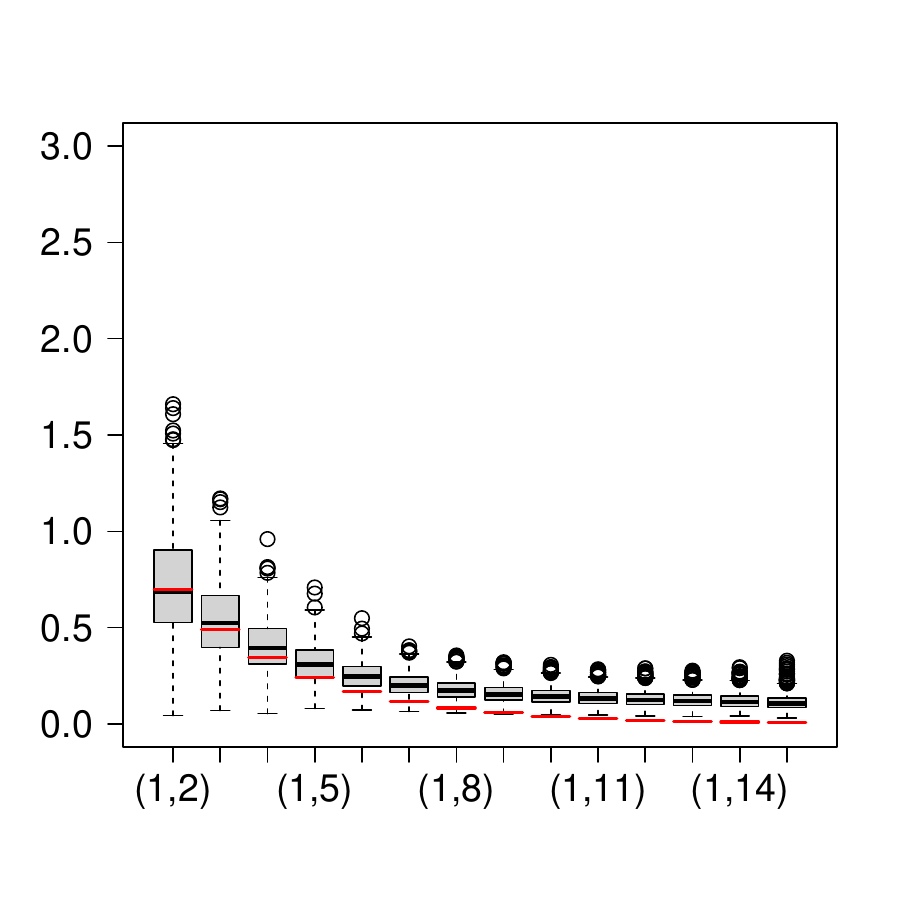}\\
\vspace{-0.3cm}
\includegraphics[width=4.5cm]{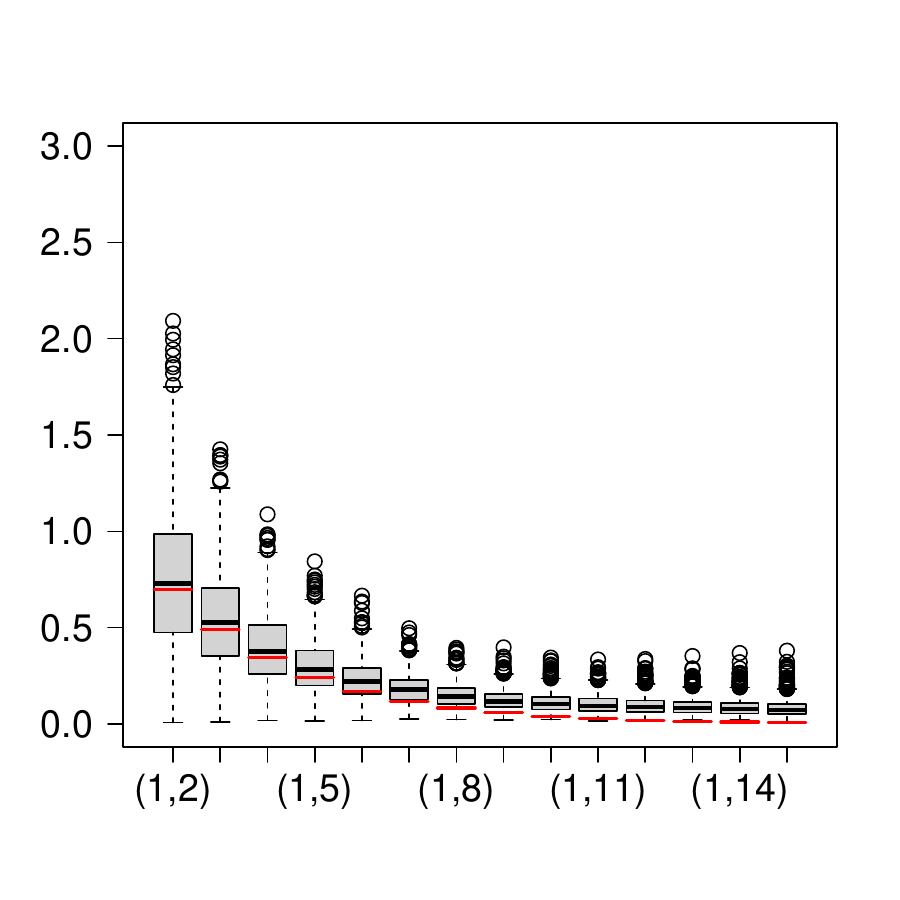}
\includegraphics[width=4.5cm]{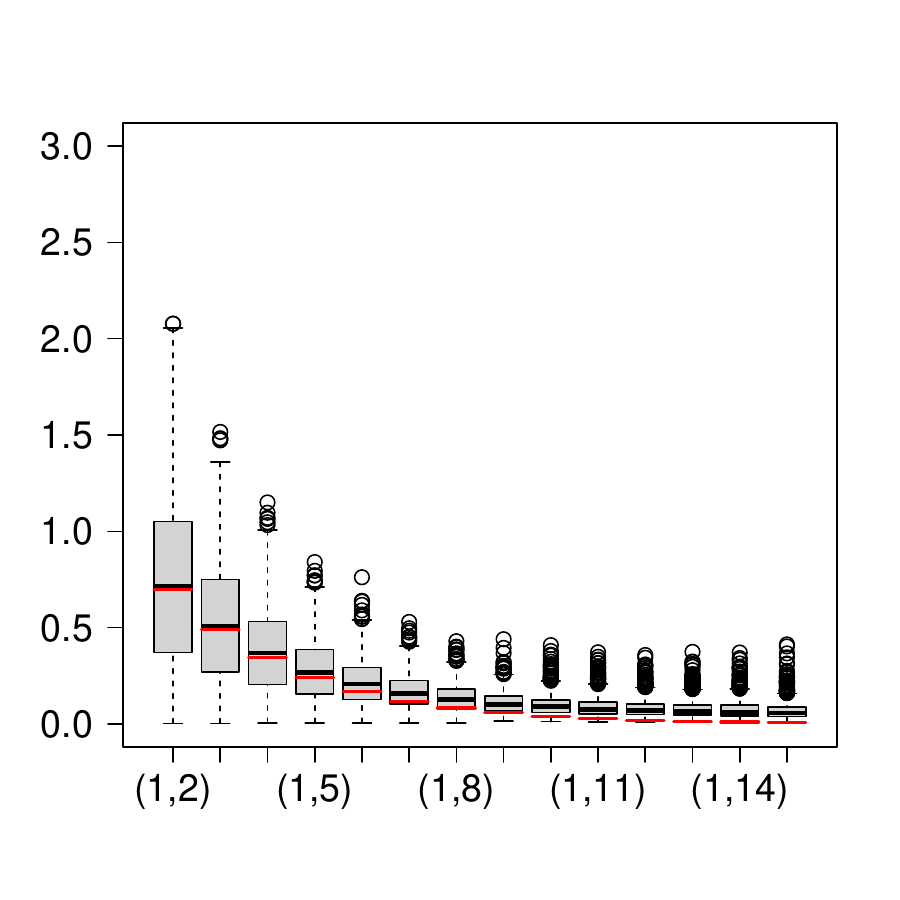}
\caption{\label{fig:TPDMest_AR_n4000}
The description is identical to that of Figure in~\ref{fig:TPDMest_AR_n1000}, with the different $n=4,000$.
}
\end{figure}

We also consider a finite version of the extremal AR(1) transformed-linear model to induce sparsity.
Specifically, we let $C\in\reals^{15\times 15}$ be a lower triangular Toeplitz matrix with elements $C_{ij}=\phi^{i-j}I_{i\ge j}$, where $\phi=0.7$ induces positive dependence among the components of $\bX$.
Figures~\ref{fig:TPDMest_AR_n1000}, \ref{fig:TPDMest_AR_n2000}, and \ref{fig:TPDMest_AR_n4000} display box-plots of the first-row off-diagonal TPDM estimates across different $n\in\cbr{1500,2500,4000}$ and $k/n\in\cbr{0.1,0.05,0.02,0.01,0.007}$.
The total mass is $m=\text{tr}(\Sigma_{\bX})=27.53$.
As expected, the bias and variability behaviors are similar to those observed where $C$ is drawn from a uniform distribution: larger effective sample sizes lead to lower variability, while smaller threshold ratios $k/n$ reduce bias.
In particular, lower thresholds $k/n$ are needed to further reduce bias for TPDM elements corresponding to larger lags, which tend to be near zero.

\begin{figure}[ht]
\centering
\includegraphics[width=4cm]{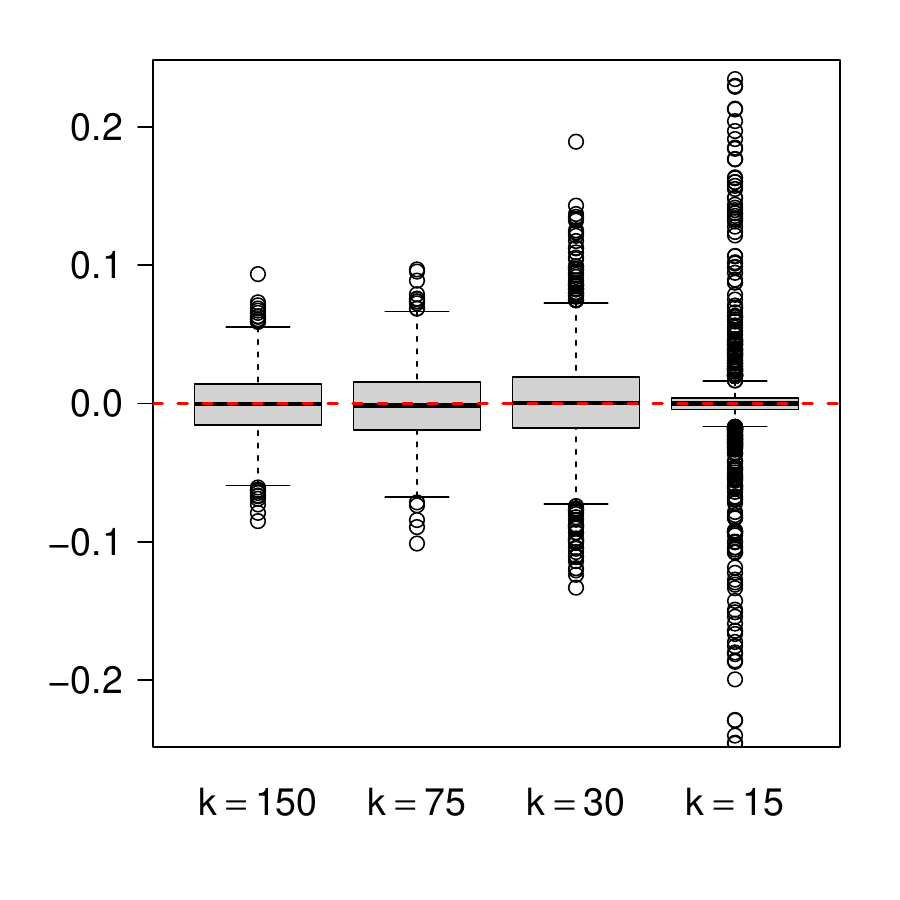}
\includegraphics[width=4cm]{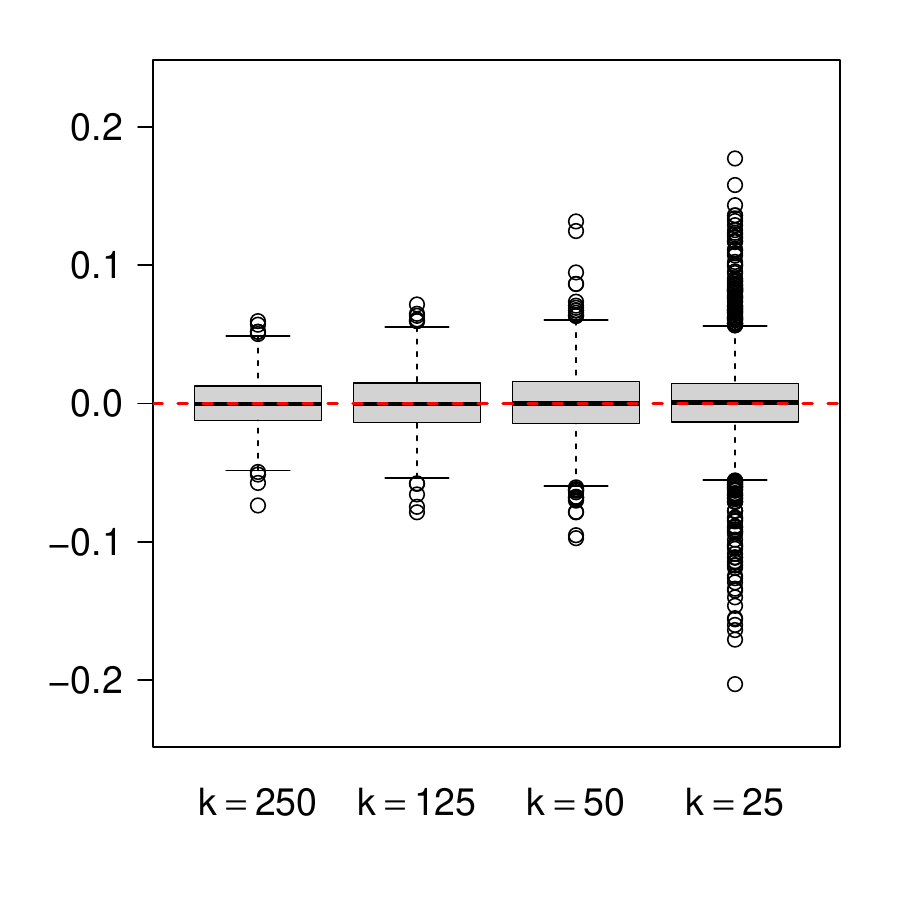}
\includegraphics[width=4cm]{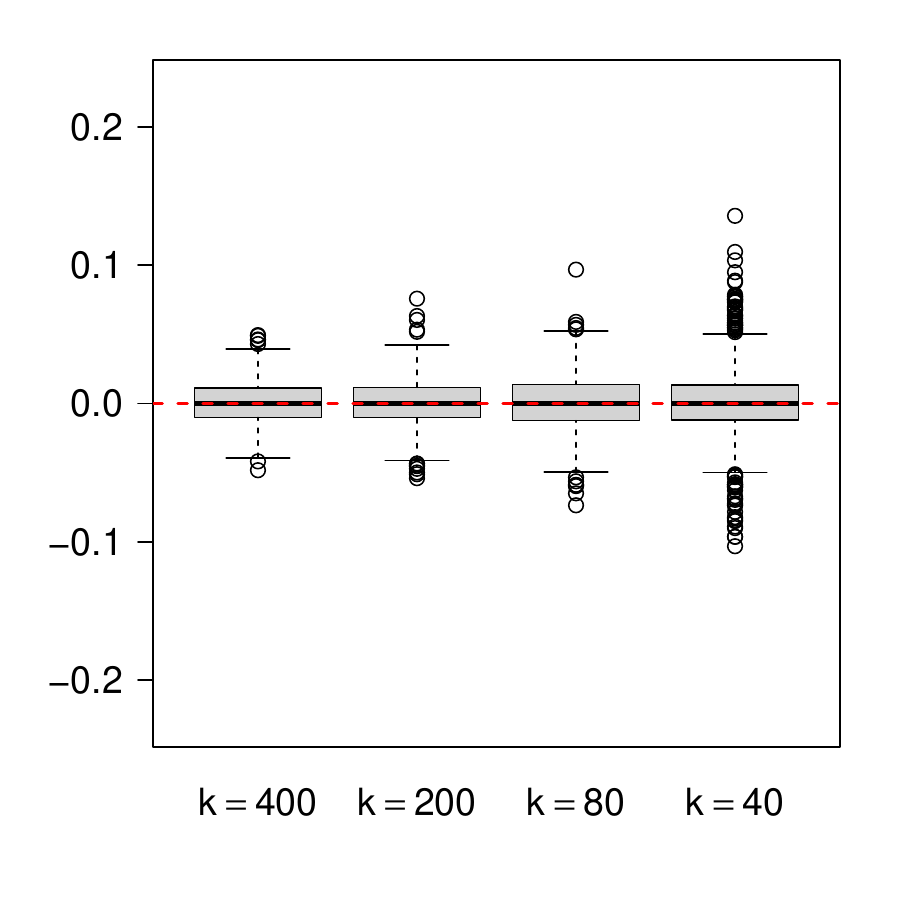}\\
\includegraphics[width=4cm]{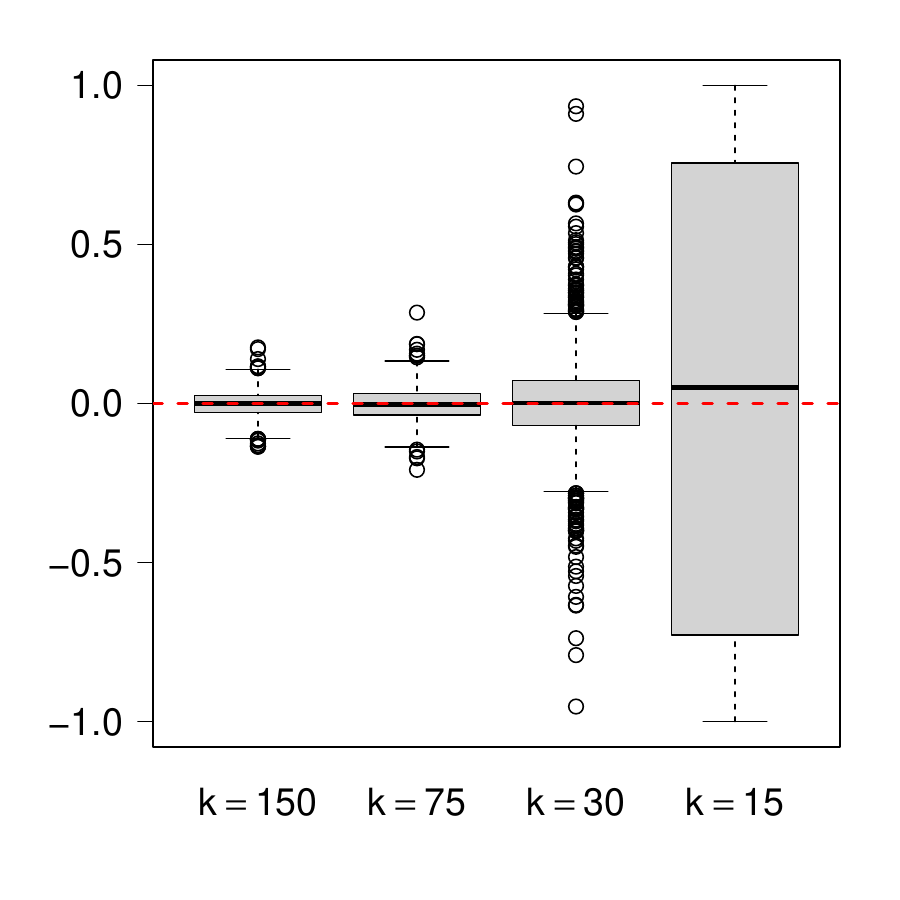}
\includegraphics[width=4cm]{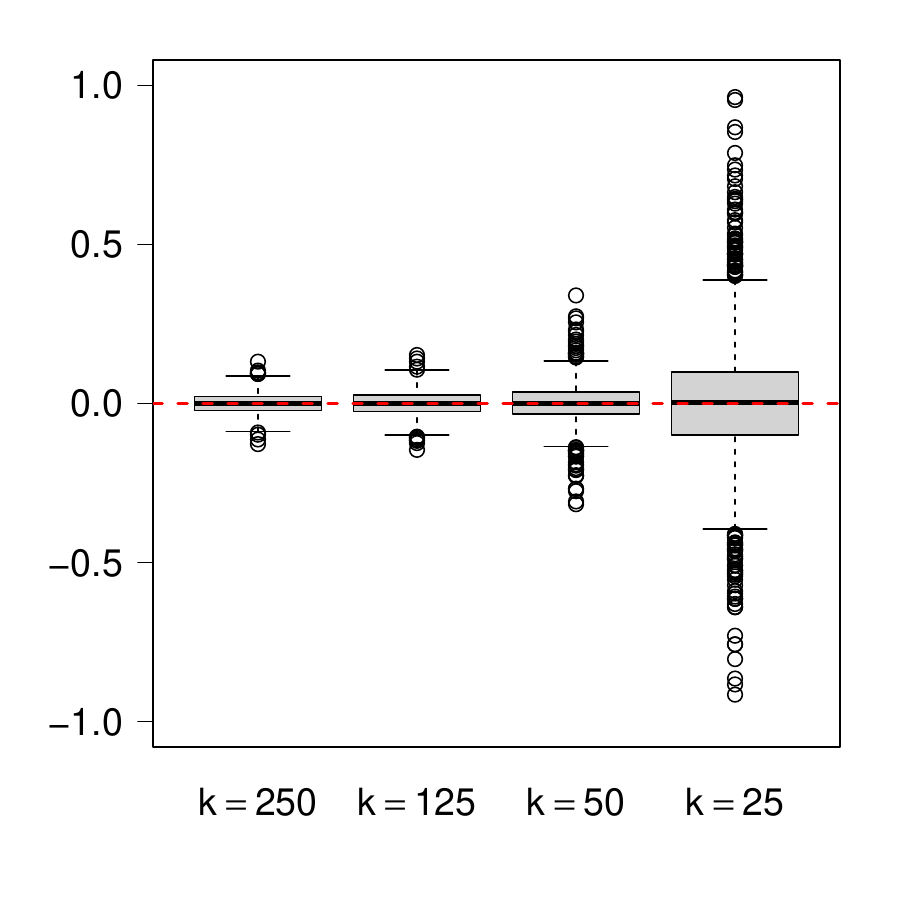}
\includegraphics[width=4cm]{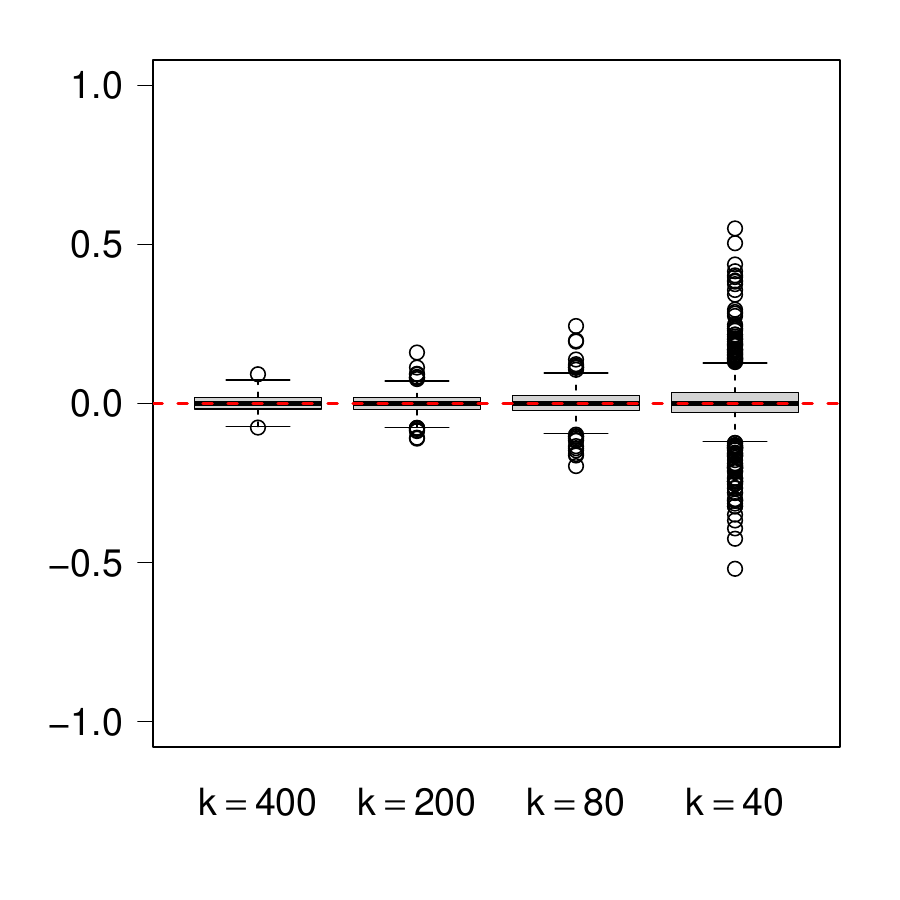}
\caption{\label{fig:condTPDM_est_AR1}
The top row shows box-plots of the off-diagonal element of the tail conditional covariance matrix estimates $[\widehat{\Sg}_{K|L}(n,k)]_{12}$, under $1,000$ iterations, across sample sizes $n\in\cbr{1500,2500,4000}$ from left to right.
The bottom row shows box-plots of the partial tail correlation estimates $\hat{\rho}(n,k)_{ij|L}$.
Red line segments indicate the specified values $[\Sg_{K|L}]_{12}=0$ and $\rho_{12|L}=0$.
}
\end{figure}

Our primary interest is in the asymptotic behavior of the off-diagonal element of $[\widehat{\Sg}_{K|L}](n,k)_{12}$ and $\hat{\rho}_{ij|L}(n,k)$.
Figure~\ref{fig:condTPDM_est_AR1} presents box-plots of these estimates across different $(n,k)$, showing reduced bias and variability for larger sample sizes.
